\documentclass[aps,prb,twocolumn,10pt,superscriptaddress,showpacs]{revtex4-2}
\usepackage{amsmath,amssymb,graphics,epsfig,epstopdf,color,verbatim,ulem,braket,tabularx,multirow,array,diagbox,colortbl,hhline}
\usepackage[colorlinks,linkcolor=blue,citecolor=blue,urlcolor=blue]{hyperref}
\usepackage{physics}
\usepackage{bm}
\usepackage{blindtext}
\graphicspath{{Figures/}}

\begin{document}

\title{Ising phase transitions and thermodynamics of correlated fermions \\
in a two-dimensional spin-dependent lattice potential}

\author{Zhuotao Xie}
\affiliation{Institute of Modern Physics, Northwest University, Xi'an 710127, China}

\author{Yu-Feng Song}
\email{jlyfsong@mail.ustc.edu.cn}
\affiliation{Institute of Modern Physics, Northwest University, Xi'an 710127, China}
\affiliation{Hefei National Laboratory for Physical Sciences at Microscale and Department of Modern Physics, University of Science and Technology of China, Hefei, Anhui 230026, China}

\author{Yuan-Yao He}
\email{heyuanyao@nwu.edu.cn}
\affiliation{Institute of Modern Physics, Northwest University, Xi'an 710127, China}
\affiliation{Shaanxi Key Laboratory for Theoretical Physics Frontiers, Xi'an 710127, China}
\affiliation{Peng Huanwu Center for Fundamental Theory, Xian 710127, China}
\affiliation{Hefei National Laboratory, Hefei 230088, China}

\begin{abstract}
We present a {\it numerically exact} study of the Hubbard model with spin-dependent anisotropic hopping on the square lattice using auxiliary-field quantum Monte Carlo method. At half filling, the system undergoes Ising phase transitions upon cooling, leading to the formation of Ising-type antiferromagnetic order for repulsive interactions and charge-density wave order for attractive interactions at finite temperatures. By elegantly implementing the sign-problem-free condition and Hubbard-Stratonovich transformations, we achieve significant improvements in precision control of the numerical calculations and obtain highly accurate results of the transition temperatures from weak to strong interactions across representative anisotropies. We further characterize the system by examining the temperature dependence of various thermodynamic properties, including the energy, double occupancy, specific heat and charge susceptibility. Specifically, we provide unbiased numerical results of the entropy map on temperature-interaction plane, the critical entropy, and the spin, singlon and doublon correlations, all of which are directly measurable in optical lattice experiments. Away from half filling, we explore the behavior of the sign problem and investigate the possible emergence of stripe spin-density wave order in the system with repulsive interaction.
\end{abstract}

\date{\today}

\maketitle

\section{Introduction}
\label{sec:intro}

Since the mid-2010s, quantum simulation with ultracold atoms in optical lattice~\cite{Anna2007,Bloch2008,Christian2017,Florian2020} has achieved remarkable strides in studying the fundamental properties of strongly correlated fermionic systems, with the Hubbard model~\cite{Hubbard1963,Kanamori1963,Gutzwiller1963} being a central focus. These advancements are gradually bridging the gap between theoretical predictions~\cite{Arovas2022,Qin2022} and experimental realizations, offering unprecedented insights into quantum many-body physics. A landmark example is the most recent progress~\cite{Shao2024} in directly observing the celebrated antiferromagnetic (AFM) phase transition in the three-dimensional (3D) Hubbard model through an optical lattice experiment. Although this N\'{e}el transition and its critical properties have been extensively investigated in numerous previous theoretical works~\cite{Kakehashi1987,Tahvildar1997,Staudt2000,Werner2005,Kent2005,Rohringer2011,Fuchs2011L,Kozik2013,Khatami2016,Katanin2017,Rampon2024,Lenihan2022,Iskakov2022,Fanjie2024,Song2024a,Song2024b}, their experimental realization holds significant importance and has garnered widespread attention~\cite{Niaz2024,Miller2024,Yuanyao2025}. It not only validates theoretical predictions but also demonstrates the crucial ability to build large-scale and nearly uniform optical lattice, reach sufficiently low temperatures and control the interaction strength and fermion filling of the system~\cite{Shao2024}. This opens up new possibilities for experimentally exploring the model with doping~\cite{Tahvildar1997,Katanin2017,Lenihan2022}, which remains considerably challenging for cutting-edge quantum many-body computations. 

In two-dimensional (2D) optical lattice experiments utilizing single-site and spin-resolved imaging techniques, the main effort has been devoted to studying AFM correlation properties of the standard Hubbard model~\cite{Daniel2013,Hart2015,Parsons2016,Boll2016,Lawrence2016,Mazurenko2017,Brown2017,Christie2019,Koepsell2019,Joannis2021,Xu2023,Prichard2024}. However, as the key distinction from its 3D counterpart, the AFM ordering and associated N\'{e}el transition which spontaneously break spin SU(2) symmetry is absent at finite temperatures in the 2D Hubbard model, as dictated by the Mermin-Wagner theorem~\cite{Mermin1966}. Although different regimes can be qualitatively distinguished upon cooling in the model~\cite{Thomas2021}, the AFM correlations remain short-ranged as bounded by the finite temperature, and the temperature-interaction $(T$-$U)$ phase diagram at half filling is characterized by smooth crossovers~\cite{LeBlanc2020,Aaram2020}. Since the atom cooling is currently the key challenge in optical lattice experiments~\cite{Christian2017,Florian2020}, the absence of AFM phase transition diminishes the prominence of experimental signatures and could present a significant obstacle in tackling more complex problems in Hubbard model, such as revealing the deep connections between quantum magnetism and high-temperature superconductivity~\cite{Arovas2022,Qin2022}. 

To address the above issue, one potential solution is to introduce a spin-dependent anisotropy in the hopping term of the 2D Hubbard model~\cite{Feiguin2009,Feiguin2011,Huang2013,Gukelberger2014,Gukelberger2017}. This modification breaks the spin SU(2) symmetry down to a discrete $\mathbb{Z}_2$ symmetry, thereby enabling a finite-temperature Ising phase transition~\cite{Gukelberger2017}. In experiment, such a lattice model can either be directly realized through spin-dependent optical lattices~\cite{Mandel2003,Vincent2004,Jotzu2015,Yang2017,Kuzmenko2019}, or equivalently built using higher orbitals in an optical lattice~\cite{Torben2007,Zhao2008,Congjun2008,Wirth2011,Hung2011}, or optical tweezer arrays~\cite{Wei2024}. Specifically for the former, the spin-dependent hopping anisotropy can be created by a magnetic gradient modulation~\cite{Jotzu2015} or a vector light shift~\cite{Yang2017}. These techniques offer the opportunities to observe the thermal phase transitions for correlated fermions in 2D optical lattice, similar to its 3D analog~\cite{Shao2024}. On the theoretical aspect, only a limited number of studies~\cite{Feiguin2009,Feiguin2011,Huang2013,Gukelberger2014,Gukelberger2017} have explored such a modified 2D Hubbard model. Most of them concentrated on the exotic paired states~\cite{Feiguin2009,Feiguin2011} or superfluidities~\cite{Huang2013,Gukelberger2014}, and the incommensurate density wave~\cite{Gukelberger2014} with attractive interactions away from half filling. More relevantly, in Ref.~\onlinecite{Gukelberger2017}, the Ising AFM phase transition in the half-filled model with repulsive interactions was investigated using quantum Monte Carlo algorithms. Nevertheless, the study only presented numerical results for a rather weak interaction of $U=3t$ and did not address any other thermodynamic properties beyond the phase transition.

In this work, we perform a systematic study for finite-temperature properties of the 2D Hubbard model with spin-dependent anisotropic hopping, implementing the {\it numerically exact} auxiliary-field quantum Monte Carlo (AFQMC) method~\cite{Blankenbecler1981,Hirsch1983,White1989,Scalettar1991,McDaniel2017,Sun2024,Yuanyao2019,Yuanyao2019L,Assaad2008,Chang2015}. At half filling, we first verify the sign-problem-free condition and test different Hubbard-Stratonovich (HS) transformations to achieve high-precision results. Then we focus on the Ising phase transitions and various thermodynamic properties of the model covering a wide range of the interaction strength. We also present numerical results of the entropy map on $T$-$U$ plane (including the critical entropy) and several kinds of site-resolved correlation functions, which can provide comparison and benchmark for future optical lattice experiments. Away from half filling, we extend our exploration to examine the potential stripe spin-density wave (SDW) ordering in the model with repulsive interactions at $1/8$-hole doping.

The rest of this paper is organized as follows. In Sec.~\ref{sec:modelmethod}, we describe the lattice model, AFQMC method with essential ingredients, and the physical quantities we compute. In Sec.~\ref{sec:HalfResults}, we concentrate on our numerical results for the Ising phase transitions and thermodynamic properties of the model at half filling. In Sec.~\ref{sec:DopeResults}, we study the sign problem and magnetic correlations for the doping case. Finally, in Sec.~\ref{sec:Summary}, we summarize our work and discuss the future opportunities for the modified Hubbard model in our study. The Appendixes contain important derivations for noninteracting spin susceptibilities, strong interaction limit, general AFQMC formalism, and supplementary data for this work.

\section{Model, Method, and physical observables}
\label{sec:modelmethod}

In this section, we first describe the lattice model, analyze its symmetry properties, and discuss its fundamental physics at half filling (Sec.~\ref{sec:TheModel}). We then introduce the key components of the AFQMC method used in our work, focusing on the sign-problem-free condition and examining the impact of various HS transformations (Sec.~\ref{sec:AFQMCmethod}). Finally, we define the physical observables and address important details in AFQMC calculations (Sec.~\ref{sec:AFQMCObs}).

\subsection{The modified Hubbard model}
\label{sec:TheModel}

\begin{figure}[b]
\includegraphics[width=0.99\columnwidth]{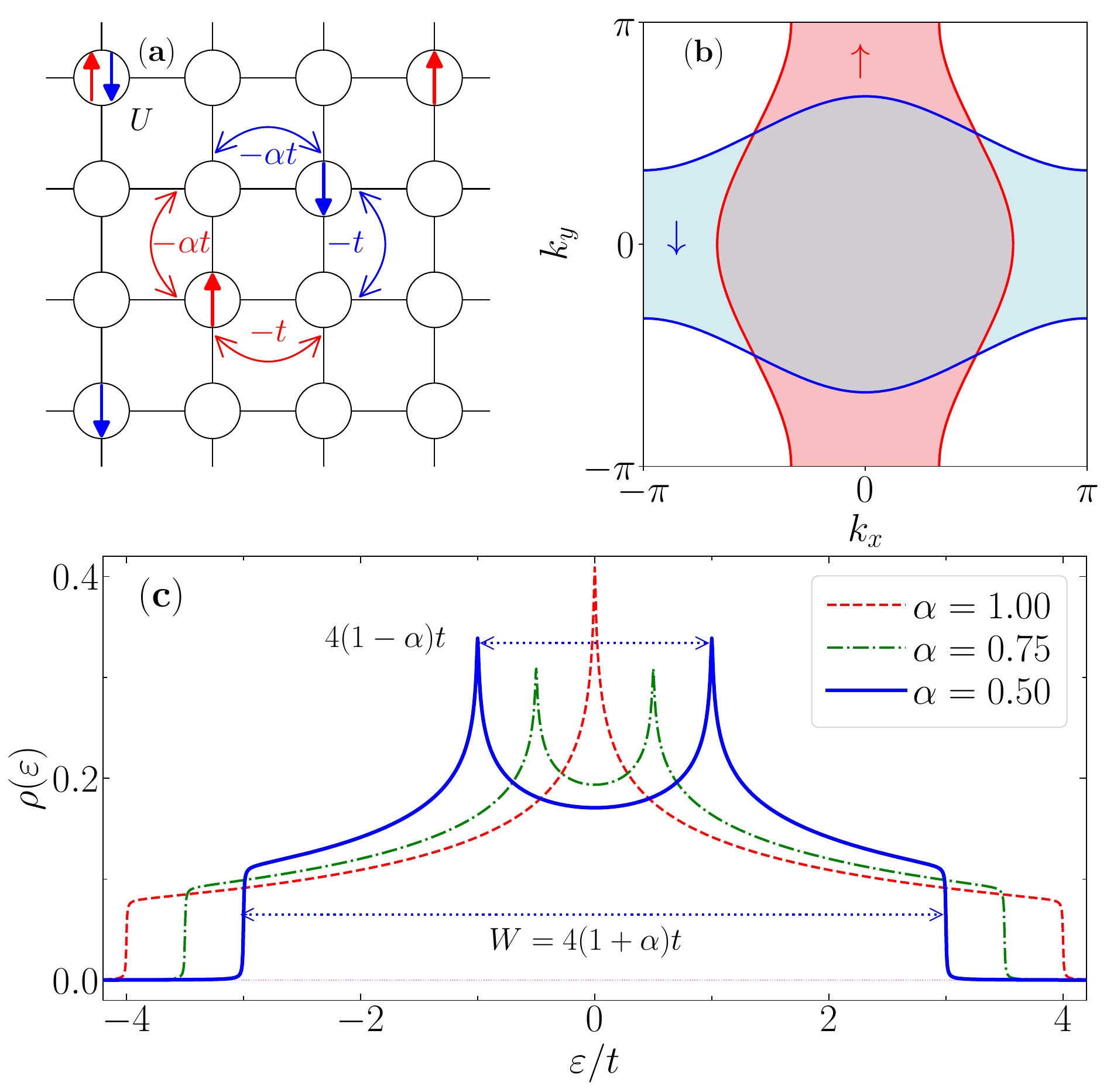}
\caption{(a) A schematic demonstration of the 2D Hubbard model with spin-dependent anisotropic hopping. (b) The spin-nematic Fermi surfaces for the two spin species with $\mu=0$ shown for $\alpha=0.50$. (c) The local density of states $\rho(\epsilon)$ for the noninteracting part of the model with $\mu=0$ shown for $\alpha=1.00,0.75$, and $0.50$. The energy distance between the two diverging peaks is $4(1-\alpha)t$, while the bandwidth is $W=4(1+\alpha)t$.}
\label{fig:ModelDemon}
\end{figure}

We study the 2D Hubbard model with spin-dependent anisotropic hopping~\cite{Feiguin2009,Feiguin2011,Huang2013,Gukelberger2014,Gukelberger2017} on square lattice described by the following Hamiltonian:
\begin{equation}\begin{aligned}
\label{eq:Model}
\hat{H}= & -\sum_{\mathbf{i},\sigma}\sum_{\mathbf{v}\in\{\mathbf{x}, \mathbf{y}\}} t_{\mathbf{v}\sigma}\big(c_{\mathbf{i},\sigma}^+ c_{\mathbf{i}+\mathbf{v},\sigma}^{} + c_{\mathbf{i}+\mathbf{v},\sigma}^+ c_{\mathbf{i}, \sigma}^{}\big) \\
& +U \sum_{\mathbf{i}}\Big(\hat{n}_{\mathbf{i}\uparrow}\hat{n}_{\mathbf{i}\downarrow}-\frac{\hat{n}_{\mathbf{i}\uparrow}+\hat{n}_{\mathbf{i}\downarrow}}{2}\Big) + \mu\sum_{\mathbf{i}}\hat{n}_{\mathbf{i}},
\end{aligned}\end{equation}
where $\sigma$ ($=\uparrow$ or $\downarrow$) denotes spin, $\mathbf{i}=(i_x,i_y)$ are the coordinates of the lattice site, and $\hat{n}_{\mathbf{i}\sigma}=c_{\mathbf{i},\sigma}^+c_{\mathbf{i},\sigma}^{}$ is the density operator with $\hat{n}_{\mathbf{i}}=\hat{n}_{\mathbf{i}\uparrow}+\hat{n}_{\mathbf{i}\downarrow}$. The chemical potential term $\mu$ is a pure doping, and the system is half-filled as the fermion filling $n=1$ for $\mu=0$. The on-site interaction can be repulsive ($U>0$) or attractive ($U<0$). The spin-dependent anisotropy in nearest-neighbor (NN) hopping term is defined via the hopping amplitudes as $\alpha = t_{\mathbf{x}\downarrow}/t_{\mathbf{x}\uparrow} = t_{\mathbf{y}\uparrow}/t_{\mathbf{y}\downarrow}$, and $\alpha\in[0,1]$ serves as another tuning parameter of the above model. Then $\alpha=1$ recovers the isotropic model, and $\alpha=0$ corresponds to the fully anisotropic case for which the kinetic part is purely one dimensional. We take $t_{\mathbf{x}\uparrow} = t_{\mathbf{y}\downarrow} = t$ and $t_{\mathbf{x}\downarrow}=t_{\mathbf{y}\uparrow}=\alpha t$ [see the hopping pattern in Fig.~\ref{fig:ModelDemon}(a)] and reach the kinetic energy dispersions as
\begin{equation}\begin{aligned}
\label{eq:EkDisp}
&\varepsilon_{\mathbf{k},\uparrow}   = -2t(\cos k_x + \alpha\cos k_y) \\
&\varepsilon_{\mathbf{k},\downarrow} = -2t(\alpha\cos k_x + \cos k_y),
\end{aligned}\end{equation}
where the momentum $k_x,k_y$ are defined in units of $2\pi/L$ with the system size $N_s=L^2$. This leads to the spin-nematic Fermi surfaces~\cite{Gukelberger2014} (or mismatched Fermi surfaces~\cite{Gukelberger2017}) as shown in Fig.~\ref{fig:ModelDemon}(b) for $\alpha=0.5$. Moreover, comparing to the isotropic case, the hopping anisotropy reduces the noninteracting bandwidth to $W=4(1+\alpha)t$ (with $\alpha\in[0,1]$), and splits the single diverging peak in local density of states into two symmetric peaks with energy distance $4(1-\alpha)t$, which are illustrated in Fig.~\ref{fig:ModelDemon}(c). Throughout this work, we set $t$ as the energy unit. 

The symmetry properties of the model described in Eq.~(\ref{eq:Model}) are quite intriguing. {\it First}, the spin SU(2) symmetry is clearly broken by the spin-dependent hopping anisotropy, but the conservation of the $z$-component total magnetization $\hat{S}^z=\sum_{\mathbf{i}}\hat{s}_{\mathbf{i}}^z$ is still preserved. This results in a residual spin $U(1)$ symmetry corresponding to the spin rotation in the $x$-$y$ plane. Besides, the model Hamiltonian is invariant under the following $\mathbb{Z}_2$ transformation $\hat{Q}=\hat{R}_{\rm spin}\times\hat{C}_{x\leftrightarrow y}$, where $\hat{R}_{\rm spin}$ stands for the spin inversion and $\hat{C}_{x\leftrightarrow y}$ represents the spatial exchange of $x$ and $y$ axes (which can be space rotation by 90$^\circ$~\cite{Gukelberger2014} or the mirror reflection about the diagonal of the square lattice). We note that this $\mathbb{Z}_2$ symmetry is preserved by the Hamiltonian regardless of the fermion filling. {\it Second}, the hopping anisotropy also breaks the charge SU(2) symmetry of the isotropic model ($\alpha=1$) at half filling (see Appendix E of Ref.~\onlinecite{Cornelia2021}), which guarantees the degeneracy between charge-density wave (CDW) order and $s$-wave spin-singlet pairing order. The lifting of the degeneracy is clear since the mismatched Fermi surfaces of the two spin species suppress $s$-wave spin-singlet pairing. {\it Third}, the hopping anisotropy further reduces the spatial symmetry to the point group of a rectangle which can pose constraints on the fermion pairing symmetry~\cite{Gukelberger2014} regarding the attractive interaction away from half filling. The translational and spatial inversion symmetries remain unchanged. {\it Fourth}, the particle-hole symmetry of the isotropic model at half filling is still preserved with the hopping anisotropy, as indicated by the symmetric local density of states shown in Fig.~\ref{fig:ModelDemon}(c).

At half filling, the ground state of the model (\ref{eq:Model}) with $U>0$ can be analyzed from both the weakly and strongly interacting limits. At $U=0$, although the hopping anisotropy removes the divergence in the density of states at the Fermi energy [see Fig.~\ref{fig:ModelDemon} (c)], the perfect Fermi surface nesting still holds as $\varepsilon_{\mathbf{k}+\mathbf{M},\sigma}=-\varepsilon_{\mathbf{k},\sigma}$ with the vector $\boldsymbol{\mathbf{M}}=(\pi,\pi)$. As a result, the longitudinal spin susceptibility $\chi^{zz}(\mathbf{M})$ diverges logarithmically as approaching $T=0$ while the transverse spin susceptibility $\chi^{xy}(\mathbf{M})$ saturates to a constant (see Appendix~\ref{sec:AppendixA}). Thus, the system can develop the AFM long-range order in spin-$\hat{s}^z$ channel (namely, the Ising AFM order) at infinitesimal repulsive interaction in the ground state. In strongly interacting limit, the charge degree of freedom is frozen and the model (\ref{eq:Model}) reduces to a spin-$1/2$ XXZ model with Ising anisotropy (see Appendix~\ref{sec:AppendixB}) as 
\begin{equation}\begin{aligned}
\label{eq:XXZModel}
\hat{H}_{\rm XXZ} = \sum_{\langle\mathbf{i}\mathbf{j}\rangle}\Big[ J_{xx}\big(\hat{S}^x_{\mathbf{i}}\hat{S}^x_{\mathbf{j}}+\hat{S}^y_{\mathbf{i}}\hat{S}^y_{\mathbf{j}}\big)+J_{z}\hat{S}^z_{\mathbf{i}}\hat{S}^z_{\mathbf{j}}\Big],
\end{aligned}\end{equation}
where $J_{xx}=4\alpha t^2/U$ and $J_z=2(1+\alpha^2)t^2/U$, resulting in $J_z/J_{xx}=(1+\alpha^2)/(2\alpha)\ge1$. Thus, with the hopping anisotropy $\alpha\ne1$, the above XXZ model favors an Ising AFM ordered ground state. At intermediate $U$, the system undergoes a smooth crossover between the two limits. Combining these results, the ground state of the half-filled model in Eq.~(\ref{eq:Model}) with $U>0$ always has Ising AFM long-range order, which spontaneously breaks the aforementioned $\mathbb{Z}_2$ symmetry and hence allows for finite-temperature phase transitions even in two dimensions.

We next connect the repulsive model discussed above to the attractive case at half filling ($\mu=0$). Let us consider a partial particle-hole transformation $\hat{P}$ which only involves spin-down fermions as
\begin{equation}\begin{aligned}
\label{eq:PartialPH}
\hat{P}^+ c_{\mathbf{i},\downarrow}^+\hat{P} &= (-1)^{\mathbf{i}} c_{\mathbf{i},\downarrow} \\
\hat{P}^+ c_{\mathbf{i},\downarrow}\hat{P} &= (-1)^{\mathbf{i}} c_{\mathbf{i},\downarrow}^+,
\end{aligned}\end{equation}
where $(-1)^{\mathbf{i}}=(-1)^{i_x+i_y}$ meaning the site parity. Note that this is different from the particle-hole symmetry for which the symmetry operation acts on fermions of both spin species. Then it is straightforward to verify that the operator $\hat{P}$ maps the repulsive model into an attractive one as $\hat{P}^+\hat{H}(U)\hat{P}=\hat{H}(-U)$ (with the hopping term unchanged)~\cite{Mingpu2016}. Correspondingly, the operator $\hat{P}$ transforms the Ising AFM order $\hat{m}=N_s^{-1}\sum_{\mathbf{i}}(-1)^{\mathbf{i}}\hat{s}_{\mathbf{i}}^z$ [with $\hat{s}_{\mathbf{i}}^z=(\hat{n}_{\mathbf{i}\uparrow}-\hat{n}_{\mathbf{i}\downarrow})/2$] in the repulsive model into the CDW order $\hat{d}=N_s^{-1}\sum_{\mathbf{i}}(-1)^{\mathbf{i}}(\hat{n}_{\mathbf{i}\uparrow}+\hat{n}_{\mathbf{i}\downarrow}-1)/2$ of the attractive model. Thus, the ground state of the model (\ref{eq:Model}) with $U<0$ has the CDW long-range order, which spontaneously breaks the translational symmetry (and also the spatial inversion symmetry), which is also a $\mathbb{Z}_2$ symmetry. We have used this mapping property between the $U>0$ and $U<0$ situations in our AFQMC simulations.

An alternative approach to break the spin SU(2) symmetry and realize the finite-temperature Ising phase transition is to introduce the spin-anisotropic hopping, where $t_{x\uparrow}=t_{y\uparrow}\ne t_{x\downarrow}=t_{y\downarrow}$, known as mass-imbalanced Hubbard model~\cite{LiuYeHua2015}. At half filling, its essential physics closely resemble those of the model (\ref{eq:Model}), including the Ising phase transition and the strong-coupling limit corresponding to the XXZ model. However, once with doping ($\mu\ne0$), the spin-up and spin-down Fermi surfaces become distinct, and a spin imbalance naturally arises, with the fermion fillings $n_{\uparrow}\ne n_{\downarrow}$. This imbalance is likely to suppress many intriguing phenomena present in the spin-balanced system. Moreover, in optical lattice experiments, spin imbalance can arise even near half filling due to the realistic lattice density disorder~\cite{Shao2024}, which effectively acts as local doping~\cite{Song2024c}. This, in turn, poses additional challenges for experiments aiming to observe the Ising AFM phase transition. In contrast, the model (\ref{eq:Model}) maintains spin balance even with doping, thereby avoiding the aforementioned issues. Therefore, we simply bypass the mass-imbalanced Hubbard model, and focus on the model (\ref{eq:Model}) in this work.

\subsection{Finite-temperature AFQMC algorithm}
\label{sec:AFQMCmethod}

The core principle of AFQMC algorithm~\cite{Blankenbecler1981,Hirsch1983,White1989,Scalettar1991,McDaniel2017,Sun2024,Yuanyao2019,Yuanyao2019L,Assaad2008,Chang2015} is to decompose the two-body interaction into noninteracting fermions coupled to auxiliary fields using HS transformations~\cite{Hirsch1983}, and then to compute the fermionic observables through importance sampling of these auxiliary fields. Among all ingredients of AFQMC algorithm, the HS transformation plays a central role, as it is directly linked to both the sign problem and the statistical uncertainty of specific observables. This point is particularly prominent in the AFQMC simulations for the 2D Hubbard modified model in Eq.~(\ref{eq:Model}).

The workflow of AFQMC algorithm starts from discretizing the inverse temperature as $\beta=M\Delta\tau$ for partition function $Z={\rm Tr}(e^{-\beta\hat{H}})={\rm Tr}[(e^{-\Delta\tau\hat{H}})^M]$ and then applies the Trotter-Suzuki decomposition for $e^{-\Delta\tau\hat{H}}$, such as the third-order formula
\begin{equation}\begin{aligned}
\label{eq:SymTrot}
e^{-\Delta\tau\hat{H}}=e^{-\Delta\tau\hat{H}_0/2}e^{-\Delta\tau\hat{H}_I}e^{-\Delta\tau\hat{H}_0/2}+O[(\Delta\tau)^3],
\end{aligned}\end{equation}
where $\hat{H}=\hat{H}_0+\hat{H}_I$ and $\hat{H}_0$ and $\hat{H}_I=U\sum_{\mathbf{i}}[\hat{n}_{\mathbf{i}\uparrow}\hat{n}_{\mathbf{i}\downarrow}-(\hat{n}_{\mathbf{i}\uparrow}+\hat{n}_{\mathbf{i}\downarrow})/2]$ are the kinetic and interaction parts, respectively. The Trotter error is practically eliminated by extrapolating results towards the $\Delta\tau\to0$ limit. Then the HS transformation is further used to deal with the $e^{-\Delta\tau\hat{H}_I}$ term. The available choices include the continuous-field Gaussian transformation~\cite{Shihao2013} and the discrete transformation with four-component~\cite{Assaad1998,WangDa2014} and two-component fields~\cite{Hirsch1983}. For the on-site Hubbard interaction, all these HS transformations are closely related to the symmetries of $\hat{H}_I$. More specifically, it has the spin SU(2) symmetry rotating the spin vector $\boldsymbol{\hat{s}}_\mathbf{i}=(\hat{s}_{\mathbf{i}}^x,\hat{s}_{\mathbf{i}}^y,\hat{s}_{\mathbf{i}}^z)$, and the charge SU(2) symmetry~\cite{Cornelia2021} for the vector $\boldsymbol{\hat{N}}_{\mathbf{i}}=({\rm Re}\hat{\Delta}_{\mathbf{i}}, {\rm Im}\hat{\Delta}_{\mathbf{i}}, \hat{T}_{\mathbf{i}})$ consisting of spin-singlet pairing $\hat{\Delta}_{\mathbf{i}}=c_{\mathbf{i},\uparrow}^+c_{\mathbf{i},\downarrow}^+$ and charge-density $\hat{T}_{\mathbf{i}}=\frac{1}{2}(-1)^{\mathbf{i}}(\hat{n}_{\mathbf{i}\uparrow}+\hat{n}_{\mathbf{i}\downarrow}-1)$. Thus, corresponding to the spin symmetry, there exists HS transformations into spin-$\hat{s}^z$, $\hat{s}^x$, and $\hat{s}^y$ channels (denoted as HS-$\hat{s}^z$, HS-$\hat{s}^x$, and HS-$\hat{s}^y$) reading
\begin{equation}\begin{aligned}
\label{eq:HSspinDecomp}
& e^{-\Delta\tau U \big(\hat{n}_{\mathbf{i}\uparrow} \hat{n}_{\mathbf{i}\downarrow} - \frac{\hat{n}_{\mathbf{i}\uparrow} + \hat{n}_{\mathbf{i}\downarrow}}{2}\big) }
= C_s\sum_{x_{\mathbf{i}}=\pm1}e^{\gamma_s x_{\mathbf{i}}(\hat{n}_{\mathbf{i}\uparrow}-\hat{n}_{\mathbf{i}\downarrow})} \\
&\hspace{2.5cm} = C_s\sum_{x_{\mathbf{i}}=\pm1}e^{\gamma_s x_{\mathbf{i}}(c_{\mathbf{i},\uparrow}^+ c_{\mathbf{i},\downarrow}^{} + c_{\mathbf{i},\downarrow}^+ c_{\mathbf{i},\uparrow}^{} )} \\
&\hspace{2.5cm} = C_s\sum_{x_{\mathbf{i}}=\pm1}e^{i\gamma_s x_{\mathbf{i}}(c_{\mathbf{i},\uparrow}^+ c_{\mathbf{i},\downarrow}^{} - c_{\mathbf{i},\downarrow}^+ c_{\mathbf{i},\uparrow}^{})},
\end{aligned}\end{equation}
with the constant $C_s=1/2$, and the coupling coefficient $\gamma_s=\cosh^{-1}(e^{+\Delta\tau U/2})$ for $U>0$ and $\gamma_s=i\cos(e^{+\Delta\tau U/2})$ for $U<0$. Related to the charge symmetry, the Hubbard interaction can instead be decomposed into free fermions in charge-density and pairing channels coupled to auxiliary fields as
\begin{equation}\begin{aligned}
\label{eq:HScharge}
& e^{-\Delta\tau U \big(\hat{n}_{\mathbf{i}\uparrow} \hat{n}_{\mathbf{i}\downarrow} - \frac{\hat{n}_{\mathbf{i}\uparrow} + \hat{n}_{\mathbf{i}\downarrow}}{2}\big) }
= C_c \sum_{x_{\mathbf{i}} = \pm 1} e^{\gamma_c x_{\mathbf{i}} (\hat{n}_{\mathbf{i}\uparrow} + \hat{n}_{\mathbf{i}\downarrow} - 1)} \\
&\hspace{1.9cm} = C_c \sum_{x_{\mathbf{i}} = \pm 1} e^{\gamma_c x_{\mathbf{i}} (c_{\mathbf{i},\uparrow}^+c_{\mathbf{i},\downarrow}^+ + c_{\mathbf{i},\downarrow}^{} c_{\mathbf{i},\uparrow}^{})} \\
&\hspace{1.9cm} = C_c \sum_{x_{\mathbf{i}} = \pm 1} e^{i\gamma_c x_{\mathbf{i}} (c_{\mathbf{i},\uparrow}^+c_{\mathbf{i},\downarrow}^+ - c_{\mathbf{i},\downarrow}^{} c_{\mathbf{i},\uparrow}^{})},
\end{aligned}\end{equation}
with the constant $C_c=e^{+\Delta\tau U/2}/2$, and the coupling coefficient $\gamma_c=i\cos^{-1}(e^{-\Delta\tau U/2})$ for $U>0$ and $\gamma_c=\cosh^{-1}(e^{-\Delta\tau U/2})$ for $U<0$. Nevertheless, the HS transformations into the pairing channels explicitly violate the charge conservation, and thus necessitates the extension of conventional AFQMC framework~\cite{Blankenbecler1981,Assaad2008} to either the Majorana representation~\cite{ZiXiang2015,Yao2024} or the Hartree-Fock-Bogoliubov space~\cite{Shihao2017}. Therefore, in the following discussion, we only focus on the HS transformation into charge-density channel (denoted as HS-$\hat{n}$) in Eq.~(\ref{eq:HScharge}). After all the above procedures, the partition function can be evaluated as $Z\simeq\sum_{\mathbf{X}}W(\mathbf{X})$, where $\mathbf{X}$ is the auxiliary-field configuration and $W(\mathbf{X})=\det[\mathbf{M}(\mathbf{X})]$ is the corresponding weight with $\mathbf{M}(\mathbf{X})$ generally as a $2N_s\times 2N_s$ matrix (see more details in Appendix~\ref{sec:AppendixC}). 

\begin{table}[t]
\centering
\caption{Summary of the data structures and sign problem situations at half filling for AFQMC simulations of the model (\ref{eq:Model}) with hopping anisotropy ($\alpha\ne1$). The two lines marked by yellow color are actually used in the large-scale AFQMC simulations in this work.}
\begin{tabular}{|c|c|c|c|c|}
\hline  
{HS type} & $U$ & Data Type & Spin & Sign problem  \\ 
\hline
\multirow{2}{*}{HS-$\hat{n}$} & $U>0$ & Complex & Decoupled & Phase problem \\ \cline{2-5}
& $U<0$ & Real & Decoupled & Sign problem \\ \hline
\multirow{2}{*}{HS-$\hat{s}^{x}$} & \cellcolor{yellow}$U>0$ & \cellcolor{yellow}Real & \cellcolor{yellow} Coupled & \cellcolor{yellow} No sign problem \\ \cline{2-5}
& $U<0$ & Complex & Coupled & No sign problem \\ \hline
\multirow{2}{*}{HS-$\hat{s}^{y}$} & $U>0$ & Complex & Coupled & No sign problem \\ \cline{2-5}
& \cellcolor{yellow}$U<0$ & \cellcolor{yellow}Real & \cellcolor{yellow}Coupled & \cellcolor{yellow}No sign problem \\ \hline
\multirow{2}{*}{HS-$\hat{s}^{z}$} & $U>0$ & Real & Decoupled & Sign problem \\ \cline{2-5}
& $U<0$ & Complex & Decoupled & Phase problem \\
\hline
\end{tabular}
\label{Table:A1}
\end{table}

In Table~\ref{Table:A1}, we summarize the data structures and sign problem situations for AFQMC simulations of the model (\ref{eq:Model}) {\it at half filling} with hopping anisotropy ($\alpha\ne1$), using four different HS transformations as HS-$\hat{s}^z$, HS-$\hat{s}^x$, HS-$\hat{s}^y$, and HS-$\hat{n}$. According to Eqs.~(\ref{eq:HSspinDecomp}) and (\ref{eq:HScharge}), the AFQMC calculations for $U>0$ must involve complex numbers with HS-$\hat{s}^y$ or HS-$\hat{n}$ due to the imaginary coupling $i\gamma_s$ or $i\gamma_c$, while only real numbers appear when using HS-$\hat{s}^z$ or HS-$\hat{s}^x$. For $U<0$, the conditions are reversed. Besides, we note the $\hat{H}_0$ term in Eq.~(\ref{eq:Model}) is spin decoupled. Within HS-$\hat{s}^z$ or HS-$\hat{n}$, all the quantities used in AFQMC calculations are spin decoupled such as $W(\mathbf{X})=W_{\uparrow}(\mathbf{X})\cdot W_{\downarrow}(\mathbf{X})$ with $W_{\sigma}(\mathbf{X})=\det[\mathbf{M}_{\sigma}(\mathbf{X})]$, where two $N_s\times N_s$ matrices $\mathbf{M}_{\uparrow}$ and $\mathbf{M}_{\downarrow}$ are involved. However, when HS-$\hat{s}^x$ or HS-$\hat{s}^y$ are applied, the two spin species are coupled, and all the matrices used in AFQMC should be $2N_s\times 2N_s$, which are exactly the same as that of systems with spin-orbit coupling~\cite{Shihao2016,Peter2017,Deng2024}. This increases the computational effort by a factor of 2 to 4 in different portions of the algorithm, compared to the spin-decoupled case.

\begin{figure}[t]
\centering
\includegraphics[width=0.99\columnwidth]{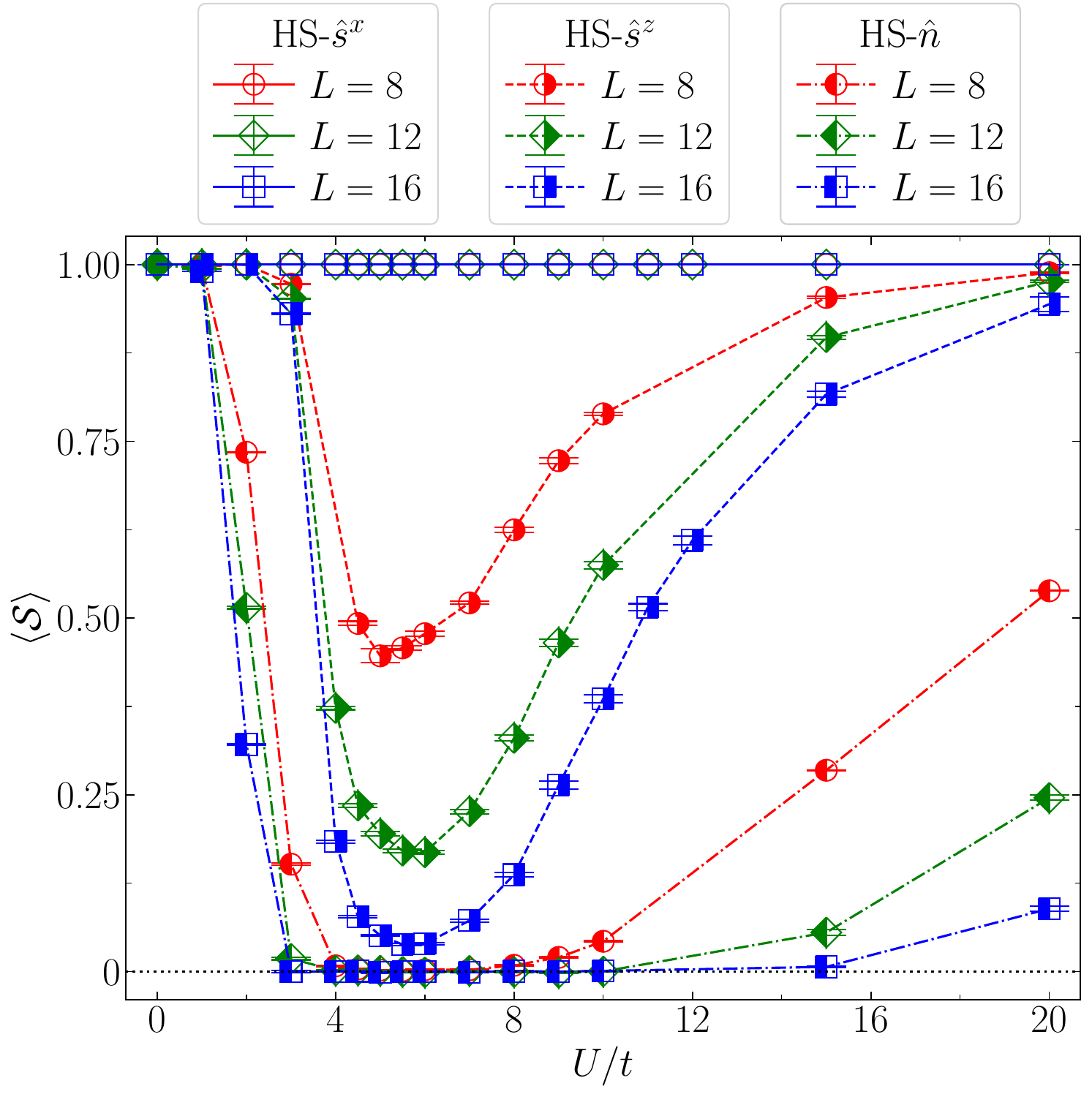}
\caption{The sign average $\langle\mathcal{S}\rangle$ as a function of interaction strength $U/t$ for the model (\ref{eq:Model}) at half filling with a hopping anisotropy $\alpha=0.5$, from AFQMC simulations with three different HS transformations, HS-$\hat{s}^x$, HS-$\hat{s}^z$, and HS-$\hat{n}$. For HS-$\hat{n}$, the real part of $\langle\mathcal{S}\rangle$ is plotted. Results from finite systems with $L=8$, $12$, and $16$ are plotted.}
\label{fig:SignHalfFill}
\end{figure}

Regarding the sign problem, the original Hubbard model [with $\alpha=1$ in Eq.~(\ref{eq:Model})] is sign-problem-free {\it at half filling} with HS-$\hat{s}^z$ or HS-$\hat{n}$ in AFQMC simulations, due to the particle-hole symmetry (for $U>0$) or the time-reversal symmetry (for $U<0$)~\cite{Wu2005}. With the hopping anisotropy ($\alpha\ne1$), the sign problem presents if HS-$\hat{s}^z$ or HS-$\hat{n}$ is still used. Specifically, for $U>0$ with HS-$\hat{n}$ and $U<0$ with HS-$\hat{s}^z$, the configuration weight $W(\mathbf{X})$ is complex and thus the phase problem (which is typically a more severe sign problem) appears. Moreover, another guiding principle of split orthogonal group~\cite{WangLei2015} proves that the modified model in Eq.~(\ref{eq:Model}) {\it at half filling} is also free of sign problem if HS-$\hat{s}^x$ or HS-$\hat{s}^y$ is used. This holds for both $\alpha=1$ and $\alpha\ne1$, and is irrespective of $U>0$ or $U<0$. To illustrate these sign problem situations, we present the numerical results of the sign average $\langle\mathcal{S}\rangle$ as a function of $U/t$ for $\alpha=0.5$ in Fig.~\ref{fig:SignHalfFill}, using HS-$\hat{s}^x$, HS-$\hat{s}^z$, and HS-$\hat{n}$ in AFQMC calculations. As expected, we observe $\langle\mathcal{S}\rangle=1$ for HS-$\hat{s}^x$ at arbitrary $U/t$, indicating no sign problem. Alternatively, the phase problem for HS-$\hat{n}$ is clearly more significant than the sign problem for HS-$\hat{s}^z$ as manifested by the smaller $\langle\mathcal{S}\rangle$ for the former case. Interestingly, for both HS-$\hat{n}$ and HS-$\hat{s}^z$, the sign/phase problem is most severe in the range of $4\le U/t\le 8$ as $\langle\mathcal{S}\rangle$ reaching the minimum, and after that $\langle\mathcal{S}\rangle$ anomalously increases with $U/t\ge 8$. This might be attributed to the fact that the contribution from the hopping anisotropy, as the origin of the sign/phase problem, is gradually diminished with increasing $U$. Away from half filling, there generally exists the sign/phase problem for all the HS transformations in Eqs.~(\ref{eq:HSspinDecomp}) and (\ref{eq:HScharge}). 

\begin{figure}[t]
\centering
\includegraphics[width=0.99\columnwidth]{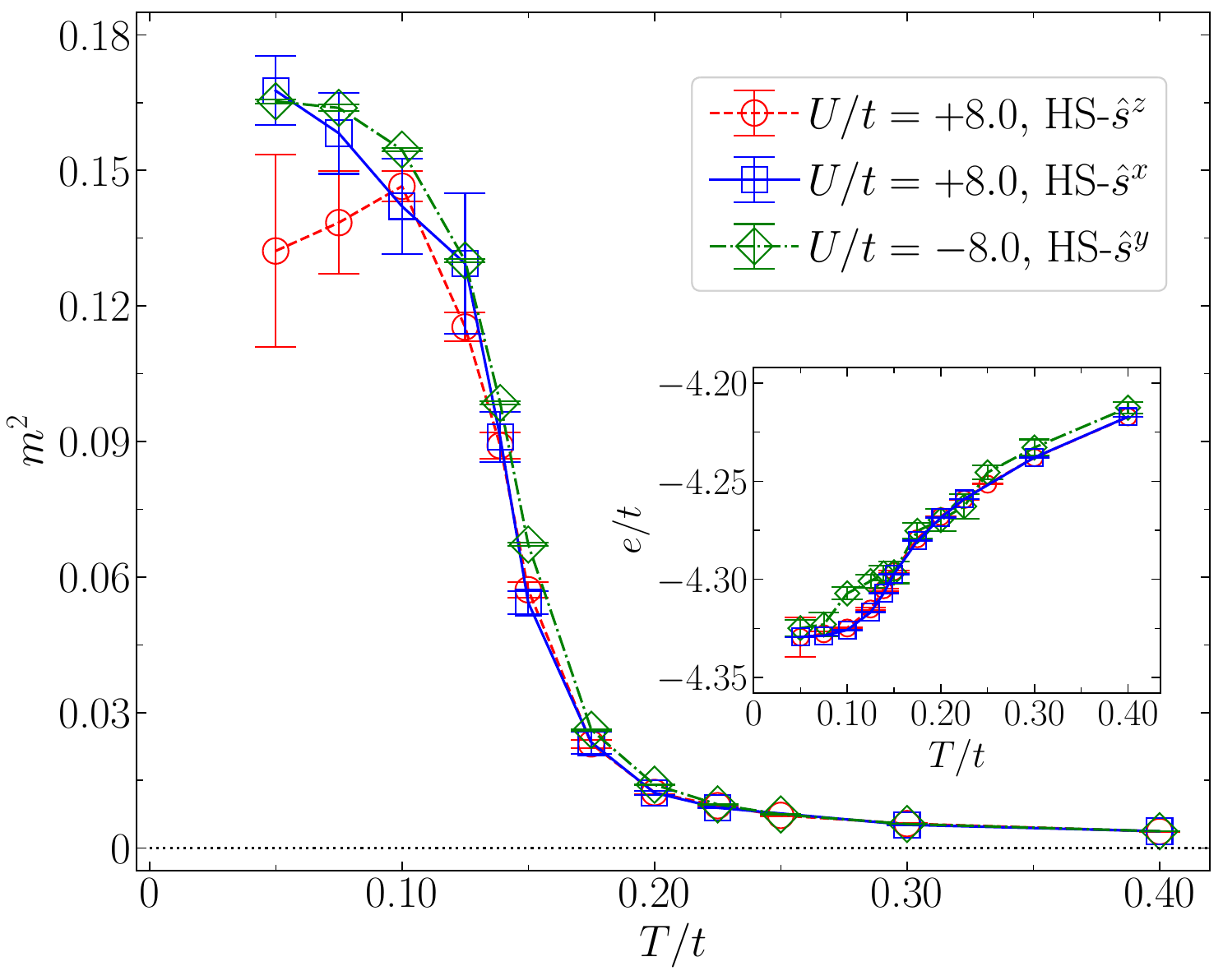} 
\caption{Comparisons of AFQMC results of the mean-squared magnetization $m^2=\langle\hat{m}^2\rangle$ for the Ising AFM order between different HS transformations for $\alpha=0.5$ and $U/t=8$ in the $L=12$ system. The inset plots the corresponding results of total energy per site $e/t$. For the third case [$U/t=-8$, HS-$\hat{s}^y$], both results of $m^2$ and $e/t$ for $U/t=8$ are obtained from those of $U/t=-8$ using HS-$\hat{s}^y$ via the partial particle-hole transformation (see discussions in the main text).}
\label{fig:CompareM2HS}
\end{figure}

Beyond the sign problem, the choice of HS transformation can significantly impact the precision of numerical results for physical observables~\cite{Assaad1998,Song2024b}. The HS-$\hat{s}^z$, HS-$\hat{s}^x$, and HS-$\hat{s}^y$ in Eq.~(\ref{eq:HSspinDecomp}) explicitly breaks the spin SU(2) symmetry, leading to substantial statistical fluctuations in spin-related observables. Similarly, the HS-$\hat{n}$ in Eq.~(\ref{eq:HScharge}) can induce similar effect for charge-density and pairing-related properties due to the charge symmetry breaking. As a result, it is preferable to use the HS transformation preserving the symmetry of the relevant observable in AFQMC simulations to achieve high-precision results. In Fig.~\ref{fig:CompareM2HS}, we demonstrate the above feature, and present the AFQMC results of the mean-squared magnetization $m^2=\langle\hat{m}^2\rangle$ for the Ising AFM order (and total energy per site $e/t$ in the inset) with different HS transformations, for $\alpha=0.5$ and $U/t=8$. It is, first and foremost, as expected that these calculations yield consistent results within the statistical uncertainties. For HS-$\hat{s}^z$, the growing error bars of the results with lowering temperature are indicative of the sign problem. For HS-$\hat{s}^x$, we observe notably noisy results of $m^2$, despite the absence of the sign problem in this case. This is attributed to the breaking of spin SU(2) symmetry in HS-$\hat{s}^x$ as discussed above. To obtain highly accurate results of AFM spin correlations and $m^2$, we implement the mapping property between the $U>0$ and $U<0$ cases (as described in Sec.~\ref{sec:TheModel}). As connected by the partial particle-hole transformation in Eq.~(\ref{eq:PartialPH}), the Ising AFM order $\hat{m}$ for $U>0$ is equivalent to the CDW order $\hat{d}$ for $U<0$. Consequently, the measurements of $\hat{m}^2$ in the repulsive model and $\hat{d}^2$ in the attractive model should be exactly the same, i.e., $\langle\hat{m}^2\rangle_{\hat{H}(U)}=\langle\hat{d}^2\rangle_{\hat{H}(-U)}$. A similar relation can be obtained for the energy as $e(U)=e(-U)-U/2$. Since $\langle\hat{d}^2\rangle$ is a density related observable, its precise results can be obtained in AFQMC simulations for the attractive model using HS-$\hat{s}^x$ or HS-$\hat{s}^y$ since they preserve the charge symmetry. For simplicity, we choose HS-$\hat{s}^y$ in the simulation as it only involves real numbers for the attractive model. Then we directly take the $\langle\hat{d}^2\rangle_{\hat{H}(-U)}$ results as $m^2=\langle\hat{m}^2\rangle_{\hat{H}(U)}$ with $U>0$. This indeed produces high-precision results for $m^2$, as shown in Fig.~\ref{fig:CompareM2HS} by the green diamonds. On the other hand, we observe that the AFQMC with HS-$\hat{s}^x$ for $U>0$ already presents precise results for total energy and double occupancy as they belong to the density-related observable. For the large-scale simulations for the half filling case of the model (\ref{eq:Model}), we combine advantages of HS-$\hat{s}^x$ for $U>0$ and HS-$\hat{s}^y$ for $U<0$, and reach the numerical results with satisfying relative errors for all concerned quantities. 

In our calculations, we have removed the Trotter error by choosing small-enough $\Delta\tau$ from systematic tests with different parameters. Specifically, we adopt decreasing $\Delta\tau$ values for increasing $U/t$, i.e., from $\Delta\tau t=0.05$ for $U/t=2$ to $\Delta\tau t=0.02$ for $U/t=12$. Other efficient techniques implemented in our AFQMC algorithm include fast Fourier transform between the real and momentum spaces~\cite{Yuanyao2019,Yuanyao2019L}, the delayed update~\cite{McDaniel2017,Sun2024}, and $\tau$-line type of global update~\cite{Scalettar1991}. With these techniques, our AFQMC simulations for the model (\ref{eq:Model}) reach a large system with $L=28$. Moreover, we have benchmarked our calculations against the lattice continuous-time quantum Monte Carlo (LCT-QMC) algorithm, as reported in Ref.~\cite{Gukelberger2017}. Beside the good agreement of the numerical results, we also note that our AFQMC method can handle stronger interactions and provide more precise results in the intermediate to low temperature range, compared to the LCT-QMC method based on interaction expansion (see Appendix~\ref{sec:AppendixD}). 

\subsection{Physical observables}
\label{sec:AFQMCObs}

In this work, we concentrate on the magnetic and thermodynamic properties of the 2D modified Hubbard model in Eq.~(\ref{eq:Model}). Here we define the related observables and discuss their calculations in AFQMC method. 

We characterize the magnetic properties using the real-space spin-spin correlation function
\begin{equation}
\label{eq:RspSpinCrFt}
C_{\rm spin}(\mathbf{r}) = \frac{1}{N_s} \sum_{\mathbf{i}} \langle \hat{s}_{\mathbf{i}}^z \hat{s}_{\mathbf{i}+\mathbf{r}}^z \rangle,
\end{equation}
and its Fourier transform as the spin structure factor
\begin{equation}
\label{eq:Safmzz}
S_{zz}(\mathbf{q}) = \sum_{\mathbf{r}} C_{\rm spin}(\mathbf{r}) e^{i\mathbf{q} \cdot \mathbf{r}}.
\end{equation}
Then the mean-squared magnetization $m^2=\langle\hat{m}^2\rangle$ for the Ising AFM order can be expressed as $m^2=S_{zz}(\mathbf{M})/N_s$ with $\mathbf{M}=(\pi,\pi)$. The corresponding correlation length can be computed from $S_{zz}(\mathbf{q})$ as~\cite{Sandvik2010,Hofmann2023}
\begin{equation}
\label{eq:CorrLength}
\xi_{\rm AFM} = \big[2\sin(\pi/L)\big]^{-1} \sqrt{\frac{S_{zz}(\mathbf{M})}{S_{zz}(\mathbf{M}+\delta\mathbf{q})}-1},
\end{equation}
with $\delta\mathbf{q}=(2\pi/L,0)$ or $(0,2\pi/L)$ as the smallest momentum on a $N_s=L\times L$ lattice. For thermodynamic properties, we focus on total energy per site $e=\langle\hat{H}\rangle/N_s$, double occupancy $D=N_s^{-1}\sum_{\mathbf{i}}\langle \hat{n}_{\mathbf{i}\uparrow} \hat{n}_{\mathbf{i}\downarrow}\rangle$, and the charge compressibility as
\begin{equation}\begin{aligned}
\label{eq:ChiCharge}
\chi_e = -\frac{dn}{d\mu} = \frac{\beta}{N_s}\sum_{\mathbf{ij}}\Big(\langle \hat{n}_{\mathbf{i}} \hat{n}_{\mathbf{j}} \rangle - \langle \hat{n}_{\mathbf{i}} \rangle\langle \hat{n}_{\mathbf{j}} \rangle\Big),
\end{aligned}\end{equation}
with $n=N_s^{-1}\sum_{\mathbf{i}}\langle\hat{n}_{\mathbf{i}}\rangle$ as the fermion filling. From the energy $e(T)$, we can compute the specific heat as $C_v={\rm d}e/{\rm d}T$ and the thermal entropy density $\boldsymbol{s}(T)=S(T)/N_s$ via the improved formula proposed in Ref.~\onlinecite{Song2024b} as
\begin{equation}\begin{aligned}
\label{eq:Entropy}
\boldsymbol{s}(T)
= \ln 4 + \frac{e(T)}{T} - \int_T^{T_0}\frac{e(T^{\prime})}{{T^{\prime}}^2} dT^{\prime} - \int_0^{\beta_0} e(\beta^{\prime}) d\beta^{\prime},
\end{aligned}\end{equation}
with $T_0=1/\beta_0\in(T,\infty)$ as an intermediate temperature which serves as a free tuning parameter. Different choices of $T_0$ should produce consistent results for $S(T)$. At a fixed temperature $T$, the entropy as a function of $U$ for the half-filled system ($\mu=0$) can be computed via a more efficient scheme~\cite{Song2024b} from the total energy $E(U)$ and double occupancy $D(U)$ as
\begin{equation}\begin{aligned}
\label{eq:EntropyVsU}
\boldsymbol{s}(U) = \frac{1}{T}\Big[\frac{E(U)-F_0}{N_s} - \int_{0}^{U}D(U^{\prime})dU^{\prime} + \frac{U}{2}\Big],
\end{aligned}\end{equation}
where $F_0=-T\sum_{\mathbf{k}\sigma}\ln(1+e^{-\beta \varepsilon_{\mathbf{k},\sigma}})$ is the free energy of the noninteracting system. This method only involves AFQMC simulations for a bunch of $U$ at the fixed $T$. In practical calculations, we first perform the fitting (such as cubic-spline) for $e(T)$ and then evaluate the derivative in $C_v$ using the fitting curve. For Eq.~(\ref{eq:Entropy}), we separately fit $e(T^{\prime})$ in $T^{\prime}\in[T,T_0]$ range and $e(\beta^{\prime})$ in $\beta^{\prime}\in[0,\beta_0]$ range and then compute the integrals with the fitting functions. Similar technique is applied to evaluate the integral over $D(U^{\prime})$ in Eq.~(\ref{eq:EntropyVsU}). The uncertainties of $C_v$, $\boldsymbol{s}(T)$, and $\boldsymbol{s}(U)$ are further estimated by the bootstrapping technique~\cite{Deng2024}. We also investigate the singlon-singlon and doublon-doublon correlation functions defined as
\begin{equation}\begin{aligned}
\label{eq:SinglonDoublon}
C_{\rm singlon}(\mathbf{r}) &= \frac{1}{N_s}\sum_{\mathbf{i}}\Big( \langle\hat{\mathcal{S}}_{\mathbf{i}}\hat{\mathcal{S}}_{\mathbf{i}+\mathbf{r}}\rangle - \langle\hat{\mathcal{S}}_{\mathbf{i}}\rangle\langle\hat{\mathcal{S}}_{\mathbf{i}+\mathbf{r}}\rangle \Big)  \\
C_{\rm doublon}(\mathbf{r}) &= \frac{1}{N_s}\sum_{\mathbf{i}}\Big( \langle\hat{\mathcal{D}}_{\mathbf{i}}\hat{\mathcal{D}}_{\mathbf{i}+\mathbf{r}}\rangle - \langle\hat{\mathcal{D}}_{\bf{i}}\rangle\langle\hat{\mathcal{D}}_{\mathbf{i}+\mathbf{r}}\rangle \Big),
\end{aligned}\end{equation}
with the singlon operator $\hat{\mathcal{S}}_{\mathbf{i}}=\sum_{\sigma}\hat{n}_{\mathbf{i}\sigma}(1-\hat{n}_{\mathbf{i}\bar{\sigma}})$ and the doublon operator $\hat{\mathcal{D}}_{\bf{i}}=\hat{n}_{\bf{i}\uparrow}\hat{n}_{\bf{i}\downarrow}$. 

In the following (Sec.~\ref{sec:HalfResults}), we present the AFQMC results for the Ising AFM phase transition and thermodynamic properties of the half-filled model in Eq.~(\ref{eq:Model}), only for repulsive case ($U>0$). Nevertheless, most of these results can be mapped to the attractive model via the partial particle-hole transformation in Eq.~(\ref{eq:PartialPH}). First, since the $\hat{s}_{\mathbf{i}}^z$ operator is transformed to the density operator $\hat{d}_\mathbf{i}=(\hat{n}_{\mathbf{i}}-1)/2$, the Ising AFM order in the repulsive model corresponds to the CDW order of the attractive system and other related quantities should also satisfy the equivalence, i.e., $[C_{\rm spin}(\mathbf{r})]_{\hat{H}(U)}=[C_{\rm density}(\mathbf{r})]_{\hat{H}(-U)}$, the structure factor $[S_{zz}(\mathbf{q})]_{\hat{H}(U)}=[S_{\rm density}(\mathbf{q})]_{\hat{H}(-U)}$ with $S_{\rm density}(\mathbf{q})=\sum_{\mathbf{r}} C_{\rm density}(\mathbf{r}) e^{i\mathbf{q} \cdot \mathbf{r}}$, and the correlation length $(\xi_{\rm AFM})_{\hat{H}(U)}=(\xi_{\rm CDW})_{\hat{H}(-U)}$. The density-density correlation function $C_{\rm density}(\mathbf{r})$ is defined as
\begin{equation}\begin{aligned}
\label{eq:RspDenCrFt}
C_{\rm density}(\mathbf{r}) 
= \frac{1}{N_s} \sum_{\mathbf{i}} \langle \hat{d}_{\mathbf{i}} \hat{d}_{\mathbf{i}+\mathbf{r}} \rangle
= \frac{1}{4N_s} \sum_{\mathbf{i}}\langle\hat{n}_{\mathbf{i}}\hat{n}_{\mathbf{i}+\mathbf{r}}\rangle - \frac{1}{4},
\end{aligned}\end{equation}
where the half filling condition $N_s^{-1}\sum_{\mathbf{i}}\langle\hat{n}_{\mathbf{i}}\rangle=1$ is applied for the second equality in Eq.~(\ref{eq:RspDenCrFt}). $\xi_{\rm CDW}$ can be computed using the similar formula as Eq.~(\ref{eq:CorrLength}) but with $S_{\rm density}(\mathbf{q})$. Furthermore, the critical temperatures for the Ising phase transitions of the AFM order (with $U$) and CDW order (with $-U$) should be identical, as both belong to the 2D Ising universality class~\cite{Gukelberger2017}. Second, we can obtain $D(U)=1/2-D(-U)$ [with the spin-balanced condition $N_s^{-1}\sum_{\mathbf{i}}\langle\hat{n}_{\mathbf{i}\sigma}\rangle=1/2$] for the double occupancy and $e(U)=e(-U)-U/2$ for the energy. Then the formula in Eq.~(\ref{eq:Entropy}) verifies that the entropy for the repulsive and attractive models are exactly the same, i.e., $\boldsymbol{s}(U,T)=\boldsymbol{s}(-U,T)$. Thus, the results for all these observables in the attractive model (with $-U$) at half filling can be easily derived from our AFQMC results for the repulsive case.

\section{Numerical results at half filling}
\label{sec:HalfResults}

In this section, we concentrate on AFQMC results for the finite-temperature properties of the half-filled 2D modified Hubbard model in Eq.~(\ref{eq:Model}). They include the phase diagram and Ising phase transitions (Sec.~\ref{sec:IsingTrans}), the temperature dependence of thermodynamic quantities (Sec.~\ref{sec:ThermalObs}), thermal entropy map and the critical entropy (Sec.~\ref{sec:Entropy}), and spin, singlon, and doublon correlations (Sec.~\ref{sec:Correlation}). All of these results are for the repulsive interactions, and most of them can be mapped to the attractive model (see the discussions in Sec.~\ref{sec:AFQMCObs}).

\begin{figure}[t]
\centering
\includegraphics[width=0.99\columnwidth]{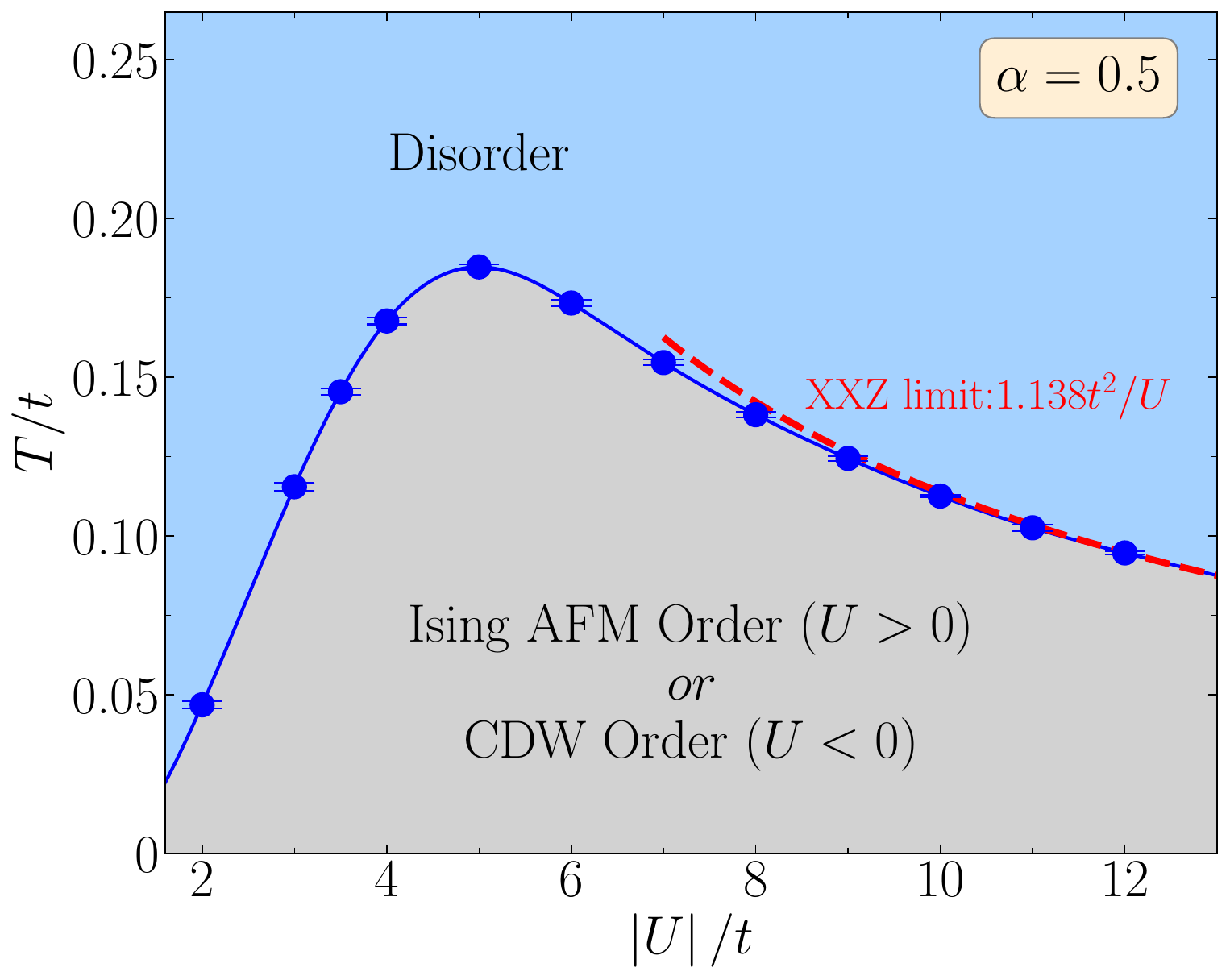}
\caption{Phase diagram of the half-filled 2D modified Hubbard model in Eq.~(\ref{eq:Model}) with the hopping anisotropy $\alpha=0.5$ from our AFQMC calculations. The low-temperature regime (gray shaded area) is covered by the Ising AFM ordered phase for $U>0$ or the CDW ordered phase for $U<0$, while the disordered phase occupies the high-temperature regime (blue shaded area). Blue circles and the corresponding interpolating connection (solid blue line) mark the critical temperatures of the Ising phase transitions. The result of XXZ limit $T_{\rm N}/t=1.138t/U$~\cite{XXZnote} is also included (dashed red line).}
\label{fig:PhaseDiagram}
\end{figure}

\begin{figure*}
\includegraphics[width=1.98\columnwidth]{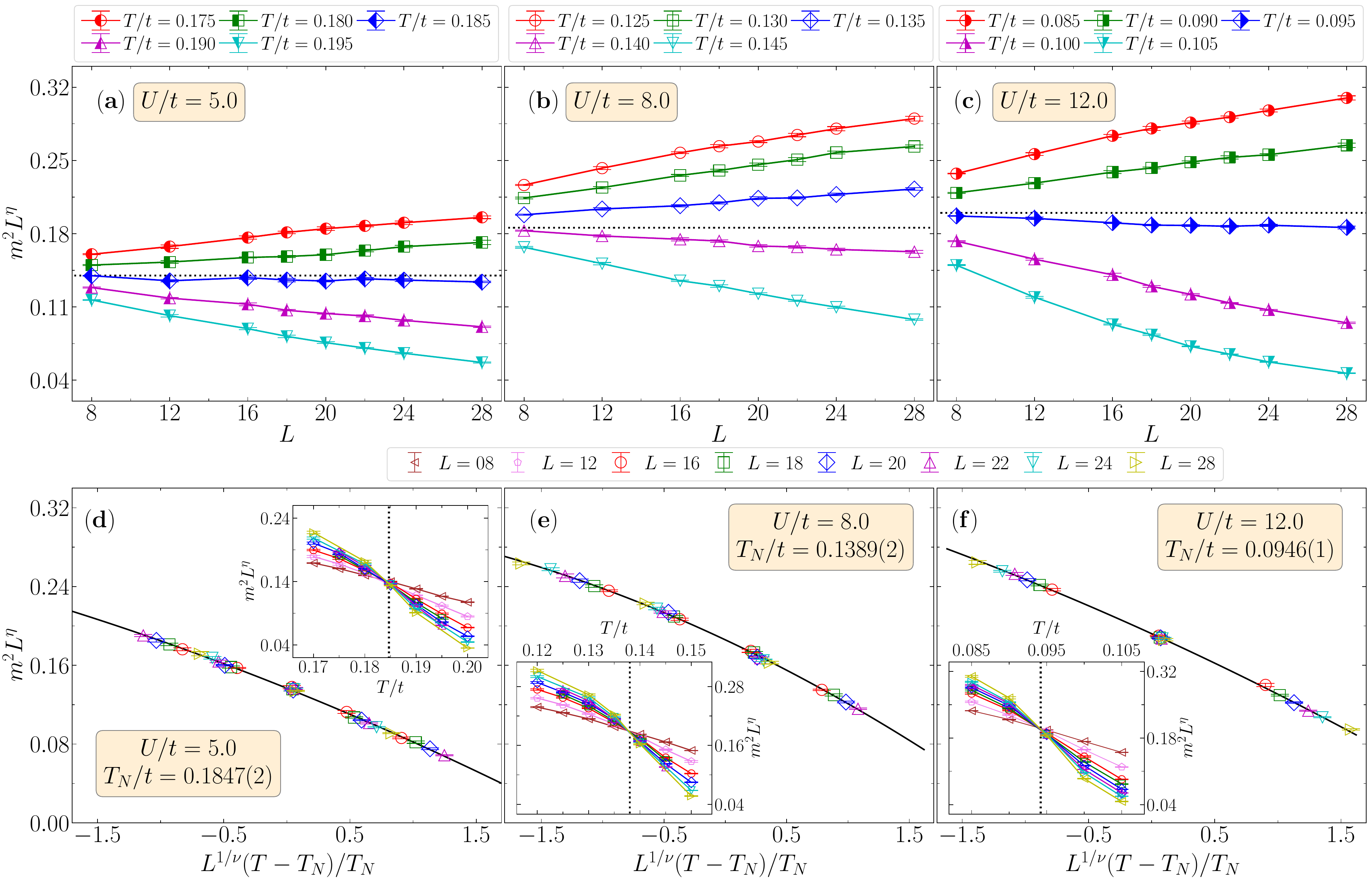}
\caption{Finite-size scalings for the mean-squared magnetization $m^2$ across the Ising AFM phase transition in the half-filled 2D modified Hubbard model with $\alpha=0.5$, using the critical exponents $\eta=1/4,\nu=1$ from the 2D Ising universality class~\cite{Gukelberger2017}. [(a), (b), and (c)] The rescaled quantity $m^2L^{\eta}$ versus linear system size $L$ at temperatures close to the transition, for $U/t=5,8$, and $12$. Correspondingly, (d)-(f) illustrate the data collapse of $m^2L^{\eta}$ versus $L^{1/\nu}(T-T_N)/T_N$ with the black solid lines as the second-order polynomials in Eq.~(\ref{eq:DataCollapse}). The transition temperatures obtained from the fitting are also included. The insets replot $m^2L^{\eta}$ versus $T/t$ for different system sizes, with the crossings well consistent with the $T_{\rm N}$ results (vertical, black dotted lines). }
\label{fig:Tn1}
\end{figure*}

\subsection{Phase diagram and Ising phase transitions}
\label{sec:IsingTrans}

In Fig.~\ref{fig:PhaseDiagram}, we plot the $T$-$U$ phase diagram of the 2D modified Hubbard model in Eq.~(\ref{eq:Model}) with the hopping anisotropy $\alpha=0.5$. The Ising phase transitions separate the high-$T$ disordered phase and the low-$T$ phase with long-range order, which is either the Ising AFM order for $U>0$ or CDW order for $U<0$. As mentioned in Sec.~\ref{sec:TheModel}, the long-range order develops at infinitesimal $|U|$ in the ground state, which renders the critical temperature scaling as $\propto t\exp[-1/(\rho_0|U|)]$ (with $\rho_0$ as the noninteracting density of states at Fermi level) in weakly interacting regime. Thus, our AFQMC calculations skip this regime, and instead cover a wide range of $|U|/t$ from $2$ to $12$. We observe that the transition temperature first increases with $|U|/t$, then reaches the highest, and finally decreases towards the strongly interacting regime. This nonmonotonic behavior can be well explained for the Ising AFM order with $U>0$. In the weak to intermediate interaction regimes, the increase in the critical temperature $T_{\rm N}/t$ is simply attributed to the enhancement of AFM spin correlations. On the other hand, towards $U/t\to\infty$ limit, the system evolves into the XXZ model in Eq.~(\ref{eq:XXZModel}) with the critical temperature $T_{\rm N}/J_{xx}=0.569(1)$~\cite{XXZnote}, which transforms to $T_{\rm N}/t=1.138t/U$ as decreasing versus $U/t$ in the context of Hubbard model. As illustrated in Fig.~\ref{fig:PhaseDiagram}, our results of $T_{\rm N}/t$ is already quite consistent with that of the XXZ model for $U/t\ge10$. These two distinct behaviors then result in the maximal $T_{\rm N}/t$$\sim$$0.185$ around $U/t=5$. These two numbers are smaller than those of 3D Hubbard model as $(T_{\rm N}/t)_{\rm max}$$\sim$$0.334$ around $U/t=8$~\cite{Song2024a,Song2024b}, for which the reduced bandwidth might be responsible as $W=6t$ for our 2D system comparing to $W=12t$ for the 3D case. 

We then turn to the determination of the critical temperature of the Ising phase transitions, using the numerical results of mean-squared magnetization $m^2$ for $U>0$. The most commonly used method is via the data collapse of $m^2$ results, which is essentially a multi-parameter fitting. It is based on the finite-size scaling relation $m^2L^{\eta}=f(x)$, with $x=L^{1/\nu}(T-T_{\rm N})/T_{\rm N}$ and $f(\cdot)$ as a scaling invariant function. This implies that, at the transition ($T=T_{\rm N}$), the quantity $m^2L^{\eta}$ should saturate to the constant $f(x=0)$ with increasing $L$. Moreover, $m^2L^{\eta}$ exhibits monotonically increasing versus $L$ in the ordered phase as $T<T_{\rm N}$, while it decreases with growing $L$ in the disordered phase. These behaviors are demonstrated in panels Fig.~\ref{fig:Tn1}(a)-\ref{fig:Tn1}(c) for $U/t=5$, $8$, and $12$. These results help to pin down $T_{\rm N}$ to a rather small temperature interval of $\Delta T=0.005t$. We then perform the data collapse for $m^2L^{\eta}$ to achieve the high-precision results of $T_{\rm N}$, via the least-squares fitting using
\begin{equation}\begin{aligned}
m^2L^{\eta} = \sum_{k=0}^2 a_k \Big[L^{1/\nu}\Big(\frac{T-T_{\rm N}}{T_{\rm N}}\Big)\Big]^k,
\label{eq:DataCollapse}
\end{aligned}\end{equation}
where $a_k$ and $T_{\rm N}$ are determined from the fitting. We directly use the exact critical exponents $\eta=1/4,\nu=1$ from the 2D Ising universality class~\cite{Gukelberger2017}. In Fig.~\ref{fig:Tn1}(d)-\ref{fig:Tn1}(f), we plot the data collapse results of $m^2L^{\eta}$ versus $L^{1/\nu}(T-T_N)/T_N$ for the same $U/t$ as presented in Figs.~\ref{fig:Tn1}(a)-\ref{fig:Tn1}(c). This procedure indeed produces highly accurate results of $T_{\rm N}$ with relative errors about $0.1\%$, as included in the plots. We furthermore verify these results with the finite-size crossings of $m^2L^{\eta}$ versus $T/t$ and find nice consistency as shown in the insets of Figs.~\ref{fig:Tn1}(d)-\ref{fig:Tn1}(f). 

\begin{figure}
\includegraphics[width=0.98\columnwidth]{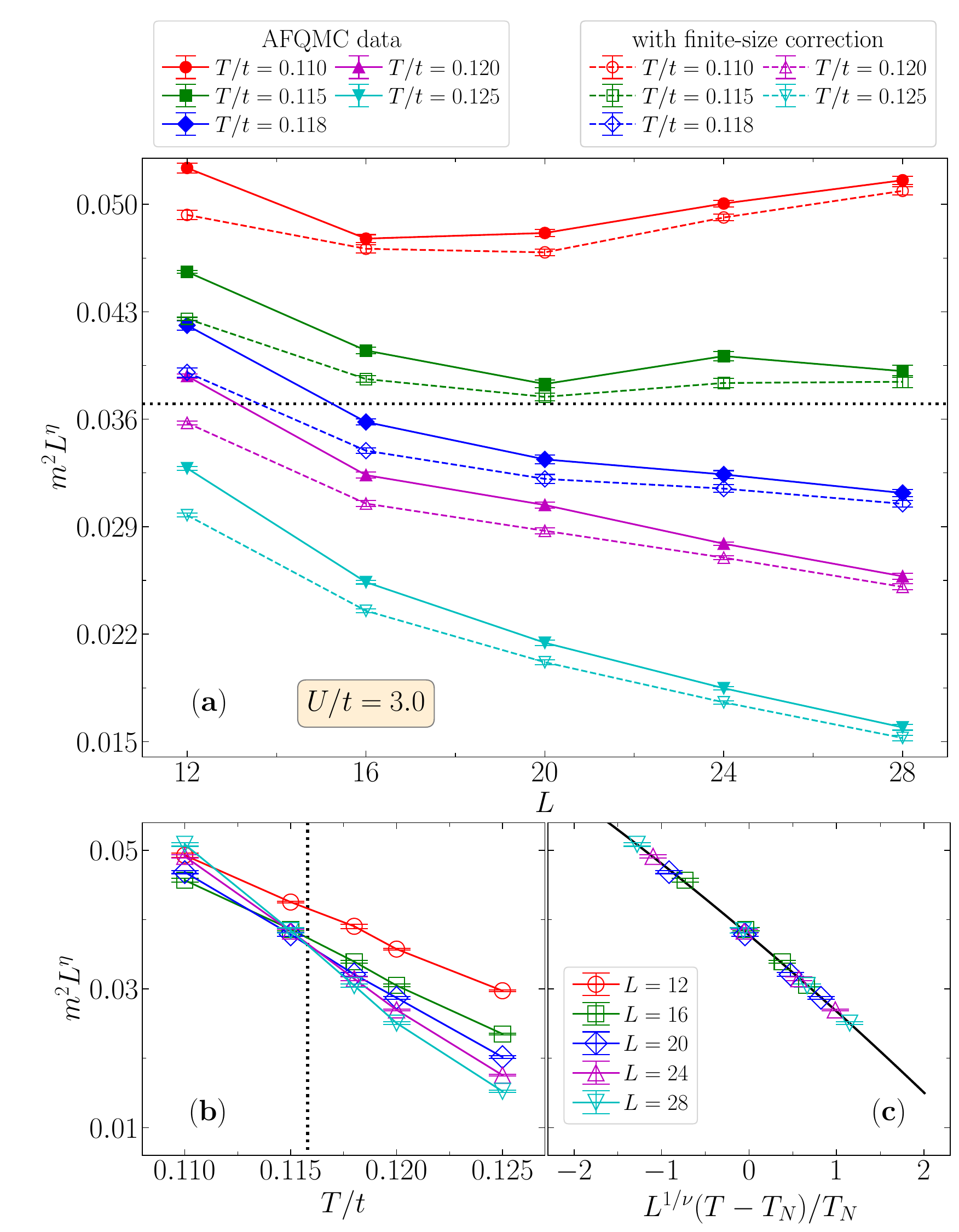}
\caption{Finite-size scaling analysis of the $m^2L^{\eta}$ results for $U/t=3$ (with $\alpha=0.5$ in the model). (a) The raw data from AFQMC and those with finite-size correction (using the noninteracting Hamiltonian as the reference system) of $m^2L^{\eta}$ versus $L$. With the corrected results of $m^2$, panel (b) illustrates the finite-size crossings of $m^2L^{\eta}$ versus temperature, and panel (c) shows the data collapse of $m^2L^{\eta}$ versus $L^{1/\nu}(T-T_N)/T_N$ with the black solid line as the polynomial in Eq.~(\ref{eq:DataCollapse}). The vertical, black dotted line in (b) plots the transition temperature $T_N/t = 0.1155(2)$ obtained from the fitting in panel (c).}
\label{fig:Tn2}
\end{figure}

In our calculations, we find that the above finite-size scaling analysis works well for $U/t\ge4$. In weakly interacting regime ($U/t\le3$), we observe strong oscillations in the numerical results of $m^2L^{\eta}$ versus $L$, which is caused by the single-particle finite-size effect of the noninteracting system. This effect simply disappears when the interaction term dominates. In Fig.~\ref{fig:Tn2}(a), we plot $m^2L^{\eta}$ versus $L$ around $T_{\rm N}$ for $U/t=3$ as an example. These visible oscillations pose a difficulty for extracting $T_{\rm N}$ from the direct data collapse calculations. In previous studies, it was demonstrated that such single-particle finite-size effect can be effectively overcome using finite-size correction techniques~\cite{Kwee2008,Freysoldt2009} or by performing finite-size simulations with the twist averaged boundary conditions~\cite{Mingpu2016,Haoxu2024}. Both methods are generally efficient in minimizing the one-body finite-size effect and thus accelerating the convergence towards the thermodynamic limit (TDL). Here we adopt the finite-size correction method, which replaces the observable $O(L)$ by the corrected one as
\begin{equation}\begin{aligned}
O_{\rm Corr}(L) = O(L) - O_{\rm Ref}(L) + O_{\rm Ref}(L=\infty),
\label{eq:FSCorrection}
\end{aligned}\end{equation}
where $O_{\rm Ref}(L)$ and $O_{\rm Ref}(L=\infty)$ are results of a specific reference system that can be solved exactly. For lattice models, the commonly used reference systems include the noninteracting Hamiltonian and the mean-field solution. It is obvious that $O(L)$ and $O_{\rm Corr}(L)$ converge to the same TDL result approaching $L=\infty$. For our case, we apply the noninteracting Hamiltonian as the reference. As shown in Fig.~\ref{fig:Tn2}(a), the corrected results of $m^2$ indeed show better $L$ dependence without oscillations comparing to the raw data. For $T/t=0.110$ and $0.115$, the nonmonotonic $m^2L^{\eta}$ after the correction actually reveals the residual finite-size effect beyond the capability of the noninteracting system, as these two temperatures are below the transition for $U/t=3$. Then the corrected results of $m^2L^{\eta}$ exhibit quite nice data collapse (within $L\ge16$) as shown in Fig.~\ref{fig:Tn2}(c), and the critical temperature $T_N/t=0.1155(2)$ obtained from the fitting also conforms with the crossings of $m^2L^{\eta}$ versus $T/t$ [see Fig.~\ref{fig:Tn2}(b)]. We apply the same procedure for $U/t=2$, and also reach satisfying result of $T_N/t$.

Beside the results in Figs.~\ref{fig:Tn1} and \ref{fig:Tn2} for the hopping anisotropy $\alpha=0.5$, we have also investigated the dependence of $T_{\rm N}/t$ on $\alpha$. Theoretically, the critical temperature is zero at the isotropic point $\alpha=1$~\cite{Mermin1966}, and it should be highest at $\alpha=0$ as the fully anisotropic case (with intermediate to large $U/t$). The latter can be understood from the facts that, the model (\ref{eq:Model}) with $\alpha=0$ degenerates to the classical Ising model~\cite{Gukelberger2017} according to Eq.~(\ref{eq:XXZModel}), and the lack of quantum fluctuation clearly enhances the Ising AFM order and thus the transition temperature. We have computed the critical temperatures for a different hopping anisotropy $\alpha=0.2$ and have indeed observed slightly higher $T_N/t$ with $U/t\ge5$ comparing to that of $\alpha=0.5$. However, the situation can be different for weak interactions. In Ref.~\onlinecite{Gukelberger2017}, it was shown that, for $U/t=3$, the results of $T_N/t$ from LCT-QMC are almost identical in the range of $0\le\alpha\le0.75$ considering the quite large error bars. Moreover, for $\alpha=0.2$, we observe that the maximum value of $T_N/t=0.1868(2)$ also occurs at $U/t=5$, which is only slightly higher than the corresponding result of $T_N/t=0.1847(2)$ for $\alpha=0.5$. All the $T_{\rm N}/t$ results from our AFQMC calculations for $\alpha=0.2$ and $\alpha=0.5$ are summarized in Appendix~\ref{sec:AppendixD}.

\subsection{Temperature dependences of thermodynamic quantities}
\label{sec:ThermalObs}

\begin{figure*}
\includegraphics[width=1.90\columnwidth]{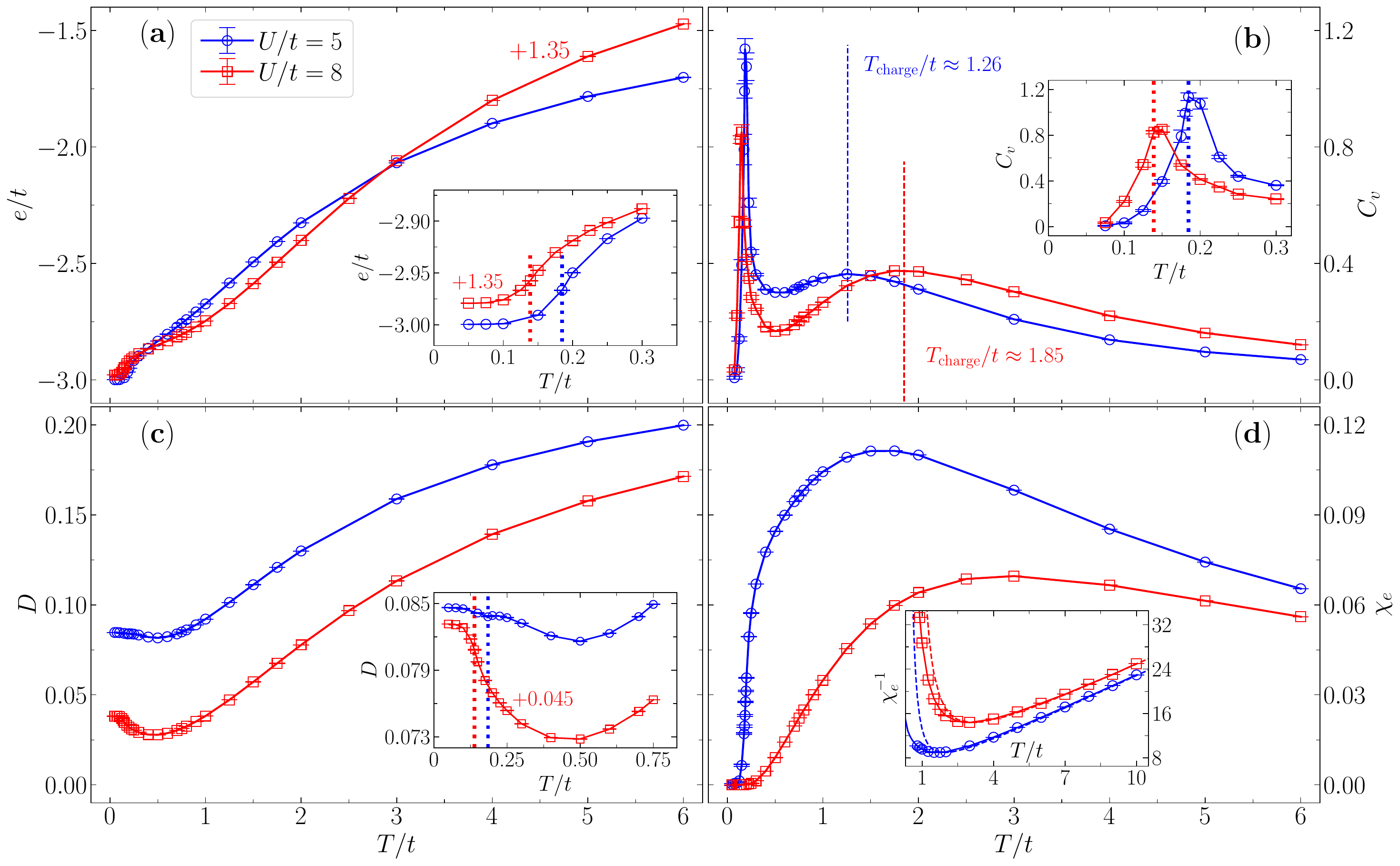}
\caption{AFQMC results of (a) total energy per site $e/t$, (b) specific heat $C_v$, (c) double occupancy $D$, and (d) charge susceptibility $\chi_e$, for $U/t=5$ and $8$ with the hopping anisotropy $\alpha=0.5$. In (a), (b) and (c), the insets plot the zoom-in results in a smaller temperature range with the vertical dotted lines marking the corresponding Ising AFM phase transition. The inset of (d) presents the $\chi_e^{-1}$ results with the dashed lines representing the atomic limit. The results of $e/t$ in (a), and results of $D$ in the inset of (c) for $U/t=8$ are shifted by $+1.35$ and $+0.045$, respectively, to fit into the plots. In the main plot of (b), the vertical dashed lines signify the high-temperature peak (charge peak) location of $C_v$. These results are from $L=12$ system, and the residual finite-size effect is mostly negligible (see discussions in Appendix~\ref{sec:AppendixD}). }
\label{fig:ThermalResults}
\end{figure*}

In addition to driving phase transitions, the interplay between quantum and thermal fluctuations in strongly correlated fermionic systems can lead to a wealth of phenomena at finite temperatures, as manifested by thermodynamic quantities. Here we focus on the commonly used thermodynamics (see Sec.~\ref{sec:AFQMCObs}) as a function of temperature for the half-filled 2D modified Hubbard model in Eq.~(\ref{eq:Model}) with $\alpha=0.5$. In Fig.~\ref{fig:ThermalResults}, we summarize the AFQMC results of the energy $e/t$, specific heat $C_v$, double occupancy $D$, and charge susceptibility $\chi_e$ for two representative interactions $U/t=5$ and $8$. These results span a wide temperature range, as $T/t\in[0.05,6]$, containing the Ising phase transitions.

For the half-filled Hubbard model, the energy density $e/t$ is equal to $-0.25U/t$ at $T/t\to\infty$, regardless of the hopping anisotropy. For the low-temperature limit, we find that, at $T/t\le0.075$, $e/t$ already converges to its ground-state value for both $U/t=5$ and $8$ [inset of Fig.~\ref{fig:ThermalResults}(a)]. In the middle, $e/t$ decreases and develops inflection points (most rapid decaying) with lowering temperature, which is quite prominent around the phase transition [inset of Fig.~\ref{fig:ThermalResults}(a)]. These inflection points in $e/t$ thus corresponds to the peaks in specific heat as $C_v={\rm d}e/{\rm d}T$. As shown in Fig.~\ref{fig:ThermalResults}(b), $C_v$ exhibits a double-peak structure, which has also been confirmed in both 2D and 3D standard Hubbard models~\cite{Duffy1997,Paiva2001,Qiaoyi2023,Song2024b}. Similar to the 3D case~\cite{Song2024b}, the sharp low-$T$ peak corresponds to the phase transition [inset of Fig.~\ref{fig:ThermalResults}(b)], and thus it is called spin peak. Their difference lies in the fact that, this spin peak in our 2D system should possess a logarithmic divergence in the TDL (as $\propto\ln|T-T_{\rm N}|$, according to the exact solution of 2D Ising model), while it is just a finite peak around the transition without divergence in the 3D case~\cite{Song2024b}. The broadened high-$T$ peak in $C_v$ is called charge peak since it originates from the charge (fermionic) excitations across the gap between the upper and lower Hubbard bands. The peak locations (as $T_{\rm charge}$) illustrated in Fig.~\ref{fig:ThermalResults}(b) are also comparable with those in 3D. Moreover, as discussed in Ref.~\onlinecite{Song2024b}, this charge peak behavior can be addressed by the atomic limit (under the condition of $\beta U\gg1$) of Hubbard model, which predicts the peak location at $T\simeq 0.208U$. The drifting towards this limit with increasing $U$ is clear from our results, as $T_{\rm charge}\simeq 1.26t=0.252U$ for $U/t=5$ and $T_{\rm charge}\simeq 1.85t=0.231U$ for $U/t=8$. 

Double occupancy $D$ can characterize the localization of fermions in repulsive systems. It approaches $D=1/4$ in the infinite-$T$ limit at half filling for arbitrary $U$, and this number decays as the system is cooled to the energy scale $T$$\sim$$U$. As shown in Fig.~\ref{fig:ThermalResults}(c), $D$ reaches the minimum around $T/t=0.5$, then anomalously increases, and saturates to the ground-state results for both interactions. The anomalous increase of $D$ upon cooling as $(\partial D/\partial T)_{U}<0$, resembling the Pomeranchuk effect in liquid $^3$He~\cite{Richardson1997}, has also been verified to exist in both 2D and 3D Hubbard models by many previous studies, i.e., Refs.~\onlinecite{Thomas2021,Qiaoyi2023,Song2024b} and the references therein. These features can be first understood from a mathematical aspect via the Maxwell's relation~\cite{Werner2005,Song2024b} between $D$ and thermal entropy $\boldsymbol{s}$ as
\begin{equation}\begin{aligned}
\label{eq:Maxwell}
\Big(\frac{\partial\boldsymbol{s}}{\partial U}\Big)_{T} 
= -\Big(\frac{\partial D}{\partial T}\Big)_{U}.
\end{aligned}\end{equation}
The minimum of $D(T)$ with fixed $U$ as $(\partial D/\partial T)_{U}=0$ can be associated with the extreme points of $\boldsymbol{s}(U)$ at fixed $T$ as $(\partial\boldsymbol{s}/\partial U)_{T}=0$ and can be further related to the extrema of the isentropic lines (see Fig.~\ref{fig:EntropyLines} and the discussions in Sec.~\ref{sec:Entropy}). Similarly, the anomalous region with $(\partial D/\partial T)_{U}<0$ for both interactions can be explained via $(\partial\boldsymbol{s}/\partial U)_{T}>0$ in the same temperature range for the corresponding $U$ (which can be directly observed from the entropy map in Fig.~\ref{fig:EntropyLines}). From the physical perspective, the increase of $D$ with decreasing $T$ within $T/t\le0.45$ for $U/t=8$ [inset of Fig.~\ref{fig:ThermalResults}(c)] is driven by the virtual hopping of fermions, a consequence of spin-exchange interactions described by the XXZ model. This process slightly delocalizes the fermions, and thereby promotes the double occupancy. However, the situation for $U/t=5$ is more complicated, for which we actually find a rather shallow kink in our AFQMC results of $D$ around the Ising phase transition [see inset of Fig.~\ref{fig:ThermalResults}(c)]. In Appendix~\ref{sec:AppendixD}, we furthermore present zoom-in results from $L=12$ and $16$ for $U/t=5$ to unambiguously show that this unobvious kink persists to larger systems and is indeed an intrinsic behavior. This kink then results in a local minimum at $T/t\simeq0.185$ and a local maximum at $T/t\simeq0.25$, in addition to the prominent minimum around $T/t=0.50$. Subsequently, the increase of $D$ upon cooling at $T/t\le 0.185$ is also induced by the spin-exchange physics, while it is due to the entrance into Fermi liquid from a bad metal state~\cite{Song2024b} in the temperature range $0.25\le T/t\le 0.50$. Similar physics of $D$ versus $T$ has also been thoroughly studied and discussed for the 3D Hubbard model in Ref.~\onlinecite{Song2024b}.

Charge compressibility $\chi_e$ can qualitatively distinguish the metallic and insulating behavior of correlated fermion systems. In Fig.~\ref{fig:ThermalResults}(d), we present the results of $\chi_e$ and its inverse $\chi_e^{-1}$ (the inset) versus $T$. For both $U/t=5$ and $8$, upon cooling, $\chi_e$ initially increases, and reach a maximum roughly around $T_{\rm charge}$ as marked in $C_v$ [see Fig.~\ref{fig:ThermalResults}(b)], after which quantum fluctuations induced by the hopping term take effect. It then decays monotonically to zero as approaching $T=0$, indicating a gapped ground state. In high-temperature regime, we observe the linear-$T$ dependence of $\chi_e^{-1}$ [inset of Fig.~\ref{fig:ThermalResults}(d)], which coincides with the atomic limit result as $\chi_e^{-1} = T/(n - n^2/2)+ U/2$ with $\beta U\gg1$. This $\chi_e^{-1}$ behavior has also been found in both 2D and 3D Hubbard models~\cite{Edwin2019,Qiaoyi2023,Song2024b}. Moreover, this linear-$T$ dependence can induce the similar linear resistivity (as $\rho$) as observed in 2D standard Hubbard model~\cite{Edwin2019,Peter2019}, considering the Nernst-Einstein relation $\rho=\chi_e^{-1}/D_{\rm diff}$, where $D_{\rm diff}$ is the diffusivity and it only slightly decreases at high temperatures and saturates to the so-called Mott-Ioffe-Regel limit $ta^2/\hbar$ (with $a$ as the lattice constant)~\cite{Peter2019}. The deviation from the atomic limit of $\chi_e^{-1}$ around $T=T_{\rm charge}$ again addresses the onset of quantum fluctuations. 

Combining the above discussions, we conclude that many prominent features in the thermodynamic quantities presented in Fig.~\ref{fig:ThermalResults} might be shared in various Hubbard models with different ingredients, including dimensionality, hopping anisotropy, and lattice geometry. They include the spin and charge peaks in $C_v$, the minimum and anomalous increase with cooling in $D$, and the linear-$T$ dependence of $\chi_e$ in high-$T$ regime. The formulas of $e/t$, $C_v$ and $\chi_e$ in the atomic limit can be found in Appendix C of Ref.~\onlinecite{Song2024b}. Besides, we present only the $L=12$ results in Fig.~\ref{fig:ThermalResults} for two reasons. First, the finite system size of $L=12$ is comparable to the optical lattice scale of existing experiments for the 2D Hubbard model~\cite{Parsons2016,Boll2016,Lawrence2016,Mazurenko2017,Christie2019}. Second, we have verified that the residual finite-size effects are mostly negligible, except in a small region around the phase transition (mainly for $e/t$ and $C_v$). Further results and discussions on finite-size effects can be found in Appendix~\ref{sec:AppendixD}.

\subsection{Thermal entropy map and the critical entropy}
\label{sec:Entropy}

\begin{figure*}
\includegraphics[width=2.07\columnwidth]{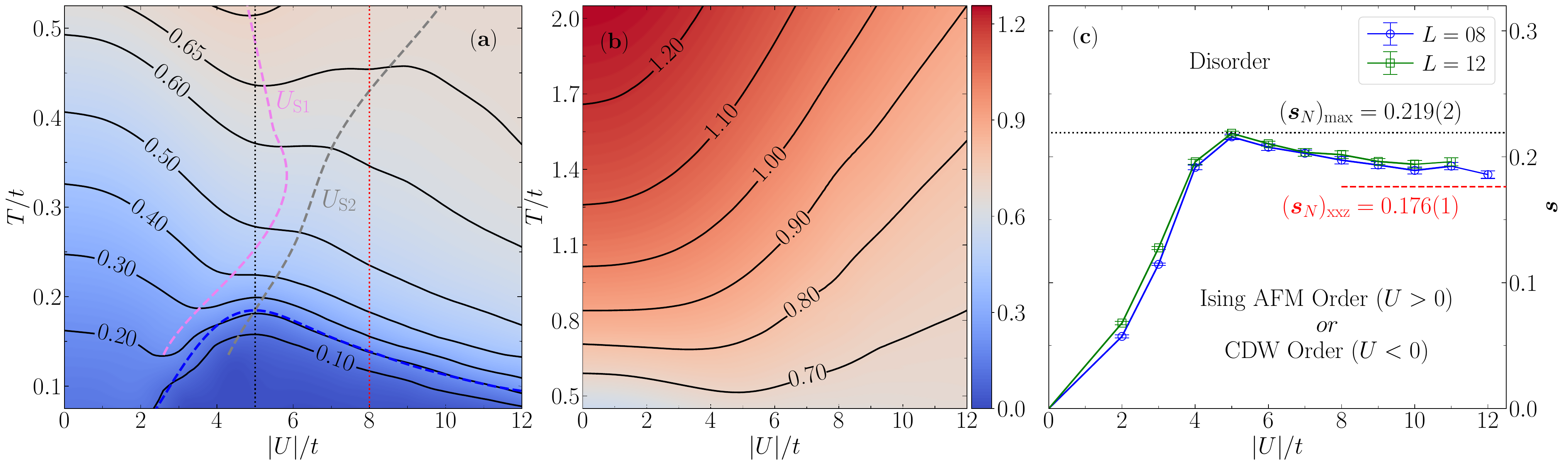}
\caption{Thermal entropy map ($\boldsymbol{s}$) on $T$-$U$ plane for the half-filled 2D modified Hubbard model with $\alpha=0.5$. Panels (a) and (b) illustrate the heatmap of $\boldsymbol{s}$ from $L=12$ system in the temperature ranges $0.075\leq T/t\leq0.5$ and $0.50\leq T/t\leq 2.0$, respectively. Black solid lines are the isentropic curves with the constant $\boldsymbol{s}_i$ values marked on the lines. In (a), the connecting lines for the local minimum ($U_{\rm S1}$, violet dashed line) and the local maximum ($U_{\rm S2}$, gray dashed line) of the isentropic curves versus $|U|/t$, and the Ising phase transition temperatures (blue dashed line) are illustrated. (c) Plots the critical entropy $\boldsymbol{s}_{\rm N}$ for the Ising phase transitions from both $L=8$ and $12$, with the maximum $(\boldsymbol{s}_{\rm N})_{\rm max}=0.219(2)$ (black dotted line) achieved at $|U|/t=5$. Result of the corresponding XXZ model as $(\boldsymbol{s}_{\rm N})_{\rm XXZ}=0.176(1)$~\cite{XXZnote} is also included (horizontal, red dashed line). All these results apply for both repulsive and attractive interactions.}
\label{fig:EntropyLines}
\end{figure*}

In optical lattice experiments, it is the thermal entropy that is directly controlled and measured~\cite{Shao2024}, while the temperature is usually obtained via the benchmark with known results~\cite{Parsons2016,Boll2016,Lawrence2016,Mazurenko2017}, such as those from AFQMC. Here we present the numerical results of entropy map on $U$-$T$ plane and the critical entropy of the Ising phase transitions for the half-filled 2D modified Hubbard model with $\alpha=0.5$. All the entropy data are obtained from our AFQMC calculations for $U>0$ case, and the attractive system has exactly the same results as $\boldsymbol{s}(U,T)=\boldsymbol{s}(-U,T)$ (see Sec.~\ref{sec:AFQMCObs}). We simply omit the unit of $\boldsymbol{s}$ (which is $k_B$) in the plot and discussions. 

We first compute the entropy $\boldsymbol{s}$ using Eq.~(\ref{eq:Entropy}) for a fixed $U/t$ in the temperature range $0.075\le T/t\le 2.0$, repeat the calculation for a group of $U/t$ values (with an interval $\Delta U/t=1.0$), and then obtain the complete entropy map on the full $U$-$T$ plane via an interpolation for the numerical data. In Figs.~\ref{fig:EntropyLines}(a) and \ref{fig:EntropyLines}(b), we present the heatmap of $\boldsymbol{s}$ alongside an ensemble of isentropic curves, $T_i(U)$, along which $\boldsymbol{s}(T_i(U), U)=\boldsymbol{s}_i$ is a constant for each curve. We further obtain the locations of the local minimum (as $U_{\rm S1}$) and local maximum (as $U_{\rm S2}$) of all these $T_i(U)$ curves via ${\rm d}T_i(U)/{\rm d}U=0$, and include them in Fig.~\ref{fig:EntropyLines}(a). Moreover, the slope of $T_i(U)$ can be connected to $U$-derivative of $\boldsymbol{s}$ at $T=T_i$ as~\cite{Song2024c}
\begin{equation}\begin{aligned}
\frac{\partial\boldsymbol{s}}{\partial U}\Big|_{T=T_i} = 
-\frac{C_v(T_i)}{T_i}\frac{{\rm d}T_i(U)}{{\rm d}U},
\label{eq:DiffSOvU}
\end{aligned}\end{equation}
where $C_v(T)=T\times(\partial\boldsymbol{s}/\partial T)$ is the specific heat, being positive for $T>0$. This equality explicitly demonstrates that the slopes of $T_i(U)$ and $\boldsymbol{s}(U,T)$ with respect to $U$ have opposite signs, and that their extreme points, where ${\rm d}T_i/{\rm d}U=0$ and $(\partial\boldsymbol{s}/\partial U)|_{T=T_i}=0$, occur at the same $U$. Thus, the $U_{\rm S1}$ and $U_{\rm S2}$ curves are also the local maximum and local minimum of $\boldsymbol{s}(U,T)$ as a function of $U$ at fixed $T$, respectively. Subsequently, the Maxwell's relation in Eq.~(\ref{eq:Maxwell}) translates $U_{\rm S1}$ and $U_{\rm S2}$ into the extreme points of double occupancy $D(T)$ at fixed $U$. This point is clearly confirmed by our numerical results for both $U/t=5$ and $8$. As shown in Fig.~\ref{fig:EntropyLines}(a), upon cooling, the $U/t=5$ vertical line intersects the $U_{\rm S1}$ curve for twice and the $U_{\rm S2}$ curve for once. These intersections correspond to the local minimum at $T/t\simeq0.50$, local maximum at $T/t\simeq 0.25$ and local minimum at $T/t\simeq0.185$ in the $D(T)$ curve, as illustrated in Fig.~\ref{fig:ThermalResults}(c) and discussed in Sec.~\ref{sec:ThermalObs}. Similarly, the single intersection between the $U/t=8$ vertical line and $U_{\rm S2}$ curve corresponds to the local minimum of $D(T)$ at $T/t\simeq0.45$. Note that these intersections in Fig.~\ref{fig:EntropyLines}(a) may exhibit small discrepancies compared to the extreme points of double occupancy $D(T)$ in Fig.~\ref{fig:ThermalResults}(c), due to the limited precision of $U_{\rm S1}$ and $U_{\rm S2}$, which are derived from the interpolated data. 

Another important feature of the entropy map is the {\it interaction-induced adiabatic cooling}, resembling that observed in the 3D Hubbard model~\cite{Song2024c}. As illustrated in Fig.~\ref{fig:EntropyLines}(a), in both $U<U_{\rm S1}$ and $U>U_{\rm S2}$ regimes, the temperature decreases significantly along the isentropic curves as $U/t$ increases. For example, for the curve with $\boldsymbol{s}(T_i(U), U)=0.50$, the temperature drops from $T_i(U=0)=0.40t$ to $T_i(U=12t)=0.17t$. This feature could act as an internal mechanism for cooling fermions in optical lattice experiments. Besides, for $U>U_{\rm S2}$, the isentropic curves with $\boldsymbol{s}_i=\boldsymbol{s}(T_i(U), U)<\ln 2$ should follow $T_i(U)=cJ_{xx}\propto 1/U$ [with $J_{xx}=4\alpha t^2/U$ in Eq.~(\ref{eq:XXZModel}), and $c$ as a finite constant] towards $U\to\infty$, and $\boldsymbol{s}_i$ is also the entropy of the corresponding XXZ model at the temperature $T/J_{xx}=c$. Nevertheless, as shown in Fig.~\ref{fig:EntropyLines}(b), the $U_{\rm S1}$ curve and the corresponding {\it interaction-induced adiabatic cooling} in weakly interacting regime, disappear in high-$T$ regime at $T/t>0.85$ [or equivalently $\boldsymbol{s}(T_i(U), U)>0.8$]. The $U_{\rm S2}$ curve should instead persist and possess the asymptotic behavior $U_{\rm S2}\propto T$ in high-$T$ regime~\cite{Song2024b,Song2024c}.

The critical entropy $\boldsymbol{s}_{\rm N}$ is a key metric for realizing thermal phase transitions in optical lattice experiments. In Fig.~\ref{fig:EntropyLines}(c), we present AFQMC results of $\boldsymbol{s}_{\rm N}$ from $L=8$ and $12$ for the Ising phase transitions in Fig.~\ref{fig:PhaseDiagram}. The residual finite-size effect only exists for weak interactions as $U/t<4$. We observe that $\boldsymbol{s}_{\rm N}$ reaches its maximum value of $(\boldsymbol{s}_{\rm N})_{\rm max}=0.219(2)$ at $U/t=5$, and as $U$ increases, it gradually approaches the value of the corresponding XXZ model as $(\boldsymbol{s}_{\rm N})_{\rm XXZ}=0.176(1)$~\cite{XXZnote}. We find that the maximum critical entropy is close to the lowest single-particle entropy achieved for the 3D Hubbard model in Ref.~\onlinecite{Shao2024}. This suggests the potential for future experimental studies of the Ising phase transitions in the 2D modified Hubbard model.

\begin{figure}[t]
\includegraphics[width=0.99\columnwidth]{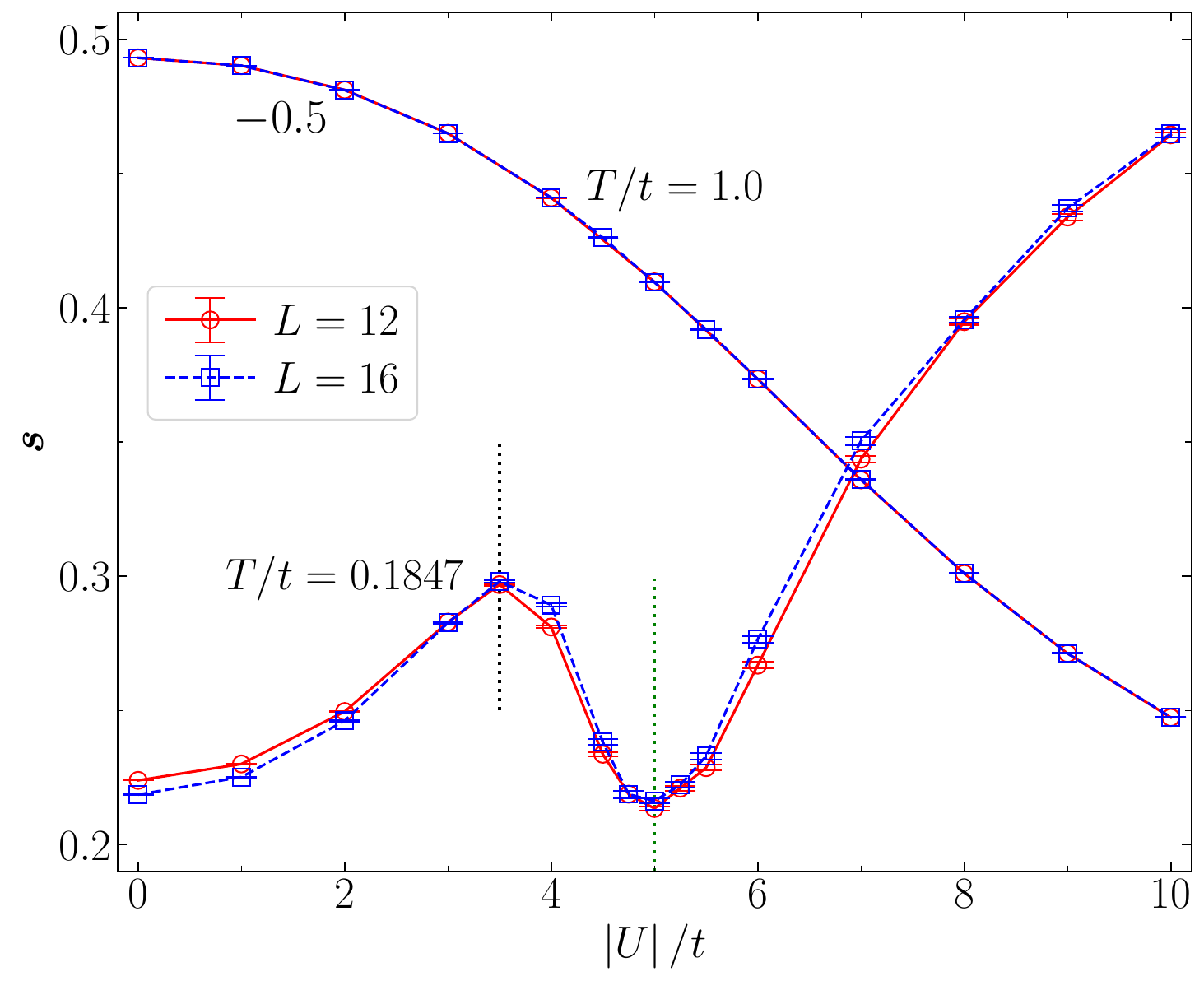}
\caption{Thermal entropy $\boldsymbol{s}$ as a function of $|U|/t$ at $T/t=1.0$ and $0.1847$, for the half-filled 2D modified Hubbard model with $\alpha=0.5$. The results for $T/t=1.0$ is shifted by $-0.5$ to fit into the plot. The vertical dotted lines mark the local maximum at $U/t\simeq 3.5$ and local minimum at $U/t\simeq 5.0$ of the $\boldsymbol{s}(U)$ results at $T/t=0.1847$. The results are from $L=12$ and $16$ systems, and they apply for both repulsive and attractive interactions. }
\label{fig:SvsU}
\end{figure}

While the entropy decreases monotonically with cooling at fixed $U$ [Figs.~\ref{fig:EntropyLines}(a) and \ref{fig:EntropyLines}(b)], its dependence on $U$ at fixed $T$ can display more complex and intriguing behaviors~\cite{Song2024a,Song2024b}. We calculate the entropy $\boldsymbol{s}$ as a function of $U/t$ using Eq.~(\ref{eq:EntropyVsU}) at the highest transition temperature $T/t=0.1847$ and at a relatively high temperature $T/t=1.0$, with the numerical results presented in Fig.~\ref{fig:SvsU}. At $T/t=0.1847$, the local maximum at $U/t\simeq 3.5$ and the local minimum at $U/t\simeq 5.0$ of $\boldsymbol{s}(U)$ actually correspond to the $U_{\rm S1}$ and $U_{\rm S2}$ at that temperature, as discussed earlier in relation to Fig.~\ref{fig:EntropyLines}(a). Moreover, following the Maxwell's relation in Eq.~(\ref{eq:Maxwell}), this local minimum at $U/t\simeq 5.0$ of $\boldsymbol{s}(U)$ provides further confirmation of the shallow minimum in $D(T)$ at $T/t\simeq 0.185$ for $U/t=5$ (as discussed in Sec.~\ref{sec:ThermalObs}). Similar to that in the 3D Hubbard model~\cite{Song2024a,Song2024b}, the local maximum of $\boldsymbol{s}(U)$ (namely, $U_{\rm S1}$) corresponds to the crossover boundary between Fermi liquid and bad metal state, while its local minimum (namely, $U_{\rm S2}$) is mainly attributed to the valley structure of spin entropy induced by the strongest AFM spin correlations around $U_{\rm S2}$. For comparison, at $T/t=1.0$, the entropy $\boldsymbol{s}(U)$ decays monotonically with $U/t$ across the plotted range, causing the local maximum $U_{\rm S1}$ to disappear. This trend aligns with the behavior depicted in Fig.~\ref{fig:EntropyLines}(b). Conversely, the local minimum $U_{\rm S2}$ in this case is observed at $U/t\sim20$ based on our numerical results (not shown in Fig.~\ref{fig:SvsU}).

\begin{figure}[t]
\includegraphics[width=0.99\columnwidth]{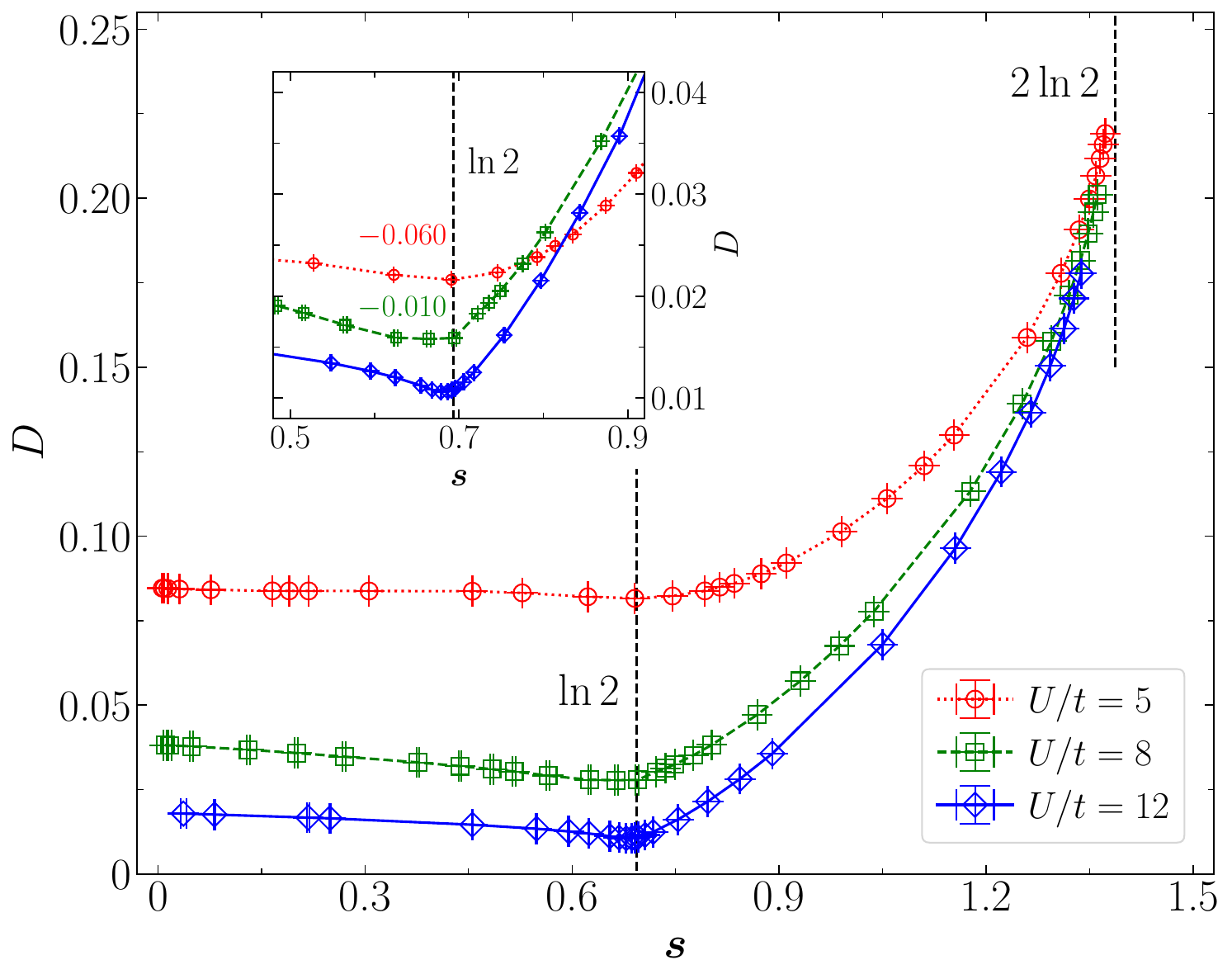}
\caption{Double occupancy $D$ as a function of the entropy $\boldsymbol{s}$ at $U/t=5$, $8$, and $12$, for the half-filled 2D modified Hubbard model with $\alpha=0.5$. The $\boldsymbol{s}=\ln 2$ and $2\ln 2$ are marked by vertical, black dashed lines. The inset illustrates the zoomin results within $\boldsymbol{s}\in[0.5,0.9]$, and the shifts of $-0.06$ for $U/t=5$, and $-0.01$ for $U/t=8$, are applied to fit into the plots. The results are from $L=12$ system. }
\label{fig:DvsS}
\end{figure}

Additionally, we examine the double occupancy $D$ as a function of the entropy $\boldsymbol{s}$, presenting the results for $U/t=5,8$, and $12$ in Fig.~\ref{fig:DvsS}. All the $D(\boldsymbol{s})$ curves begin at the point $(\boldsymbol{s}=2\ln 2,\ D=1/4)$ at $T/t=\infty$, then decreases with cooling, and reach the minimum around $\boldsymbol{s}=\ln 2$. Via the following derivative
\begin{equation}\begin{aligned}
\Big(\frac{\partial D}{\partial\boldsymbol{s}}\Big)_U = 
\Big(\frac{\partial D}{\partial T}\Big)_U \times 
\Big[\Big(\frac{\partial\boldsymbol{s}}{\partial T}\Big)_U\Big]^{-1},
\label{eq:DiffDOvS}
\end{aligned}\end{equation}
where $(\partial\boldsymbol{s}/\partial T)_U$ is positive, the minimum of $D(\boldsymbol{s})$ can be connected to the local minimum of $D$ with respect to $T$. As revealed in Ref.~\onlinecite{Gorelik2012}, the location of the minimum in $D(\boldsymbol{s})$ (denoted as $\boldsymbol{s}^{\star}$) converges to $\boldsymbol{s}=\ln 2$ as approaching $U/t=\infty$, and it separates two regimes for the strongly interacting system: one dominated by charge physics (with $\boldsymbol{s}>\boldsymbol{s}^{\star}$) and the other by spin-exchange physics (with $\boldsymbol{s}<\boldsymbol{s}^{\star}$). This is clearly supported by our results in Fig.~\ref{fig:DvsS} from two aspects. First, the minimum locations in $D(\boldsymbol{s})$ curves for $U/t=8$ and $12$ are already very close to $\boldsymbol{s}=\ln 2$. Second, as explained in Sec.~\ref{sec:ThermalObs}, the spin-exchange physics is responsible for the anomalous increase of $D$ upon cooling in the $\boldsymbol{s}<\boldsymbol{s}^{\star}$ regime, which conforms with the above two-regime physical picture. For intermediate interaction $U/t=5$, we also observe the minimum around $\boldsymbol{s}=\ln 2$, which actually corresponds to the local minimum in $D(T)$ curve at $T/t\simeq 0.50$ [inset of Fig.~\ref{fig:ThermalResults}(c)]. This is likely a coincidence, for which the two-regime picture described above does not apply. These results for $D(\boldsymbol{s})$ with $U>0$ in Fig.~\ref{fig:DvsS} can be mapped to the attractive system, i.e., $[D(\boldsymbol{s})]_{\hat{H}(-U)}=1/2-[D(\boldsymbol{s})]_{\hat{H}(U)}$ (see Sec.~\ref{sec:AFQMCObs}). As a result, the minimum of $D(\boldsymbol{s})$ for $U>0$ is flipped to the maximum in the case of attractive interactions.

We note that nearly all the features in the entropy map, as shown in Figs.~\ref{fig:EntropyLines}, \ref{fig:SvsU}, and \ref{fig:DvsS}, are also present in the 3D half-filled Hubbard model~\cite{Song2024a,Song2024b,Song2024c}. In paramagnetic phase, these features are closely tied to the metal-insulator crossover physics and the evolution of AFM spin correlations as the interaction increases. More importantly, given that these phenomena are already prominent in the intermediate to high-temperature regime, they fall well within the capabilities of current optical lattice experiments for Hubbard models. Thus, examining these entropy features might be a promising research direction for the experiments.

\subsection{Spin, singlon, and doublon correlations}
\label{sec:Correlation}

Equipped with the single-site and spin-resolved imaging techniques, 2D optical lattice experiments can access the real-space correlation functions with controllable distance in various channels~\cite{Parsons2016,Boll2016,Lawrence2016,Mazurenko2017}. Here, for the half-filled 2D modified Hubbard model, we focus on the spin, singlon, and doublon correlations, which are the primary targets of experimental studies.

\begin{figure}[t]
\includegraphics[width=0.965\columnwidth]{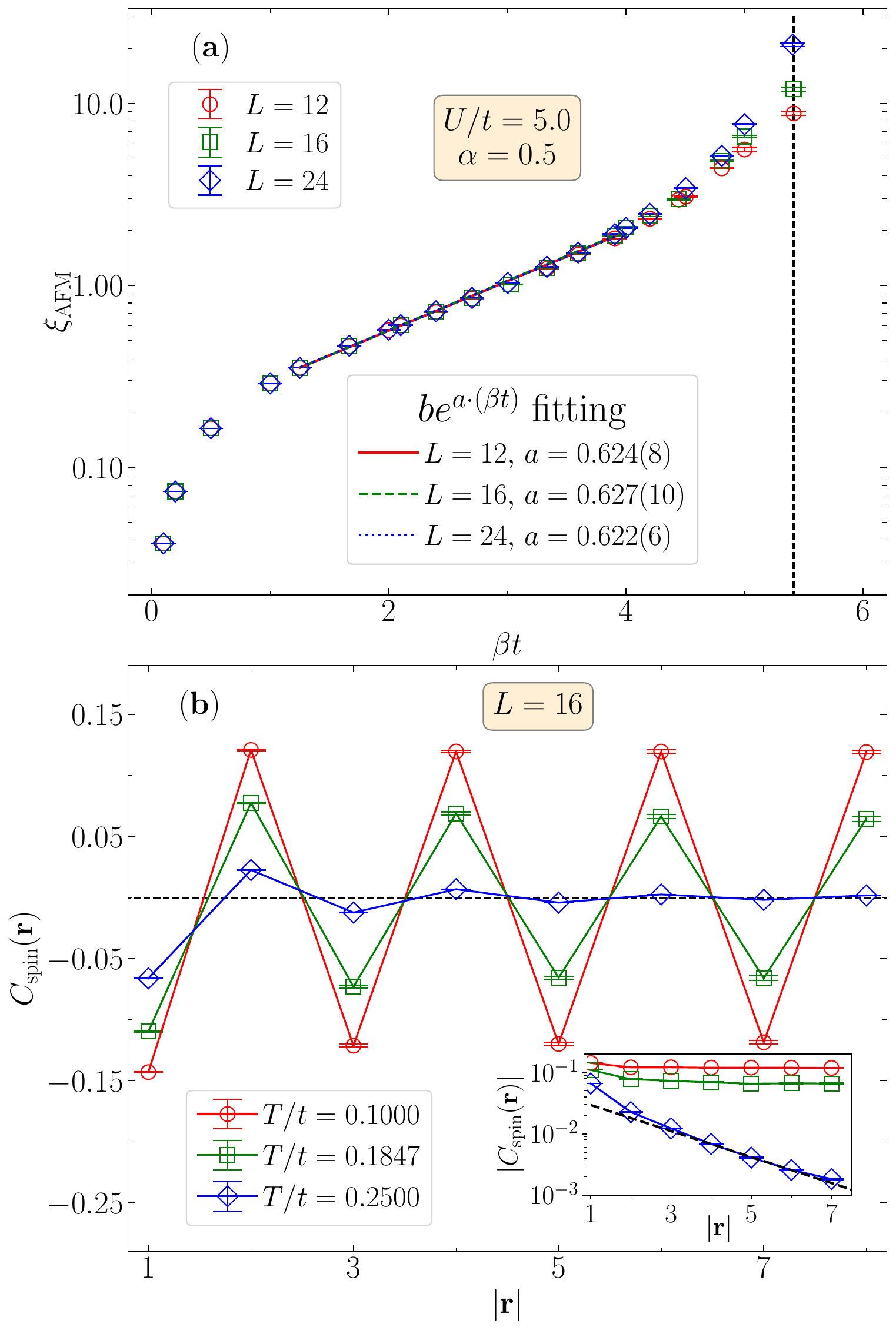}
\caption{Ising AFM spin correlation properties for the half-filled 2D modified Hubbard model with $U/t=5$ and $\alpha=0.5$. (a) Shows the semilog plot of correlation length $\xi_{\rm AFM}$ as a function of inverse temperature $\beta t$, from $L=12,16,24$ systems. The exponential fits [$\propto e^{a\cdot(\beta t)}$] in the range of $\beta t\in[1.2,4.0]$ produce well consistent coefficient $a\simeq 0.62$. (b) Presents the spin correlation $C_{\rm spin}(\mathbf{r})$ versus $|\mathbf{r}|$ along the $x$ direction [with $\mathbf{r}=(x,0)$] at $T/t=0.10$, $0.1847$, and $0.25$ from the $L=16$ system. The inset shows a semi-log plot of $|C_{\rm spin}(\mathbf{r})|$ versus $|\mathbf{r}|$. For $T/t=0.25$ from the exponential fit ($\propto e^{-|\mathbf{r}|/\xi}$) (black dashed line) produces $\xi_{\rm AFM}=2.05(2)$, well consistent with the result $\xi_{\rm AFM}=2.07(2)$ shown in (a). }
\label{fig:RSpinCrF}
\end{figure}

\begin{figure}[t]
\includegraphics[width=0.942\columnwidth]{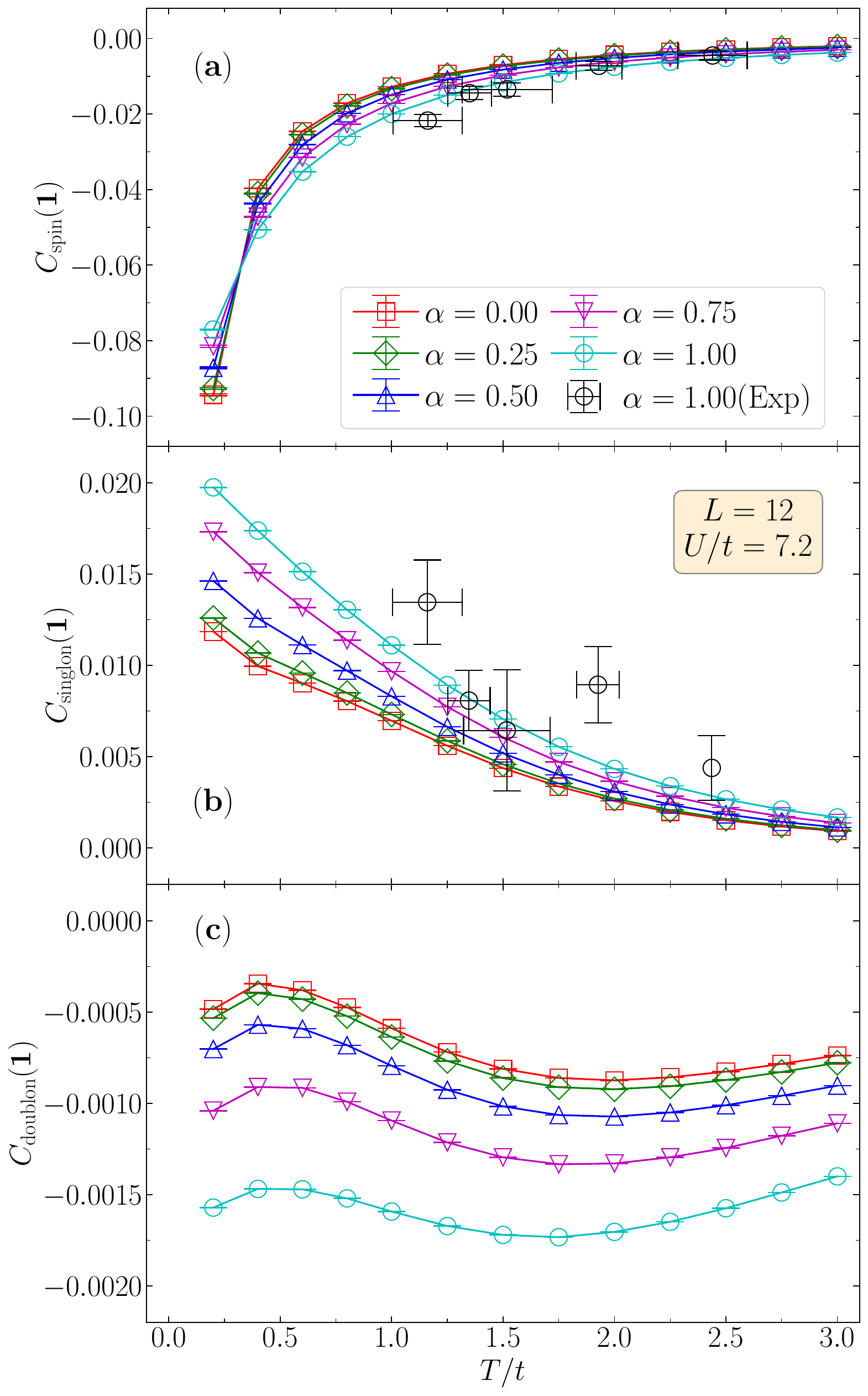}
\caption{The nearest-neighbor correlations in (a) spin channel [as $C_{\rm spin}(\mathbf{1})$], (b) singlon channal [as $C_{\rm singlon}(\mathbf{1})$], and (c) doublon channal [as $C_{\rm doublon}(\mathbf{1})$], as a function of temperature, for five different hopping anisotropies $\alpha=0$$\sim$$1$, with $U/t=7.2$ at half filling. The experimental results of $C_{\rm spin}(\mathbf{1})$ and $C_{\rm singlon}(\mathbf{1})$ for the isotropic 2D Hubbard model (with $\alpha=1$) from Ref.~\onlinecite{Lawrence2016} are also included [denoted as ``$\alpha=1.00({\rm Exp})$'']. These results are from $L=12$ system. }
\label{fig:ShortRCrFt}
\end{figure}

In Fig.~\ref{fig:RSpinCrF}, we present the results for the correlation length $\xi_{\rm AFM}$ of the AFM spin-spin correlation, calculated using Eq.~(\ref{eq:CorrLength}), along with its distance dependence, for the model with $U/t=5$ and $\alpha=0.5$. At high temperatures, the system exhibits only short-range correlations, resulting in a tiny $\xi_{\rm AFM}$, smaller than the lattice spacing. Then in the intermediate region of $\beta t\in[1.2, 4.0]$ (with $\beta=1/T$), the correlation length grows exponentially, $\xi_{\rm AFM}\propto e^{a\cdot(\beta t)}$, with $a\simeq 0.62$ from the fitting, as shown in Fig.~\ref{fig:RSpinCrF}(a). Similar results were also obtained for the 2D standard Hubbard model with $U/t=2$~\cite{Thomas2021}. With further cooling, $\xi_{\rm AFM}$ rises even more rapidly, as from $\xi_{\rm AFM}=2.07(2)$ at $\beta t=4$ to $\xi_{\rm AFM}=7.66(5)$ at $\beta t=5$ (with $L=24$). Approaching the Ising AFM phase transition, $\xi_{\rm AFM}$ in TDL should diverge, as reflected by the accelerating growth of finite-size $\xi_{\rm AFM}$ results. The correlation length can be alternatively characterized by the distance dependence of the correlation function. As shown in Fig.~\ref{fig:RSpinCrF}(b), the rapid decay of $C_{\rm spin}(\mathbf{r})$ and and its near-zero value at $|\mathbf{r}|=5$ for $T/t=0.25$ indicate short-range correlations in the disordered phase. For this case, an exponential fit ($\propto e^{-|\mathbf{r}|/\xi}$) to $|C_{\rm spin}(\mathbf{r})|$ yields $\xi_{\rm AFM}=2.05(2)$ [inset of Fig.~\ref{fig:RSpinCrF}(b)], which is well consistent with the result shown in Fig.~\ref{fig:RSpinCrF}(a), obtained using Eq.~(\ref{eq:CorrLength}). In contrast, the roughly constant $C_{\rm spin}(\mathbf{r})$ versus $|\mathbf{r}|$ at $T/t=0.10$ suggests the presence of the long-range Ising AFM order. At the transition $T/t=0.1847$, the rather slow decline of $C_{\rm spin}(\mathbf{r})$ indicates a large correlation length. The semilog plot in the inset of Fig.~\ref{fig:RSpinCrF}(b) further illustrates the exponential decay of $|C_{\rm spin}(\mathbf{r})|$ at $T/t=0.25$, and the algebraic decay at both $T/t=0.10$ and $0.1847$.

We then compute the nearest-neighbor correlations in the spin, singlon, and doublon channels, using Eq.~(\ref{eq:RspSpinCrFt}) and Eq.~(\ref{eq:SinglonDoublon}), with an additional average of $\mathbf{r}=(\pm1,0)$ and $\mathbf{r}=(0,\pm1)$ results. We plot these correlations versus $T/t$ for five different hopping anisotropies $\alpha=0$$\sim$$1$ in Fig.~\ref{fig:ShortRCrFt}. The interaction $U/t=7.2$ is chosen to align with the value that was experimentally studied in Ref.~\onlinecite{Lawrence2016} for the isotropic 2D Hubbard model ($\alpha=1$), whose results of $C_{\rm spin}(\mathbf{1})$ and $C_{\rm singlon}(\mathbf{1})$ are also included in the plots. As shown in Fig.~\ref{fig:ShortRCrFt}(a), the spin correlation $C_{\rm spin}(\mathbf{1})$ strengthens as $T$ decreases for all values of $\alpha$, with only small discrepancies observed across different $\alpha$. Interestingly, a crossing of these results occurs at $T/t\simeq 0.3$, which can be attributed to the onset of the Ising AFM phase transition for $\alpha\ne 1$, along with the enhancement of AFM spin correlations as $\alpha$ approaches the fully anisotropic limit ($\alpha=0$), as discussed in Sec.~\ref{sec:IsingTrans}. As a comparison, the presence of hopping anisotropy $\alpha\ne 1$ simply weakens the singlon and doublon correlations. While the singlon correlation $C_{\rm singlon}(\mathbf{1})$ becomes stronger at lower $T$, the doublon correlation $C_{\rm doublon}(\mathbf{1})$ shows a nonmonotonic temperature dependence, which may be related to the double occupancy. We also note that the precision of experimental measurements for these correlations must be improved in order to resolve the results for different $\alpha$ values at various temperatures.

\begin{figure}[t]
\centering
\includegraphics[width=0.98\columnwidth]{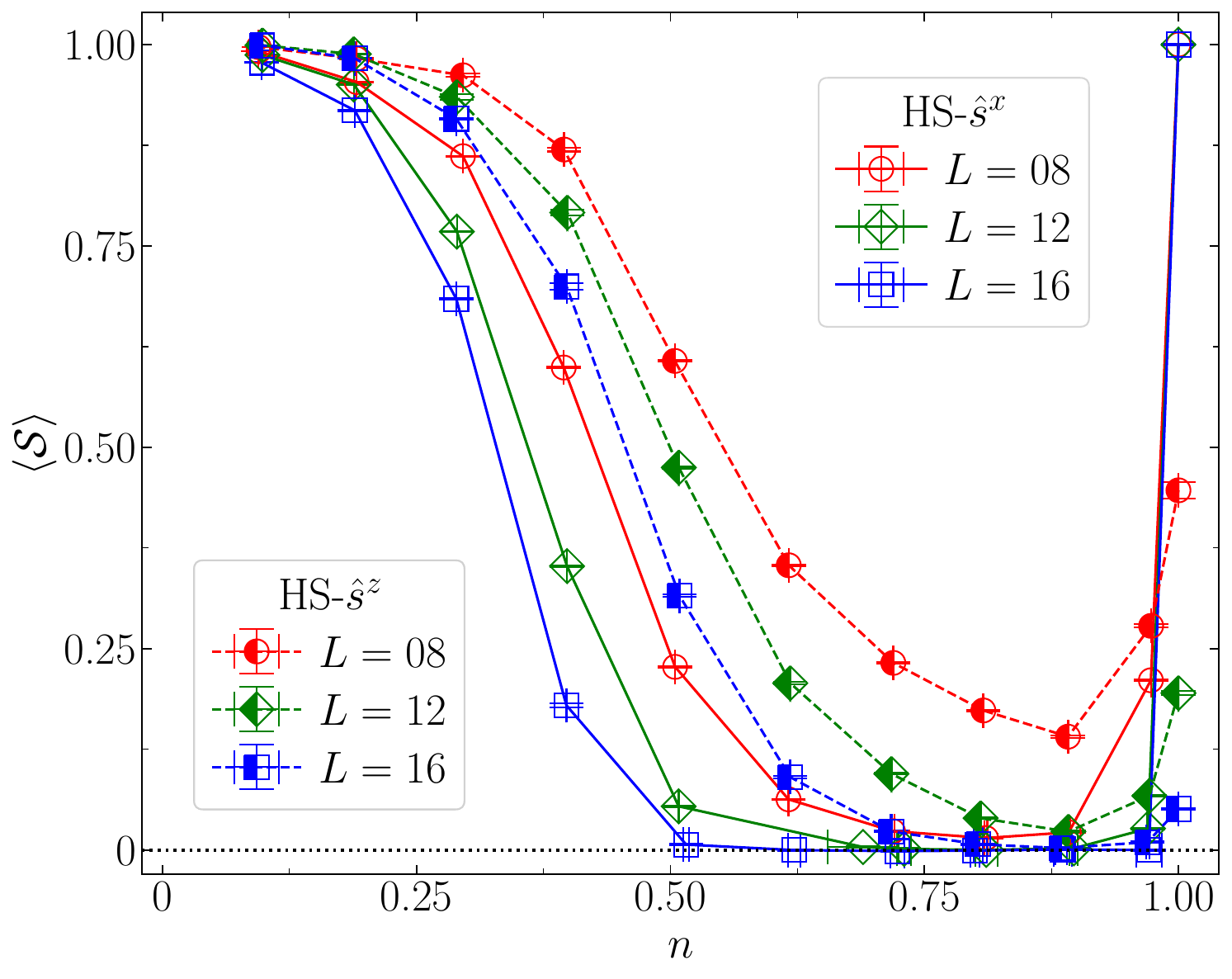}
\caption{Comparison of the sign average $\langle\mathcal{S}\rangle$ as a function of the fermion filling $n$ for $L=8,12,16$, using HS-$s^x$ (open symbols with solid lines) and HS-$s^z$ (left-filled symbols with dashed lines), for the 2D modified Hubbard model with doping. The parameters are set as $\alpha=0.5$, $U/t=5$, and $T/t=0.1847$. }
\label{fig:Signproblem}
\end{figure}

\section{Away from half filling}
\label{sec:DopeResults}

Doping, as the deviation from half filling, adds another dimension to the Hubbard model, for which the doping-induced phenomena, such as possible $d$-wave superconductivity~\cite{Qin2020,Haoxu2024} and stripe density wave ordering~\cite{Boxiao2017,Haoxu2022,Xiao2023}, have become the focus of significant interest. Nevertheless, the doped Hubbard model with repulsive interactions remains one of the most formidable challenges in theoretical condensed matter physics. This model typically encounters the minus sign problem, making it extremely challenging to obtain highly accurate results via AFQMC method. This also applies to the 2D modified Hubbard model with doping in Eq.~(\ref{eq:Model}). With hopping anisotropy $\alpha\ne 1$, the model with $\mu\ne0$ suffers from the sign problem for both $U>0$ and $U<0$, regardless of the applied HS transformation. Here, we focus on the repulsive interactions and the hole doping case ($\mu>0$), from which the electron doping ($\mu<0$) can be obtained via a particle-hole transformation, i.e., $c_{\mathbf{i}\sigma}^+\to(-1)^{\mathbf{i}}c_{\mathbf{i}\sigma}$ and $c_{\mathbf{i}\sigma}\to(-1)^{\mathbf{i}}c_{\mathbf{i}\sigma}^+$.

\begin{figure}[t]
\includegraphics[width=0.975\columnwidth]{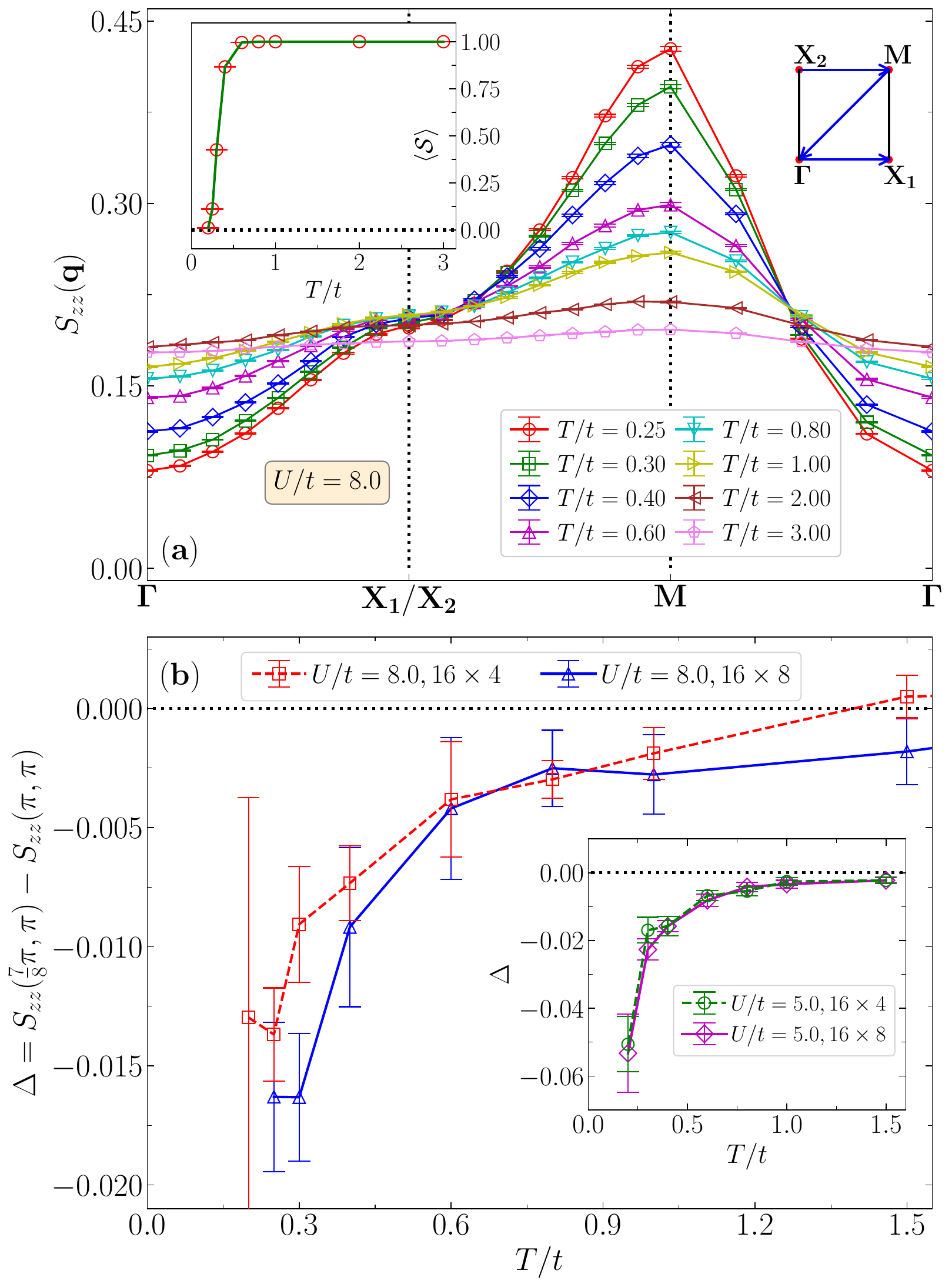}
\caption{AFQMC results of spin structure factor $S_{zz}(\mathbf{q})$ for the 2D modified Hubbard model at $1/8$-hole doping, with $\alpha=0.5$ and $U/t=8$. (a) Plots $S_{zz}(\mathbf{q})$ from $16\times 8$ system along the path $\boldsymbol{\Gamma}\to\mathbf{X}_1/\mathbf{X}_2\to\mathbf{M}\to\boldsymbol{\Gamma}$ in the first Brillouin zone (illustrated in the upper right corner) for various temperatures. The corresponding sign average $\langle\mathcal{S}\rangle$ is shown in the inset. (b) Displays the temperature dependence of the difference $\Delta=S_{zz}[\mathbf{q}_0=(7\pi/8,\pi)] - S_{zz}[\mathbf{M}=(\pi,\pi)]$ from both $16\times 4$ and $16\times 8$ systems. As a comparison, the inset shows the same quantity for $U/t=5$ (also with $\alpha=0.5$ and at $1/8$-hole doping). }
\label{fig:SzzDoping}
\end{figure}

We begin by examining the sign problem situations with different HS transformations. As discussed in Sec.~\ref{sec:AFQMCmethod}, HS-$\hat{n}$ has a phase problem for $U>0$, and therefore it is excluded from our calculations. Following the half-filled case in Sec.~\ref{sec:HalfResults}, we choose $\alpha=0.5$ and compute the sign average $\langle\mathcal{S}\rangle$ as a function of fermion filling $n$. In Fig.~\ref{fig:Signproblem}, we present the results of $\langle\mathcal{S}\rangle$ for $U/t=5$ at $T/t=0.1847$, corresponding to the highest $T_{\rm N}$ at half filling. Surprisingly, HS-$\hat{s}^x$ exhibits a more severe sign problem than HS-$\hat{s}^z$, even though the former is free of the sign problem at half filling. The decaying $\langle\mathcal{S}\rangle$ with increasing $L$ signals the inevitable fate of the sign problem when approaching TDL, including $\langle\mathcal{S}\rangle\to0$ and observables dominated by statistical noise. Moreover, we find that $\langle\mathcal{S}\rangle$ drops rapidly once with doping, and it is most severe in the regime of $0.70\le n<1.0$ for HS-$\hat{s}^z$. Based on these results, we adopt HS-$\hat{s}^z$ in our AFQMC calculations as follows.

Since the combined $\mathbb{Z}_2$ symmetry of the 2D modified Hubbard model in Eq.~(\ref{eq:Model}), as discussed in Sec.~\ref{sec:TheModel}, is preserved regardless of the fermion filling, the spontaneous breaking of this symmetry and the resulting magnetic order can, in principle, emerge with doping. Then the key question for the doped system with $U>0$ is whether the Ising AFM ordered phase can persist at finite doping. Reminiscent of the magnetic phase diagram of the 3D doped Hubbard model~\cite{Tahvildar1997,Katanin2017,Rampon2024,Lenihan2022}, it is likely that the model in our study exhibits a similar AFM dome around half filling, which consists of conventional AFM order at low doping and incommensurate SDW order at larger doping~\cite{Katanin2017}. Due to the limitations imposed by the sign problem, it is currently impossible to provide a definitive answer to the above question based on our AFQMC calculations. 

We instead turn to investigating the possible stripe SDW ordering. In Fig.~\ref{fig:SzzDoping}(a), we present the results of spin structure factor $S_{zz}(\mathbf{q})$ from $16\times8$ system (with periodic boundary conditions) for the model at $1/8$-hole doping, with $\alpha=0.5$ and $U/t=8$. Ranging from the high temperature $T/t=3.0$ to the lowest accessible temperature $T/t=0.25$, the peak of $S_{zz}(\mathbf{q})$ develops around $\mathbf{M}=(\pi,\pi)$, signaling the dominant role of AFM spin correlations. Additionally, we observe that $S_{zz}(\mathbf{q})$ at $\mathbf{q}_0=(7\pi/8,\pi)$, corresponding to the filled stripe SDW order at $1/8$-hole doping~\cite{Boxiao2017,Haoxu2022,Xiao2023}, also rises rapidly as the temperature decreases. If the stripe SDW long-range order appears at a critical temperature $T_{c}$, then $S_{zz}(\mathbf{q}_0)$ should exceed $S_{zz}(\mathbf{M})$ at a higher temperature than $T_c$. To gain further insight into this point, we compute $\Delta=S_{zz}(\mathbf{q}_0)-S_{zz}(\mathbf{M})$ versus temperature for both $16\times4$ and $16\times8$ systems. As shown in Fig.~\ref{fig:SzzDoping}(b), $\Delta$ exhibits small absolute values ($<$0.02), suggesting the competition between the Ising AFM order and the stripe SDW order. Moreover, an upturn begins to emerge in $\Delta$ around $T/t=0.25$, which could be interpreted as a positive indication that $S_{zz}(\mathbf{q}_0)$ may overtake $S_{zz}(\mathbf{M})$ at lower temperatures. But, unfortunately, the large error bars in the $\Delta$ results make this positive signature less conclusive. Nevertheless, we have also examined the same quantity $\Delta$ for $U/t=5$, as shown in the inset of Fig.~\ref{fig:SzzDoping}(b). In this case, $|\Delta|$ is approximately four times greater than that for $U/t=8$, and no upturn is observed in $\Delta$ even at the lowest accessible temperature, $T/t=0.20$. This suggests that the Ising AFM order may still dominate at $U/t=5$ with $1/8$-hole doping, while the stripe SDW order is less favored.

The systematic study for the finite-temperature properties of the model (\ref{eq:Model}) with doping requires the use of other advanced many-body numerical techniques, such as the constrained-path (CP) AFQMC algorithm~\cite{Yuanyao2019,Zhang1999,Xiao2023}. Through a bias-variance trade-off, this algorithm controls the sign problem by applying an appropriate constraint during the configuration sampling. With the self-consistent scheme of optimizing the constraint~\cite{Yuanyao2019,Xiao2023}, the CP-AFQMC method can potentially provide accurate results for both Ising AFM order and stripe SDW order. We leave the CP-AFQMC study of the model with doping to future work.

\section{summary and discussion}
\label{sec:Summary}

In summary, we have employed the numerically unbiased AFQMC simulations to study the finite-temperature properties of correlated fermions in a 2D spin-dependent optical lattice. The system is described by a 2D modified Hubbard model with spin-dependent anisotropic hopping in Eq.~(\ref{eq:Model}). Attributed to a discrete $\mathbb{Z}_2$ symmetry breaking, the model at half filling hosts Ising phase transitions at finite temperatures~\cite{Gukelberger2017}. Below the transition temperature, long-range order develops, either as the Ising antiferromagnetism ($U>0$) or the charge-density wave ($U<0$). We have presented comprehensive results on the Ising phase transitions and the thermodynamics of the system, offering a valuable benchmark and guide for future optical lattice experiments~\cite{Mandel2003,Vincent2004,Jotzu2015,Yang2017,Kuzmenko2019}.

On the methodological side, we have demonstrated how to achieve high-precision numerical results for the above model at half filling through an elegant combination of the sign-problem-free condition and various HS transformations in AFQMC simulations. The key components include the HS-$\hat{s}^x$ and HS-$\hat{s}^y$ transformations for the Hubbard interactions in Eq.~(\ref{eq:HSspinDecomp}), and the mapping between $U>0$ and $U<0$ results via the partial particle-hole transformation in Eq.~(\ref{eq:PartialPH}). We have also efficiently dealt with the finite-size effect in the weakly interacting regime via a correction technique using the noninteracting system as the reference. 

Our primary physical results focus on the half filling case, where we have obtained highly accurate estimates for the critical temperatures of the Ising phase transitions in the model, considering representative hopping anisotropies. We have also investigated the temperature dependence of widely used thermodynamic quantities and provided a thorough discussion of the underlying physics, which may be broadly applicable to various Hubbard models. To connect with optical lattice experiments, we have built the thermal entropy map on $U$-$T$ plane, computed the critical entropy, and highlighted the intriguing structures of the entropy as a function of interaction strength at intermediate temperatures. Among our key findings, we identify the highest critical temperature of $T/t=0.1847(2)$ at $U/t=5$ with a hopping anisotropy of $\alpha=0.5$, which is accompanied by a critical entropy per particle of $\boldsymbol{s}_{\rm N}/k_B=0.219(2)$. Furthermore, we have studied the correlation length of the Ising AFM order, and the short-range correlations in spin, singlon, and doublon channels. Away from half filling, we have found that the sign problem is actually alleviated by switching to the HS-$\hat{s}^z$ transformation in AFQMC simulations. With the limited numerical results, we have explored the stripe SDW ordering in the model with $U>0$ at $1/8$-hole doping, and observed positive indications for its ordering with cooling, which demands more thorough investigations using other numerical methods. 

Our work also has important implications for addressing a range of challenging problems in strongly correlated fermion systems. By extending to doped systems, the lattice model in Eq.~(\ref{eq:Model}) with repulsive interactions may serve as a 2D platform that complements the studies of the 3D Hubbard model~\cite{Shao2024,Katanin2017}. This extension enables the exploration of thermal magnetic phase transitions in 2D, and also the intricate interplay between quantum magnetism and other exotic phenomena, such as $d$-wave superconductivity. Furthermore, switching to attractive interactions, the potential for unconventional paired states and spin-triplet $p$-wave superfluidity~\cite{Feiguin2009,Feiguin2011,Huang2013,Gukelberger2014} represents long-pursued states in lattice models that warrant more attention from both theory and experiment.

\begin{acknowledgments}
We thank Yan-Cheng Wang for providing the associated numerical results of XXZ model. Y.-Y.He acknowledges Xing-Cao Yao and Bing Yang for valuable discussions. This work was supported by the National Natural Science Foundation of China (under Grants No. 12247103 and No. 12204377), the Innovation Program for Quantum Science and Technology (under Grant No. 2021ZD0301900), and the Youth Innovation Team of Shaanxi Universities.
\end{acknowledgments}

\appendix

\section{The spin susceptibilities of the noninteracting system}
\label{sec:AppendixA}

In this Appendix, we present the derivations for the spin susceptibilities of the noninteracting Hamiltonian in Eq.~(\ref{eq:Model}) at half filling ($\mu=0$), including both the longitudinal and transverse components, $\chi^{zz}(\mathbf{q})$ and $\chi^{xy}(\mathbf{q})$. We also examine their asymptotic behaviors as temperature approaches zero, considering the effects of different hopping anisotropies.

We start from the imaginary-time spin-spin correlation functions with $\tau>0$ as 
\begin{equation}\begin{aligned}
S_{zz}(\mathbf{q},\tau) = \sum_{\mathbf{r}}e^{i\mathbf{q}\cdot\mathbf{r}}C_{zz}(\mathbf{r},\tau), \\
S_{xy}(\mathbf{q},\tau) = \sum_{\mathbf{r}}e^{i\mathbf{q}\cdot\mathbf{r}}C_{xy}(\mathbf{r},\tau),
\label{eq:A1}
\end{aligned}\end{equation}
with the real-space correlations defined as
\begin{equation}\begin{aligned}
C_{zz}(\mathbf{r},\tau) &= \frac{1}{N_s}\sum_{\mathbf{i}}\Big(\langle\hat{s}^{z}_{\mathbf{i}}(\tau)\hat{s}^{z}_{\mathbf{i}+\mathbf{r}}(0)\rangle - \langle\hat{s}^{z}_{\mathbf{i}}(\tau)\rangle\langle\hat{s}^{z}_{\mathbf{i}+\mathbf{r}}(0)\rangle\Big), \\
C_{xy}(\mathbf{r},\tau) &= \frac{1}{N_s}\sum_{\mathbf{i}}\Big(\langle\hat{s}^{x}_{\mathbf{i}}(\tau)\hat{s}^{x}_{\mathbf{i}+\mathbf{r}}(0)\rangle + \langle\hat{s}^{y}_{\mathbf{i}}(\tau)\rangle\langle\hat{s}^{y}_{\mathbf{i}+\mathbf{r}}(0)\rangle\Big),
\label{eq:A2}
\end{aligned}\end{equation}
where $\hat{s}^{z}_{\mathbf{i}}=(\hat{n}_{\mathbf{i}\uparrow}-\hat{n}_{\mathbf{i}\downarrow})/2$, $\hat{s}^{x}_{\mathbf{i}}=(c_{i\uparrow}^+c_{\mathbf{i}\downarrow}+c_{\mathbf{i}\downarrow}^+c_{\mathbf{i}\uparrow})/2$, and $\hat{s}^{y}_{\mathbf{i}}=(c_{\mathbf{i}\uparrow}^+c_{\mathbf{i}\downarrow}-c_{\mathbf{i}\downarrow}^+c_{\mathbf{i}\uparrow})/(2i)$ are the spin operators of fermions. We further define the momentum space operator $\hat{s}^{z}_{\mathbf{q}}$ as
\begin{equation}\begin{aligned}
\hat{s}_{\mathbf{q}}^{z}
&=\frac{1}{\sqrt{N_s}}\sum_{\mathbf{i}}\hat{s}_{\mathbf{i}}^z e^{-i\mathbf{q}\cdot\mathbf{R_i}} \\
&=\frac{1}{2\sqrt{N_s}}\sum_{\mathbf{k}}\big(c_{\mathbf{k}\uparrow}^+ c_{\mathbf{k+q}\uparrow} - c_{\mathbf{k}\downarrow}^+ c_{\mathbf{k+q}\downarrow}\big).
\label{eq:A3}
\end{aligned}\end{equation}
Then $S_{zz}(\mathbf{q},\tau)$ in Eq.~(\ref{eq:A1}) can be simplified as
\begin{equation}\begin{aligned}
S_{zz}(\mathbf{q},\tau)
=\langle\hat{s}^{z}_{\mathbf{q}}(\tau)\hat{s}^{z}_{-\mathbf{q}}(0)\rangle-\langle \hat{s}^{z}_{\mathbf{q}}(\tau)\rangle\langle \hat{s}^{z}_{-\mathbf{q}}(0)\rangle.
\label{eq:A4}
\end{aligned}\end{equation}
The $\langle\hat{s}^{z}_{\mathbf{q}}(\tau)\hat{s}^{z}_{-\mathbf{q}}(0)\rangle$ term can computed using the Wick theorem with noninteracting single-particle Green's functions
\begin{equation}\begin{aligned}
\langle c_{\mathbf{k}\sigma}(\tau)c_{\mathbf{k}^{\prime}\sigma}^+(0)\rangle &= \delta_{\mathbf{k}\mathbf{k}^{\prime}}e^{-\tau\varepsilon_{\mathbf{k},\sigma}}[1-f_{\rm FD}(\varepsilon_{\mathbf{k},\sigma})], \\
\langle c_{\mathbf{k}\sigma}(0)c_{\mathbf{k}^{\prime}\sigma}^+(\tau)\rangle &= \delta_{\mathbf{k}\mathbf{k}^{\prime}}e^{+\tau\varepsilon_{\mathbf{k},\sigma}}[1-f_{\rm FD}(\varepsilon_{\mathbf{k},\sigma})], \\
\langle c_{\mathbf{k}\sigma}^+(0)c_{\mathbf{k}^{\prime}\sigma}(\tau)\rangle &= \delta_{\mathbf{k}\mathbf{k}^{\prime}}e^{-\tau\varepsilon_{\mathbf{k},\sigma}}f_{\rm FD}(\varepsilon_{\mathbf{k},\sigma}), \\
\langle c_{\mathbf{k}\sigma}^+(\tau)c_{\mathbf{k}^{\prime}\sigma}(0)\rangle &= \delta_{\mathbf{k}\mathbf{k}^{\prime}}e^{+\tau\varepsilon_{\mathbf{k},\sigma}}f_{\rm FD}(\varepsilon_{\mathbf{k},\sigma}),
\label{eq:A5}
\end{aligned}\end{equation}
where $f_{\rm FD}(\varepsilon)=(1+e^{\beta\varepsilon})^{-1}$ is the Fermi-Dirac distribution function, and $\varepsilon_{\mathbf{k},\sigma}$ is the kinetic energy dispersion in Eq.~(\ref{eq:EkDisp}). Thus, we reach the final result of 
\begin{equation}\begin{aligned}
&S_{zz}(\mathbf{q},\tau) \\
&=\frac{1}{4N_s}\sum_{\mathbf{k},\sigma}\big(\langle c_{\mathbf{k},\sigma}^+(\tau)c_{\mathbf{k},\sigma}(0)\rangle\langle c_{\mathbf{k+q},\sigma}(\tau)c_{\mathbf{k+q},\sigma}^+(0)\rangle \big) \\
&=\frac{1}{4N_s}\sum_{\mathbf{k},\sigma}\frac{e^{\varepsilon_{\mathbf{k},\sigma}}}{1+e^{\beta\varepsilon_{\mathbf{k},\sigma}}}\frac{e^{(\beta-\tau)\varepsilon_{\mathbf{k}+\mathbf{q},\sigma}}}{1+e^{\beta\varepsilon_{\mathbf{k}+\mathbf{q},\sigma}}}.
\label{eq:A6}
\end{aligned}\end{equation}
Then the longitudinal spin susceptibility $\chi^{zz}(\mathbf{q})$ can be obtained from the Fourier transform of $S_{zz}(\mathbf{q},\tau)$ as
\begin{equation}\begin{aligned}
\chi^{zz}(\mathbf{q}) 
&= \chi^{zz}(\mathbf{q},i\omega=0) = \int_{0}^{\beta}S_{zz}(\tau,\mathbf{q})e^{i\omega\tau}\mathrm{d}\tau \\
&=\frac{1}{4N_s}\sum_{\mathbf{k},\sigma}\frac{f_{\rm FD}(\varepsilon_{\mathbf{k}+\mathbf{q},\sigma}) - f_{\rm FD}(\varepsilon_{\mathbf{k},\sigma})}{\varepsilon_{\mathbf{k},\sigma} - \varepsilon_{\mathbf{k}+\mathbf{q},\sigma}}.
\label{eq:A7}
\end{aligned}\end{equation}
Similarly, the transverse spin susceptibility $\chi^{xy}(\mathbf{q})$ can be obtained as
\begin{equation}\begin{aligned}
\chi^{xy}(\mathbf{q})
= \frac{1}{2N_s}\sum_{\mathbf{k},\sigma}\frac{f_{\rm FD}(\varepsilon_{\mathbf{k}+\mathbf{q},\bar{\sigma}}) - f_{\rm FD}(\varepsilon_{\mathbf{k},\sigma})}{\varepsilon_{\mathbf{k},\sigma} - \varepsilon_{\mathbf{k}+\mathbf{q},\bar{\sigma}}},
\label{eq:A8}
\end{aligned}\end{equation}
where $\bar{\sigma}=-\sigma$ means the opposite spin index of $\sigma$.

\begin{figure}[h]
\centering
\includegraphics[width=0.98\columnwidth]{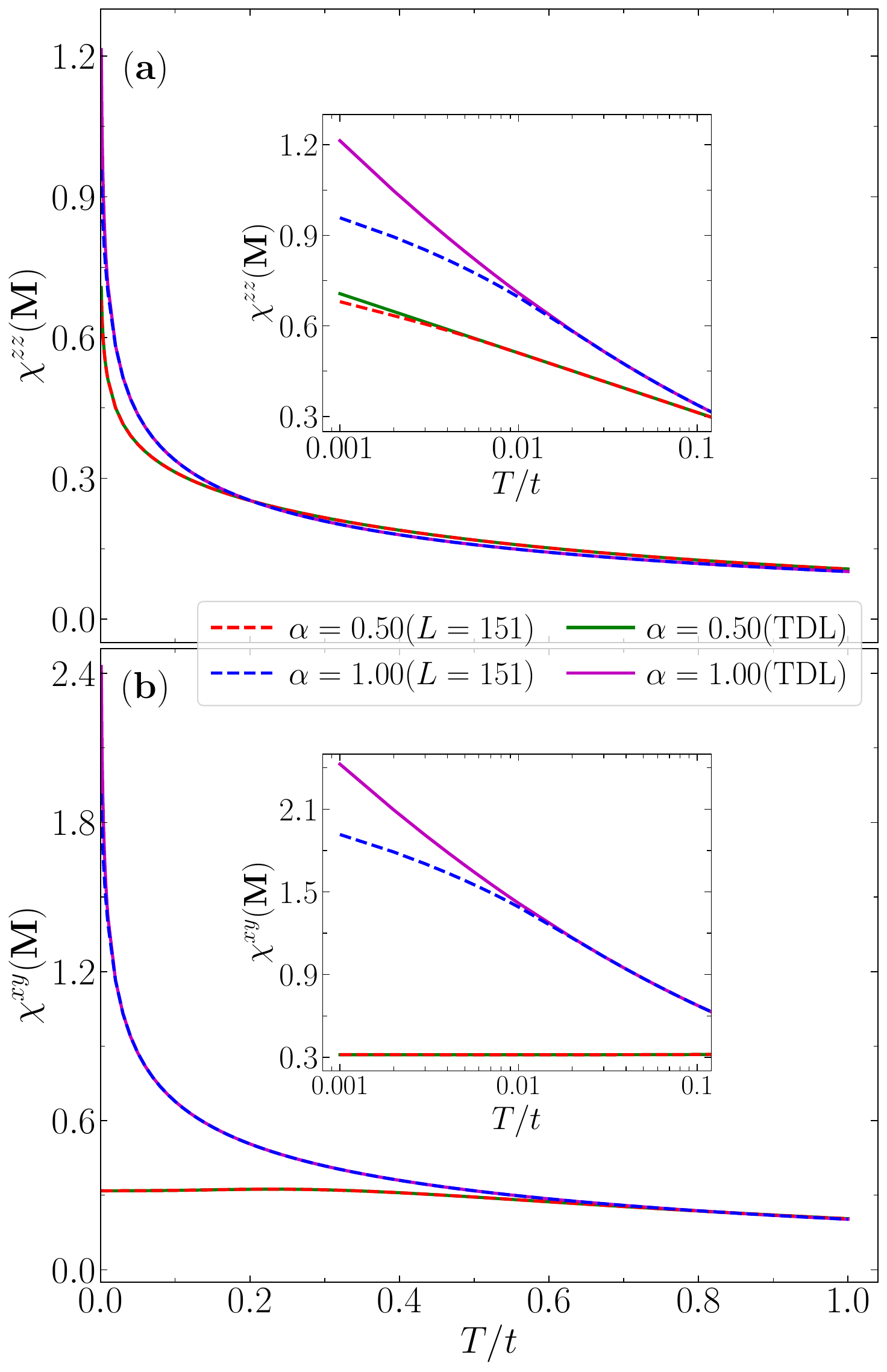}
\caption{The spin susceptibilities a function of temperature for the noninteracting Hamiltonian in Eq.~(\ref{eq:Model}) at half filling ($\mu=0$), with $\alpha=1$ and $\alpha=0.5$. [(a) and (b)] The longitudinal and transverse components, $\chi^{zz}(\mathbf{M})$ and $\chi^{xy}(\mathbf{M})$, respectively. The insets plots the zoom-in results for $T/t\le 0.1$. Results from both $L=151$ and TDL are plotted. }
\label{fig:A1_ChiSpin}
\end{figure}

For the isotropic case ($\alpha=1$), we have $\varepsilon_{\mathbf{k},\uparrow}=\varepsilon_{\mathbf{k},\downarrow}=-2t(\cos k_x + \alpha\cos k_y)$, which simply results in $\chi^{xy}(\mathbf{q})=2\chi^{zz}(\mathbf{q})$. In contrast, for the anisotropic case ($\alpha\ne1$), the twice relation between $\chi^{xy}(\mathbf{q})$ and $\chi^{zz}(\mathbf{q})$ does not hold anymore. Nevertheless, we can prove that, for arbitrary $\alpha$, $\chi^{zz}(\mathbf{q})$ at $\mathbf{q}=\mathbf{M}=(\pi,\pi)$ has a logarithmic divergence as $\propto -\ln T$. We proceed with the simple relations as $\varepsilon_{\mathbf{k},\sigma} - \varepsilon_{\mathbf{k}+\mathbf{M},\sigma} = 2\varepsilon_{\mathbf{k},\sigma}$ and 
\begin{equation}\begin{aligned}
f_{\rm FD}(\varepsilon_{\mathbf{k}+\mathbf{M},\sigma}) - f_{\rm FD}(\varepsilon_{\mathbf{k},\sigma}) 
&= \frac{e^{\beta\varepsilon_{\mathbf{k},\sigma}(\mathbf{k})}-1}{e^{\beta\varepsilon_{\mathbf{k},\sigma}(\mathbf{k})}+1}  \\
&= \tanh\frac{\beta\varepsilon_{\mathbf{k},\sigma}}{2},
\label{eq:A9}
\end{aligned}\end{equation}
and thus $\chi^{zz}(\mathbf{M})$ can be written as
\begin{equation}\begin{aligned}
\chi^{zz}(\mathbf{M})=\frac{1}{4N_s}\sum_{\mathbf{k},{\sigma}}\tanh\Big(\frac{\beta\varepsilon_{\mathbf{k},\sigma}}{2}\Big)\frac{1}{2\varepsilon_{\mathbf{k},\sigma}}.
\label{eq:A10}
\end{aligned}\end{equation}
With the local density of states (LDOS) $\rho_{\sigma}(\varepsilon)=N_s^{-1}\sum_{\mathbf{k}}\delta(\varepsilon-\varepsilon_{\mathbf{k},\sigma})$, the summation in Eq.~(\ref{eq:A10}) can be transformed into an integral over the energy as
\begin{equation}\begin{aligned}
\chi^{zz}(\mathbf{M})=\frac{1}{4}\sum_{\sigma}\int_{-\infty}^{+\infty}\rho_{\sigma}(\varepsilon)\tanh{\Big(\frac{\beta\varepsilon}{2}\Big)}\frac{1}{2\varepsilon}\mathrm{d}\varepsilon.
\label{eq:A11}
\end{aligned}\end{equation}
As shown in Sec.~\ref{sec:TheModel}, the LDOS satisfies $\rho_{\uparrow}(\varepsilon)=\rho_{\downarrow}(\varepsilon)=\rho(\varepsilon)$ for arbitrary $\alpha$, and $\rho(\varepsilon)$ is nonzero only for $\varepsilon\in[-W/2,+W/2]$ with $W$ representing the bandwidth. Thus we can obtain
\begin{equation}\begin{aligned}
\chi^{zz}(\mathbf{M})=\frac{1}{2}\int_{-W/2}^{+W/2}\rho(\varepsilon)\tanh{\Big(\frac{\beta\varepsilon}{2}\Big)}\frac{1}{2\varepsilon}\mathrm{d}\varepsilon.
\label{eq:A12}
\end{aligned}\end{equation}
We then concentrate on the calculation of this integral at the low-temperature limit ($\beta\to\infty$) and linearize the $\tanh(x)$ function such that $\tanh(x)$$\sim$$x$ for $-1<x<+1$, $\tanh(x)$$\sim$$-1$ for $x<-1$, and $\tanh(x)$$\sim$$+1$ for $x>+1$. We further approximate $\rho(\varepsilon)$ in the integral with $\rho(\varepsilon_F)$ as the LDOS at Fermi energy. Combining these simplifications, $\chi^{zz}(\mathbf{M})$ can be approximated as 
\begin{equation}\begin{aligned}
&\chi^{zz}(\mathbf{M}) \\
&\approx \bigg[-\int^{-2/\beta}_{-W/2}\,\mathrm{d}\varepsilon+\int^{+2/\beta}_{-2/\beta}\frac{\beta\varepsilon}{2}\,\mathrm{d}\varepsilon+\int_{+2/\beta}^{+W/2}\,\mathrm{d}\varepsilon\bigg]\frac{\rho(\varepsilon)}{4\varepsilon} \\
&\approx \rho(\varepsilon_F)\bigg[-\int^{-2/\beta}_{-W/2}\,\mathrm{d}\varepsilon+\int_{+2/\beta}^{+W/2}\,\mathrm{d}\varepsilon\bigg]\frac{1}{4\varepsilon} \\
&=\frac{\rho(\varepsilon_F)}{2}\ln({\frac{W}{4T}}) \propto -\ln T.
\label{eq:A13}
\end{aligned}\end{equation}
Since $\rho(\varepsilon_F)$ diverges for $\alpha=1$ and it is finite for $\alpha\ne 1$ [see Fig.~\ref{fig:ModelDemon}(c)], we can reach the conclusion that $\chi^{zz}(\mathbf{M})$ diverges logarithmically as the temperature $T=1/\beta$ goes to zero. For the isotropic case ($\alpha=1$), the relation $\chi^{xy}(\mathbf{q})=2\chi^{zz}(\mathbf{q})$ means that $\chi^{xy}(\mathbf{q})$ also has the logarithmic divergence. However, for the anisotropic case ($\alpha\ne 1$), the divergence of $\chi^{xy}(\mathbf{M})$ disappears, and instead it only saturates to a finite value as approaching $T=0$, due to the fact $\varepsilon_{\mathbf{k},\uparrow}\ne \varepsilon_{\mathbf{k},\downarrow}$. 

In Fig.~\ref{fig:A1_ChiSpin}, we plot the numerical results of $\chi^{zz}(\mathbf{M})$ and $\chi^{xy}(\mathbf{M})$, obtained from Eqs.~(\ref{eq:A7}) and (\ref{eq:A8}), for $\alpha=1$ and $\alpha=0.5$. For the $L=151$ system, we evaluate the integral in Eqs.~(\ref{eq:A7}) and (\ref{eq:A8}) using discrete summation of $\mathbf{k}$ points, while the numerical integral (such as Gauss-Legendre algorithm) is applied to replace the $\mathbf{k}$ summation for the TDL case. The logarithmic divergence of $\chi^{zz}(\mathbf{M})$ for both $\alpha=1$ and $\alpha=0.5$ is clear, as shown in the inset of Fig.~\ref{fig:A1_ChiSpin}(a). However, as shown in Fig.~\ref{fig:A1_ChiSpin}(b), $\chi^{xy}(\mathbf{M})$ only diverges for $\alpha=1$, and it clearly converges to a finite constant for $\alpha=0.5$ as approaching $T=0$.

\section{From the anisotropic Hubbard model to the XXZ model in the large-$U$ limit}
\label{sec:AppendixB}

In strongly interacting limit ($U\gg t$), the 2D modified Hubbard model in Eq.~(\ref{eq:Model}) at half filling reduces to the spin-$1/2$ XXZ model in Eq.~(\ref{eq:XXZModel}). Here we present the proof for this mapping. 

At half filling, the fermion filling is $n=1$ with $\mu=0$, so we can rewrite the model Hamiltonian in Eq.~(\ref{eq:Model}) as:
\begin{align}
\label{eq:the rewritten Hamiltonian}
\hat{H}= -\sum_{\mathbf{i},\mathbf{v},\sigma}t_{\mathbf{v} \sigma}\left(c_{\mathbf{i}, \sigma}^+ c_{\mathbf{i}+\mathbf{v}, \sigma}+c_{\mathbf{i}+\mathbf{v}, \sigma}^+ c_{\mathbf{i}, \sigma}\right)+U\sum_{\mathbf{i}}\hat{n}_{\mathbf{i}, \uparrow} \hat{n}_{\mathbf{i}, \downarrow}.
\end{align}
We take the special case of two lattice sites under this condition, and we expand the Hamiltonian in the corresponding Hilbert space as a matrix,
\begin{align}
\label{eq:B2}   
\mathbf{H} &= \begin{pmatrix}  
0 & 0 & -t_{\downarrow} & -t_{\uparrow} \\    
0 & 0 & +t_{\uparrow} & +t_{\downarrow} \\
-t_{\downarrow} & +t_{\uparrow} & U & 0 \\
-t_{\uparrow} & +t_{\downarrow} & 0 & U
\end{pmatrix},
\quad
\begin{array}{@{}c@{\;}c@{\;}c@{}c@{\;}c@{}}
| & \uparrow   & , & \downarrow & \rangle \\    
| & \downarrow & , & \uparrow   & \rangle \\   
| & \uparrow\downarrow & , & 0 & \rangle \\
| & 0 & , & \uparrow\downarrow & \rangle,
\end{array}
\end{align}
with the four Fock basis also illustrated. In Eq.~(\ref{eq:B2}), $t_{\uparrow}$ and $t_{\downarrow}$ are the hopping strengths of spin-up and -down electrons, respectively. Under the condition of $U\gg t$, the double occupancy vanishes. Thus, our goal is to project the above Hamiltonian matrix into the new Hilbert space consisting of only two basis, $|\uparrow, \downarrow\rangle$ and $|\downarrow, \uparrow\rangle$.

We first rewrite the Hamiltonian matrix Eq.~(\ref{eq:B2}) with $2\times2$ submatrices as 
\begin{align}
\label{eq:B3}
\mathbf{H} &= \begin{pmatrix}
\mathbf{H}_{00} & \mathbf{V}_{01}  \\
\mathbf{V}_{10} & \mathbf{H}_{11}
\end{pmatrix}.
\end{align}
With $U \gg t$, the sector containing the $U$ term (the $\mathbf{H}_{11}$ submatrix) belongs to the high-energy regime and should be projected out in theoretical treatments. Our primary focus is on the $\mathbf{H}_{00}$ sector, and we can reduce the dimensionality of the Hamiltonian through the downfolding method. We proceed with the eigenvalue equation for $\mathbf{H}$ as
\begin{align}   
\begin{pmatrix}
\mathbf{H}_{00} & \mathbf{V}_{01}  \\
\mathbf{V}_{10} & \mathbf{H}_{11}
\end{pmatrix}
\begin{pmatrix}
|\psi_{00}\rangle \\
|\psi_{11}\rangle
\end{pmatrix}=\varepsilon
\begin{pmatrix}
|\psi_{00}\rangle \\
|\psi_{11}\rangle
\end{pmatrix}.
\end{align}
Expand the matrix-vector multiplication, we can obtain
\begin{equation}\begin{aligned}
\label{eq:B5}
\mathbf{H}_{00}|\psi_{00}\rangle+\mathbf{V}_{01}|\psi_{11}\rangle&=\varepsilon|\psi_{00}\rangle,\\
\mathbf{V}_{10}|\psi_{00}\rangle+\mathbf{H}_{11}|\psi_{11}\rangle&=\varepsilon|\psi_{11}\rangle.
\end{aligned}\end{equation}
From the second equation of Eq.~(\ref{eq:B5}), we can formally obtain $|\psi_{11}\rangle=(\varepsilon-\mathbf{H}_{11})^{-1}\mathbf{V}_{10}|\psi_{00}\rangle$, and via substituting it into the first equation, we can reach
\begin{align}
\label{eq:B6}
[\mathbf{H}_{00}+\mathbf{V}_{01}(\varepsilon-\mathbf{H}_{11})^{-1}\mathbf{V}_{10}]|\psi_{00}\rangle&=\varepsilon|\psi_{00}\rangle.
\end{align}
Thus, we arrive at the effective Hamiltonian matrix for the $\mathbf{H}_{00}$ subspace as  
\begin{align}
\label{eq:B7}
\mathbf{H}_{\rm eff}(\varepsilon) 
= [\mathbf{H}_{00}+\mathbf{V}_{01}(\varepsilon-\mathbf{H}_{11})^{-1}\mathbf{V}_{10}],
\end{align}
which depends on the energy eigenvalue $\varepsilon$. Under the condition of $U \gg t$, the lowest eigenvalue of the $\mathbf{H}$ matrix should be $\varepsilon=0$. Therefore, we can obtain
\begin{equation}\begin{aligned}
\mathbf{H}_{\rm eff} &=\begin{pmatrix}
-t_{\downarrow} & -t_{\uparrow} \\
t_{\uparrow} & t_{\downarrow}
\end{pmatrix}
\begin{pmatrix}
-U  &  0\\
0  & -U
\end{pmatrix}^{-1}
\begin{pmatrix}
-t_{\downarrow} & t_{\uparrow} \\
-t_{\uparrow} & t_{\downarrow}
\end{pmatrix} \\
&=-\frac{2t_{\uparrow}t_{\downarrow}}{U} \begin{pmatrix}
\frac{t_{\uparrow}^2+t_{\downarrow}^2}{2t_{\uparrow}t_{\downarrow}} & -1\\
-1 & \frac{t_{\uparrow}^2+t_{\downarrow}^2}{2t_{\uparrow}t_{\downarrow}}
\end{pmatrix}.
\end{aligned}\end{equation}
Since this effective Hamiltonian matrix is expressed in the $\mathbf{H}_{00}$ sector, with the Fock basis as $|\phi_1\rangle=|\uparrow,\downarrow\rangle=c^+_{2\downarrow}c^+_{1\uparrow}|0\rangle)$ and $|\phi_2\rangle=|\downarrow,\uparrow\rangle = c^+_{2\uparrow}c^+_{1\downarrow}|0\rangle)$. Thus we can restore the Hamiltonian operator as
\begin{equation}\begin{aligned}
\hat{H}_{\rm eff}
&= \sum_{a=1}^{2}\sum_{b=1}^{2}|\phi_a\rangle\langle\phi_a|\hat{H}_{\rm eff}|\phi_b\rangle\langle\phi_b| \\
&= \sum_{a=1}^{2}\sum_{b=1}^{2}|\phi_a\rangle(\mathbf{H}_{\rm eff})_{ab}\langle\phi_b|.
\end{aligned}\end{equation}
Using the operator expressions of $|\phi_1\rangle$ and $|\phi_2\rangle$, we can obtain
\begin{equation}\begin{aligned}
\label{eq:B10}
&\hat{H}_{\rm eff}
= -\frac{2t_{\uparrow}t_{\downarrow}}{U} \Big(-c_{2\downarrow}^+c^+_{1\uparrow}c_{1\downarrow}c_{2\uparrow} - c_{2\uparrow}^+c_{1\downarrow}^+c_{1\uparrow}c_{2\downarrow} \\
&+\frac{t_{\uparrow}^{2} + t_{\downarrow}^{2}}{2t_{\uparrow}t_{\downarrow}} c_{2\uparrow}^+c_{1\downarrow}^+c_{1\downarrow}c_{2\uparrow} + \frac{t_{\uparrow}^{2} + t_{\downarrow}^{2}}{2t_{\uparrow}t_{\downarrow}} c_{2\downarrow}^+c_{1\uparrow}^+c_{1\uparrow}c_{2\downarrow} \Big).
\end{aligned}\end{equation}

Then we can transform the creation and annihilation operators of fermions to spin operators as
\begin{equation}\begin{aligned}
\label{eq:B11}
\hat{H}_{\rm eff}
=\frac{{4t_{\uparrow}}t_{\downarrow}}{U} \Big[ \frac{t_{\uparrow}^{2} + t_{\downarrow}^{2}}{2t_{\uparrow}t_{\downarrow}} \Big( \hat{S}_1^z\hat{S}_2^z - \frac{\hat{n}_1\hat{n}_2}{4} \Big) + \hat{S}_1^x\hat{S}_2^x + \hat{S}_1^y\hat{S}_2^y \Big],
\end{aligned}\end{equation}
with $\hat{S}^{z}_{\mathbf{i}}=(\hat{n}_{\mathbf{i}\uparrow}-\hat{n}_{\mathbf{i}\downarrow})/2$, $\hat{S}^{x}_{\mathbf{i}}=(c_{\mathbf{i}\uparrow}^+c_{\mathbf{i}\downarrow}+c_{\mathbf{i}\downarrow}^+c_{\mathbf{i}\uparrow})/2$, $\hat{S}^{y}_{\mathbf{i}}=(c_{\mathbf{i}\uparrow}^+c_{\mathbf{i}\downarrow}-c_{\mathbf{i}\downarrow}^+c_{\mathbf{i}\uparrow})/(2i)$, and $\hat{n}_{\mathbf{i}}=\hat{n}_{\mathbf{i}\uparrow}+\hat{n}_{\mathbf{i}\downarrow}$. Considering that the hopping strengths satisfy $t_{\mathbf{x}\uparrow} = t_{\mathbf{y}\downarrow} = t$ and $t_{\mathbf{x}\downarrow}=t_{\mathbf{y}\uparrow}=\alpha t$ in Eq.~(\ref{eq:Model}), the coefficients in Eq.~(\ref{eq:B11}) can be simplified as $4t_{\uparrow}t_{\downarrow}/U=4\alpha t^2/U$ and $(t_{\uparrow}^{2}+t_{\downarrow}^{2})/(2t_{\uparrow}t_{\downarrow})=(1+\alpha^2)/(2\alpha)$, regardless of whether the two sites in Eq.~(\ref{eq:B11}) lay in the $x$ or $y$ direction. Then the model Hamiltonian in Eq.~(\ref{eq:B11}) can be generalized to the full lattice with arbitrary number of sites as
\begin{equation}\begin{aligned}
\label{eq:B12}
\hat{H}_{\rm eff} = \sum_{\langle\mathbf{i}\mathbf{j}\rangle}\Big[ J_{xx}\big(\hat{S}^x_{\mathbf{i}}\hat{S}^x_{\mathbf{j}}+\hat{S}^y_{\mathbf{i}}\hat{S}^y_{\mathbf{j}}\big)+J_{z}\Big(\hat{S}^z_{\mathbf{i}}\hat{S}^z_{\mathbf{j}} - \frac{\hat{n}_{\mathbf{i}}\hat{n}_{\mathbf{j}}}{4}\Big)\Big],
\end{aligned}\end{equation}
with the coupling coefficients as
\begin{equation}\begin{aligned}
J_{xx}=\frac{4\alpha t^2}{U} \hspace{1.0cm} J_z=\frac{2(1+\alpha^2)t^2}{U}.
\end{aligned}\end{equation}
Furthermore, since the electrons are fully localized in the model in Eq.~(\ref{eq:B12}), the operator $\hat{n}_{\mathbf{i}}\hat{n}_{\mathbf{j}}$ can be taken as the constant $1/4$ (at half filling). So Eq.~(\ref{eq:B12}) finally reduces to
\begin{equation}\begin{aligned}
\label{eq:B14}
\hat{H}_{\rm XXZ} = \sum_{\langle\mathbf{i}\mathbf{j}\rangle}\Big[ J_{xx}\big(\hat{S}^x_{\mathbf{i}}\hat{S}^x_{\mathbf{j}}+\hat{S}^y_{\mathbf{i}}\hat{S}^y_{\mathbf{j}}\big)+J_{z}\hat{S}^z_{\mathbf{i}}\hat{S}^z_{\mathbf{j}}\Big],
\end{aligned}\end{equation}
which is exactly the spin-$1/2$ XXZ model in Eq.~(\ref{eq:XXZModel}).

\section{The general formalism of finite-temperature AFQMC}
\label{sec:AppendixC}

We consider a general lattice model with two-body interactions of spin-$1/2$ fermions, described by the model Hamiltonian $\hat{H}=\hat{H}_0+\hat{H}_I$, where $\hat{H}_0$ is the noninteracting part and $\hat{H}_I$ represents fermion-fermion interactions. $\hat{H}_0$ can be written as $\sum_{ij,\sigma\sigma^{\prime}}(\mathbf{H}_0)_{i\sigma,j\sigma^{\prime}} c_{i\sigma}^+c_{j\sigma^{\prime}}$($\sigma=\uparrow,\downarrow$ is the spin index), with $c_{i\sigma}^+$ ($c_{i\sigma}$) as creation (annihilation) operator on site $i$. The chemical potential term is included in $\hat{H}_0$ implicitly. Given a finite-size lattice (or basis) with $N_s$ sites, $\mathbf{\mathbf{H}_0}=\{(\mathbf{H}_0)_{i\sigma,j\sigma^{\prime}}\}$ is $2N_s\times 2N_s$ hopping matrix. If there are no spin-flip terms, then $\mathbf{H}_0$ is block diagonal with respect to spin species: $\mathbf{H}_0=\text{Diag}(\mathbf{H}_0^{\uparrow},\mathbf{H}_0^{\downarrow})$ with $\mathbf{H}_0^{\sigma}$ as $N_s\times N_s$ matrix. This is actually the situation of the model (\ref{eq:Model}).

The DQMC method deals with the partition function of the system as
\begin{equation}\begin{aligned}
\label{eq:PartitionFunc0}
Z=\text{Tr}(e^{-\beta\hat{H}}) = \text{Tr}(\underbrace{e^{-\Delta\tau\hat{H}}\cdots e^{-\Delta\tau\hat{H}}e^{-\Delta\tau\hat{H}}}_M),
\end{aligned}\end{equation}
where $\Delta\tau=\beta/M$ and $M$ is the number of imaginary-time slices. For a small $\Delta\tau$, the Trotter-Suzuki decomposition, such as the symmetric one in Eq.~(\ref{eq:SymTrot}), and the HS transformation, generally expressed as $e^{-\Delta\tau\hat{H}_I} = \sum_{\mathbf{x}} p(\mathbf{x})\hat{B}_I(\mathbf{x})$ [see Eqs.~(\ref{eq:HSspinDecomp}) and (\ref{eq:HScharge}) for Hubbard interaction], are applied to transform the many-body propagator $e^{-\Delta\tau\hat{H}}$ in Eq.~(\ref{eq:PartitionFunc0}) into single-particle operators expressed as free fermions coupled to auxiliary fields $\mathbf{x}=(x_{1},x_{2},\cdots,x_{N_f})$ with $N_f$ (which is equal to $N_s$ for Hubbard interaction) components. The $\hat{B}_I(\mathbf{x})$ operator can be expressed as $\hat{B}_I(\mathbf{x}) = \exp\big\{ \sum_{ij,\sigma\sigma^{\prime}}[\mathbf{H}_I(\mathbf{x})]_{i\sigma,j\sigma^{\prime}} c_{i\sigma}^+c_{j\sigma^{\prime}}\big\}$, where $\mathbf{H}_I(\mathbf{x})=\{[\mathbf{H}_I(\mathbf{x})]_{i\sigma,j\sigma^{\prime}}\}$ is a $2N_s\times 2N_s$ Hermitian or anti-Hermitian matrix. Combining with the kinetic propagator $\hat{B}_{K/2}=e^{-\Delta\tau\hat{H}_0/2}$ which has no dependence on imaginary time, we can rewrite the propagator $e^{-\Delta\tau\hat{H}(\ell)}$ at the $\ell$th time slice as
\begin{equation}\begin{aligned}
\label{eq:HSTransform}
e^{-\Delta\tau\hat{H}}
= \sum_{\mathbf{x}_{\ell}}p(\mathbf{x}_{\ell})\hat{B}_{\ell} + \mathcal{O}[(\Delta\tau)^2],
\end{aligned}\end{equation}
where $\hat{B}_{\ell}=\hat{B}_{K/2}\hat{B}_I(\mathbf{x}_{\ell})\hat{B}_{K/2}$ [with the Trotter-Suzuki decomposition in Eq.~(\ref{eq:SymTrot})]. Applying this to all time slices in Eq.~(\ref{eq:PartitionFunc0}), we arrive at $Z\simeq \sum_{\mathbf{X}}P(\mathbf{X})\text{Tr}( \hat{B}_M\cdots\hat{B}_2\hat{B}_1 )$, where $P(\mathbf{X})=\prod_{\ell=1}^{M}p(\mathbf{x}_{\ell})$ is a probability density function and the auxiliary-field configuration $\mathbf{X}=\{\mathbf{x}_M,\cdots,\mathbf{x}_2,\mathbf{x}_1\}$ contains $MN_f$ components. Since $\hat{B}_{\ell}$ is a single-particle propagator, the trace in the partition function can now be evaluated explicitly to yield
\begin{eqnarray}
\label{eq:PartitionFunc1}
Z \simeq \sum_{\mathbf{X}}P(\mathbf{X})\text{det}(\mathbf{I}_{2N_s}+\mathbf{B}_M\cdots\mathbf{B}_2\mathbf{B}_1),
\end{eqnarray}
where $\mathbf{B}_{\ell}=\mathbf{B}_{K/2}\mathbf{B}_I(\mathbf{x}_\ell)\mathbf{B}_{K/2}$ with $\mathbf{B}_I(\mathbf{x}_\ell)=e^{\mathbf{H}_I(\mathbf{x}_{\ell})}$ and $\mathbf{B}_{K/2}=e^{-\Delta\tau\mathbf{H}_0/2}$ as the propagation matrix of the interaction part and the kinetic part, respectively. We set the matrix $\mathbf{M}(\mathbf{X})=\mathbf{I}_{2N_s}+\mathbf{B}_M\cdots\mathbf{B}_2\mathbf{B}_1$. In Eq.~(\ref{eq:PartitionFunc1}), $W(\mathbf{X})=P(\mathbf{X})\text{det}[\mathbf{M}(\mathbf{X})]$ is the ``weight" of auxiliary field configuration $\mathbf{X}$. With this, we have formally mapped the study of a $D$ dimensional quantum systems for fermions with $N_s$ lattice sites into solving a $(D+1)$ classical systems with $MN_f$ classical variables (or sites on a space-time lattice).

\begin{figure}[t]
\centering
\includegraphics[width=0.99\columnwidth]{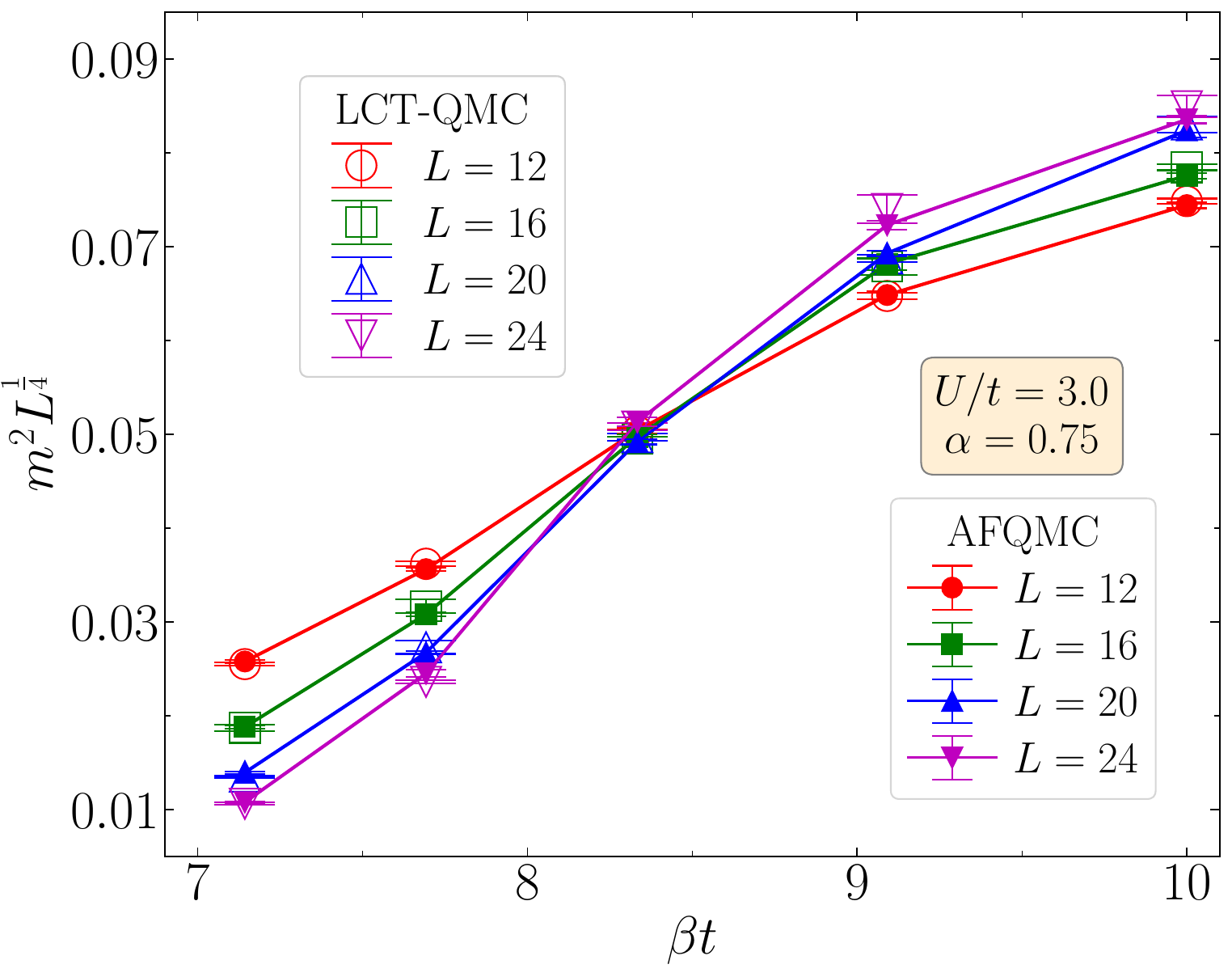} 
\caption{Benchmark between AFQMC and LCT-QMC on the numerical results of the rescaled mean-square magnetization $m^2L^{1/4}$ versus the inverse temperature $\beta t$ from $L=12,16,20,24$, for the 2D modified Hubbard model in Eq.~(\ref{eq:Model}) at half filling with $\alpha=0.75$ and $U/t=3.0$. The LCT-QMC results are from Ref.~\onlinecite{Gukelberger2017}. A finite $\Delta\tau t=0.04$ is used in AFQMC calculations. The nice agreement between these two methods is clear. }
\label{fig:A2Benchmark}
\end{figure}

Regarding the model in our study in Eq.~(\ref{eq:Model}), the noninteracting part is spin decoupled, and thus $\mathbf{H}_0=\text{Diag}(\mathbf{H}_0^{\uparrow},\mathbf{H}_0^{\downarrow})$, where $\mathbf{H}_0^{\sigma}$ is $N_s\times N_s$ matrix. If we use the HS-$\hat{n}$ or HS-$\hat{s}^z$, then interaction part is also spin decoupled with $\mathbf{H}_I(\mathbf{x}_{\ell})=\text{Diag}(\mathbf{H}_I^{\uparrow},\mathbf{H}_I^{\downarrow})$ with $\mathbf{H}_I^{\sigma}$ also as $N_s\times N_s$ matrix. As a result, the single imaginary-time slice propagation matrix $\mathbf{B}_{\ell}$ is also spin decoupled as $\mathbf{B}_{\ell}=\text{Diag}(\mathbf{B}_{\ell}^{\uparrow},\mathbf{B}_{\ell}^{\downarrow})$. Thus we can easily obtain $\mathbf{M}(\mathbf{X})=\mathbf{M}_{\uparrow}(\mathbf{X})\cdot\mathbf{M}_{\downarrow}(\mathbf{X})$ and $W(\mathbf{X})=W_{\uparrow}(\mathbf{X})\cdot W_{\downarrow}(\mathbf{X})$ with $W_{\sigma}(\mathbf{X})=\det[\mathbf{M}_{\sigma}(\mathbf{X})]$, where $\mathbf{M}_{\sigma}(\mathbf{X})$ is $N_s\times N_s$ matrix. If one applies the HS-$\hat{s}^x$ or HS-$\hat{s}^y$ decomposition for the Hubbard interaction, then $\mathbf{H}_I(\mathbf{x}_{\ell})$ and $\mathbf{B}_I(\mathbf{x}_\ell)=e^{\mathbf{H}_I(\mathbf{x}_{\ell})}$ are generally spin-coupled $2N_s\times 2N_s$ matrix. The same situation follows for $\mathbf{B}_{\ell}$ and $\mathbf{M}(\mathbf{X})$ matrices. As for the computational effort, AFQMC simulations applying HS-$\hat{s}^x$ or HS-$\hat{s}^y$ increases by a factor of 2 for the propagation of kinetic matrix, and a factor of 4 for the interaction part (including the propagation and update), comparing to that using HS-$\hat{n}$ or HS-$\hat{s}^z$.

\begin{table}[h]
\caption{Summary of the Ising phase transition temperatures, $T_{\rm N}/t$, for various interaction strengths $|U|/t$, with $\alpha=0.20$ and $0.50$. }
\setlength{\tabcolsep}{14pt}
\centering  
\begin{tabular}{|c|c|c|}  
\hline
\diagbox{$|U|/t$}{$\alpha$} & $0.20$ &$0.50$\\ 
\hline  
$2.0$ & &0.0468(2)\\  
\hline  
$3.0$ & &0.1155(2)\\  
\hline  
$3.5$ & &0.1454(2)\\  
\hline  
$4.0$ &0.1662(4)&0.1677(2)\\  
\hline  
$4.5$&0.1811(3) & \\  
\hline  
$5.0$&0.1868(2)& 0.1847(2)\\  
\hline  
$5.5$&0.1845(4) &  \\  
\hline  
$6.0$&0.1769(2) & 0.1733(2)\\  
\hline  
$7.0$& &0.1546(2) \\  
\hline  
$8.0$&&0.1389(2)\\ 
\hline  
$9.0$&&0.1244(2)\\  
\hline  
$10.0$&& 0.1126(1)\\  
\hline  
$11.0$&&0.1026(2) \\ 
\hline  
$12.0$ & &0.0946(1)\\  
\hline
\end{tabular}
\label{Table:A2}
\end{table}

\begin{table}[h]
\caption{Summary of critical entropy, $\boldsymbol{s}_{\rm N}$, for the Ising phase transitions at various $|U|/t$ with $\alpha=0.5$. We present the results from both $L=8$ and $L=12$ systems. The critical entropy of the corresponding XXZ model as $(\boldsymbol{s}_{\rm N})_{\rm XXZ}=0.1760(10)$ is also included as comparison. }
\setlength{\tabcolsep}{12pt}
\centering  
\begin{tabular}{|c|c|c|c|}  
\hline
$|U|/t$ &  $L=8$ &$L=12$& $(\boldsymbol{s}_{\rm N})_{\rm XXZ}$ \\ 
\hline  
$2.0$ &0.0573(11)&0.0679(14)&\multirow{11}{*}{$0.1760(10)$}\\  
\hhline{|---|} 
$3.0$ &0.1146(9) &0.1276(12)&\\  
\hhline{|---|}   
$4.0$ &0.1916(17)&0.1960(20)&\\  
\hhline{|---|} 
$5.0$ &0.2158(20)&0.2186(18)&\\  
\hhline{|---|} 
$6.0$ &0.2076(26)&0.2104(20)&\\  
\hhline{|---|} 
$7.0$ &0.2028(24)&0.2035(30)&\\  
\hhline{|---|} 
$8.0$ &0.1973(30)&0.2017(32)&\\ 
\hhline{|---|} 
$9.0$ &0.1935(28)&0.1963(23)&\\  
\hhline{|---|} 
$10.0$&0.1892(27)&0.1940(31)&\\  
\hhline{|---|} 
$11.0$&0.1926(29)&0.1961(32)&\\ 
\hhline{|---|}  
$12.0$&0.1850(30)& &\\ 
\hline 
\end{tabular}
\label{Table:A3}
\end{table}

\begin{figure}[t]
\centering
\includegraphics[width=0.98\columnwidth]{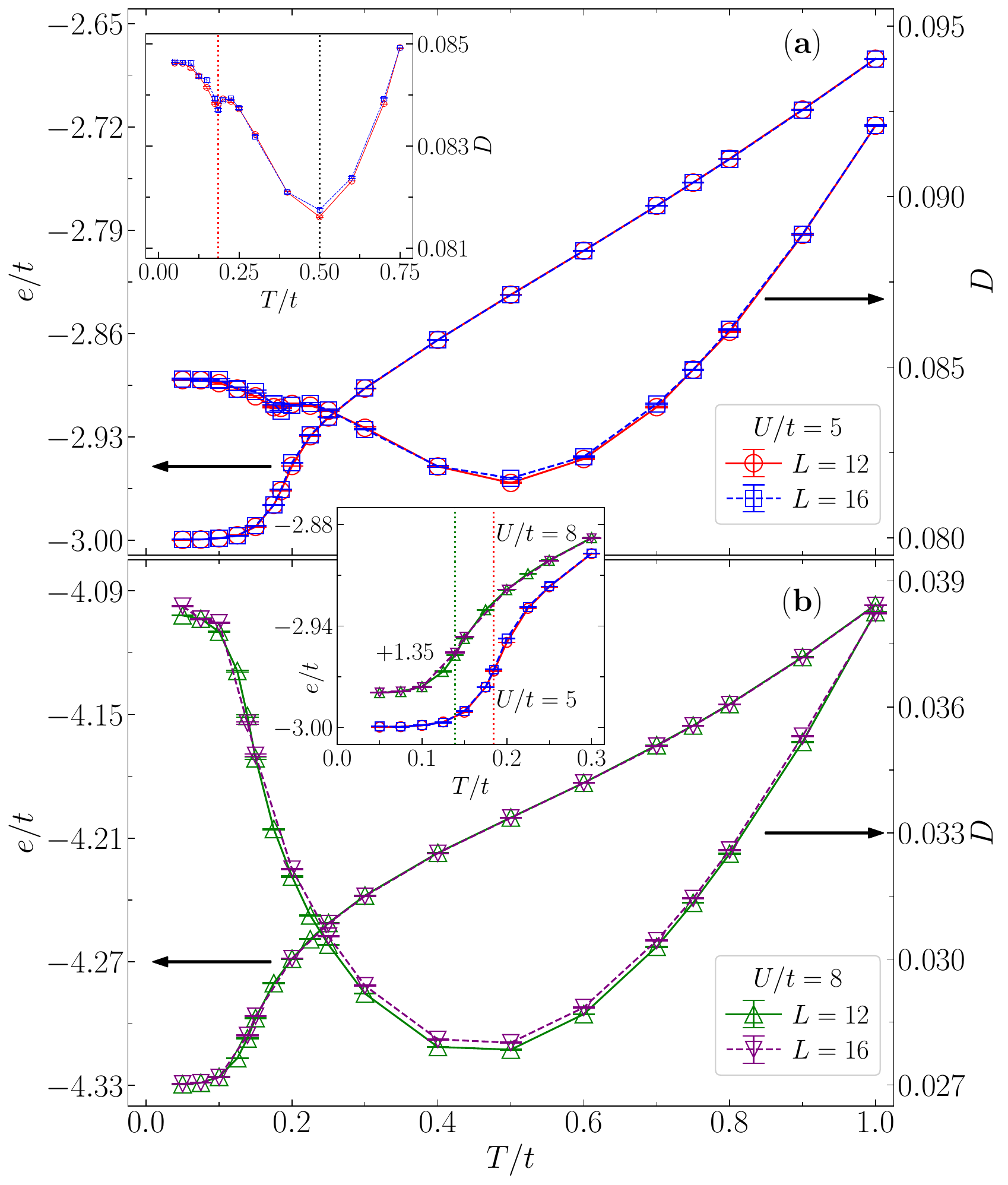}
\caption{Demonstration of the finite-size effect in the AFQMC results of the energy $e/t$ and double occupancy $D$, associated with Fig.~\ref{fig:ThermalResults}. [(a) and (b)] Results of $e/t$ and $D$ for $U/t=5$ and $U/t=8$, respectively. The insets of (a) and (b) show the zoom-in results of $D$ for $U/t=5$, and the zoom-in results of $e/t$ for both interactions, respectively. The vertical dotted lines in the inset of (a) marks the transition temperature (red) and the local minimum at $T/t=0.5$ (black). In the inset of (b), the red and green dotted lines plots $T_{\rm N}$ for $U/t=5$ and $8$, respectively. }
\label{fig:A3_finitesize}
\end{figure}

\section{Supplementary results for the main text}
\label{sec:AppendixD}

In this Appendix, we pack all the supplementary results from our AFQMC calculations for the main text of this work. They include the benchmark with the LCT-QMC results in Ref.~\onlinecite{Gukelberger2017}, the specific numbers of the critical temperature and critical entropy, and the finite-size effect for thermodynamic quantities presented in Fig.~\ref{fig:ThermalResults}.

The LCT-QMC method is free of the Trotter error, which presents in AFQMC algorithm due to the application of the Totter-Suzuki decomposition in Eq.~(\ref{eq:SymTrot}). However, this Trotter error can be easily eliminated in AFQMC calculations via extrapolating numerical results to the $\Delta\tau\to0$ limit. On the other hand, the LCT-QMC method applied in Ref.~\onlinecite{Gukelberger2017} is based on the interaction expansion, which might cause the significantly heavy computational effort to achieve satisfying results for large interactions at low temperatures. In Fig.~\ref{fig:A2Benchmark}, we plot the AFQMC results of the rescaled mean-square magnetization $m^2L^{1/4}$ on top of that from LCT-QMC method, for the 2D modified Hubbard model for the 2D modified Hubbard model in Eq.~(\ref{eq:Model}) at half filling with $\alpha=0.75$ and $U/t=3.0$. Our AFQMC calculations adopt $\Delta\tau t=0.04$ for all these results. We can observe that the results from these two methods show nice agreement for all system sizes. However, the error bars of LCT-QMC results grow significantly with increasing system size, and are also obviously larger than that of AFQMC results. Moreover, for the above parameter set, LCT-QMC predicts the transition temperature as $\beta_N t=8.5(5)$. As a comparison, our AFQMC calculations produce a more precise transition temperature as $T_N/t=0.1207(2)$, corresponding to an inverse temperature of $\beta_N t=8.29(1)$. And this result is also consistent with that from LCT-QMC, considering the statistical uncertainty. 

In Table~\ref{Table:A2}, we present the specific numbers of the critical temperature, $T_{\rm N}/t$, for the Ising phase transitions in Fig.~\ref{fig:PhaseDiagram}. The results of $T_{\rm N}/t$ for both $\alpha=0.2$ and $0.5$ are included. In Table~\ref{Table:A3}, we summarize the results of the critical entropy, $\boldsymbol{s}_{\rm N}$ (in units of $k_B$), for the Ising phase transitions with $\alpha=0.5$ from both $L=8$ and $12$ systems. These results correspond to the data plotted in Fig.~\ref{fig:EntropyLines}. The critical entropy of the corresponding spin-$1/2$ XXZ model as $(\boldsymbol{s}_{\rm N})_{\rm XXZ}=0.1760(10)$ is also included. All these results in Table~\ref{Table:A2} and Table~\ref{Table:A3} apply for both repulsive ($U>0$) and attractive ($U<0$)interactions. 

In Fig.~\ref{fig:A3_finitesize}, we show more detailed results from our AFQMC simulations to demonstrate the residual finite-size effect in all the thermodynamic quantities presented in Fig.~\ref{fig:ThermalResults}. We plot the energy $e/t$ and double occupancy $D$ from $L=12$ and $16$ systems, for both $U/t=5$ and $U/t=8$. For the results of $e/t$, it is clear that the TDL is achieved mostly, and the finite-size effect can only be observed around the Ising phase transition. As a closer look, the inset of Fig.~\ref{fig:A3_finitesize}(b) illustrates that the $e/t$ tends to show more rapid increasing  on heating with increasing $L$, for both $U/t=5$ and $8$. This is actually consistent with the fact that the specific heat $C_v$ diverges at $T=T_{\rm N}$. Consequently, this finite-size effect in $e/t$ around $T=T_{\rm N}$ is likely to persist to even larger system sizes, although it is quite weak as shown in the inset of Fig.~\ref{fig:A3_finitesize}(b). For the results of $D$, we can indeed observe the difference between $L=12$ and $16$, especially for $U/t=8$. But, note that the scale of the plot for $D$ is quite small, and the difference $\Delta D=D(L=16)-D(L=12)$ is actually tiny, i.e., $\Delta D/D(L=16)\simeq 0.6\%$ at $T/t=0.4$-$0.6$ for $U/t=8$ [see Fig.~\ref{fig:A3_finitesize}(b)]. For $U/t=5$, the residual finite-size effect in $D$ is more weak. As discussed in Sec.~\ref{sec:ThermalObs}, the $D$ curve display an unobvious kink around $T=T_{\rm N}$, resulting in a local minimum at $T/t\simeq 0.185$. This point is furthermore confirmed by the results of entropy map in Fig.~\ref{fig:EntropyLines} and the entropy variation with $U/t$ at $T/t=0.1847$ in Fig.~\ref{fig:SvsU}. As shown in the inset of Fig.~\ref{fig:A3_finitesize}(a), the local minimum at $T/t\simeq 0.185$ in $D$ for $U/t=5$ becomes more prominent in $L=16$ result than that of $L=12$. This illustrate that the kink, and the resulting local minimum and maximum around $T/t\simeq 0.185$ are intrinsic behaviors rather than some unknown finite-size effect.

\bibliography{SpinNematicRef}

\begin{thebibliography}{104}
\bibitem [{XXZ()}]{XXZnote}%
  \BibitemOpen
  \href@noop {} {}\bibinfo {note} {These results are provided by Yan-Cheng Wang
  via the stochastic series expansion QMC simulation of spin-$1/2$ XXZ model.
  The interaction ratio corresponding to our case (the hopping anisotropy
  $\alpha=0.5$) is $J_{z}/J_{xx}=1.25$. The critical temperature for the AFM
  Ising phase transition from the simulation is $T_{\rm N}/J_{z}=0.569(1)$,
  with the critical entropy per site as $S/N_s=0.176(1)$.}\BibitemShut {Stop}%
\end{thebibliography}
\end{document}